\documentclass[12pt,preprint]{emulateapj}

\def\Msun{\rm M_\odot}
\def\msun{\rm M_\odot}
\newcommand\code[1]{\textsc{\MakeLowercase{#1}}}

\begin{document}
\title{$z\sim 2-9$ Galaxies magnified by the Hubble Frontier Field Clusters II:
  Luminosity Functions and Constraints on a Faint End Turnover}
\author{R.J. Bouwens}
\affiliation{Leiden Observatory, Leiden University, NL-2300 RA Leiden, Netherlands}
\author{G. Illingworth}
\affiliation{UCO/Lick Observatory, University of California, Santa Cruz, CA 95064}
\author{R.S. Ellis}
\affiliation{Department of Physics and Astronomy, University College London, Gower Street, London WC1E 6BT, UK}
\author{P. Oesch}
\affiliation{Department of Astronomy, University of Geneva, Chemin Pegasi 51, 1290 Versoix, Switzerland}
\affiliation{Cosmic Dawn Center (DAWN), Niels Bohr Institute, University of Copenhagen, Jagtvej 128, K\o benhavn N, DK-2200, Denmark}
\author{M. Stefanon}
\affiliation{Leiden Observatory, Leiden University, NL-2300 RA Leiden, Netherlands}
\begin{abstract}
We present new determinations of the rest-UV luminosity functions
(LFs) at $z=2$-9 to extremely low luminosities ($>$$-$14 mag) from a
sample of $>$2500 lensed galaxies found behind the HFF clusters.  For
the first time, we present faint-end slope results from lensed samples
that are fully consistent with blank-field results over the redshift
range $z=2$-9, while reaching to much lower luminosities than possible
from the blank-field studies.  Combining the deep lensed sample with
the large blank-field samples allows us to set the tight constraints
on the faint-end slope $\alpha$ of the $z=2$-9 UV LFs and its
evolution.  We find a smooth flattening in $\alpha$ from
$-$2.28$\pm$0.10 ($z=9$) to $-$1.53$\pm$0.03 ($z=2$) with cosmic time
(d$\alpha$/dz$=$$-$0.11$\pm$0.01), fully consistent with dark matter
halo buildup.  We utilize these new results to present new
measurements of the evolution in the $UV$ luminosity density
$\rho_{UV}$ brightward of $-$13 mag from $z\sim9$ to $z\sim2$.
Accounting for the SFR densities to faint luminosities implied by our
LF results, we find that unobscured star formation dominates the SFR
density at $z\gtrsim4$, with obscured star formation dominant
thereafter.  Having shown we can quantify the faint-end slope $\alpha$
of the LF accurately with our lensed HFF samples, we also quantify the
apparent curvature in the shape of the UV LF through a curvature
parameter $\delta$. The constraints on the curvature $\delta$ strongly
rule out the presence of a turn-over brightward of $-$13.1 mag at
$z\sim3$, $-$14.3 mag at $z\sim6$, and $-$15.5 mag at all other
redshifts between $z\sim9$ to $z\sim2$.
\end{abstract}

\section{Introduction}

One key frontier in extragalactic astronomy is the study of lower
luminosity faint galaxies in the early universe.  Lower luminosity
galaxies in the $z\geq 3$ universe are the plausible progenitors to
several different variety of stellar systems in the nearby universe.
This has included both dwarf galaxies (Weisz et al.\ 2014;
Boylan-Kolchin et al.\ 2015) and globular clusters (Bouwens et
al.\ 2017a,c; Bouwens et al.\ 2021b; Vanzella et al.\ 2017a,b, 2019,
2020).  Through resolved stellar population analyses and abundance
matching, it is possible to set constraints on the form of the $UV$ LF
at $z\geq 2$ (Weisz et al.\ 2014; Boylan-Kolchin et al.\ 2015), with
evidence found for there being a flattening in the $UV$ LF faintward
of $-13$ mag at $z\sim7$ (Boylan-Kolchin et al.\ 2015).

Characterization of lower luminosity galaxies can give us insight into
the efficiency of star formation in very low mass galaxies in the
universe.  There has been significant debate on whether this
efficiency evolves with cosmic time since an influential analysis by
Behroozi et al.\ (2013), with some studies favoring efficient early
star formation (e.g., Harikane et al.\ 2016; Marrone et al.\ 2018) and
others disfavoring it (e.g., Harikane et al.\ 2018, 2021; Stefanon et
al.\ 2021, 2022).  Lensed galaxies behind the HFF clusters allow for a
direct look into what the efficiency of star formation is in low mass
galaxies either from a direct look at the $UV$ LF (e.g., Mu{\~n}oz \&
Loeb 2011; Finlator et al.\ 2017), the star-forming main sequence
(Santini et al.\ 2018), galaxy stellar mass function (Bhatawdekar et
al.\ 2019; Kikuchihara et al. 2020; Furtak et al. 2021), or evolution
of the star-formation rate density itself to $z\sim9$-10 (Oesch et
al.\ 2015, 2018a; McLeod et al.\ 2016; Ishigaki et al.\ 2018;
Bhatawdekar et al.\ 2019).  Insight into the star formation efficiency
of the lowest mass systems at $z\sim7$-8 provide us with clues
regarding similar star formation processes in galaxies at the earliest
times ($z \geq 12$: Wise et al.\ 2014; Barrow et al.\ 2017; Harikane
et al.\ 2021), while also constraining the nature of dark matter
(e.g., Dayal et al.\ 2017; Menci et al.\ 2018).

Finally, lower luminosity galaxies have long been speculated to
provide an important contribution to cosmic reionization (e.g., Bunker
et al.\ 2004; Yan \& Windhorst 2004; Bouwens et al.\ 2007; Ouchi et
al.\ 2009; Robertson et al.\ 2013; Atek et al.\ 2015; Livermore et
al.\ 2017; Bouwens et al.\ 2017b).  As such, there has been great
interest in quantifying their prevalence in the $z\geq 6$ universe as
well as the escape fraction in these systems and their Lyman-continuum
production efficiency (e.g., Lam et al.\ 2019; Robertson 2021).  These
lower luminosity galaxies are likely important contributors to the
ionizing flux at early times.  For LFs with a faint-end slope of $-$2,
similar to what is observed at $z\geq 6$, and a turn-over in the LF
faintward of $-$12 mag, $\geq$50\% of the ionizing photons arise from
galaxies fainter than $-$16.5 mag (e.g., Bouwens 2016).  Yet $-$16.5
mag is as faint as one can probe at $z\geq 6$ in deep fields like the
Hubble Ultra Deep Field (e.g., Schenker et al.\ 2013; McLure et
al.\ 2013; Bouwens et al.\ 2015a, 2021a; Finkelstein et al.\ 2015).

The entire enterprise of directly searching for extremely low
luminosity galaxies in the early universe took a major step forwards
with the planning and execution of the ambitious 840-orbit Hubble
Frontier Fields campaign (Coe et al.\ 2015; Lotz et al.\ 2017).  This
campaign combined the power of very long exposures with the Hubble
Space Telescope with gravitational lensing from massive galaxy
clusters to probe to unprecedented flux levels in the distant
universe.  Sensitive optical and near-IR observations were obtained of
six clusters and six parallel fields, and soon complemented by
observations in the near-UV with WFC3/UVIS (Alavi et al.\ 2016), in
the K-band with HAWK-I/MOSFIRE (Brammer et al.\ 2016), in the mid-IR
with Spitzer/IRAC (Capak et al.\ 2022, in prep) as well as near-IR
grism observations (Schmidt et al.\ 2014) and optical spectroscopy
with MUSE (Karman et al.\ 2015; Caminha et al.\ 2016; Mahler et
al.\ 2018).

There were immediate attempts to take advantage of the great potential
of deep HST observations over lensing clusters to probe the faint end
of the $UV$ luminosity functions (LFs) at high redshift.  In some
early pioneering work, constraints were set on the prevalence of
lower-luminosity $z\sim6$-7 galaxies to $-$15 mag (Atek et al.\ 2014,
2015a, 2015b; Coe et al.\ 2015; Ishigaki et al.\ 2015; Laporte et
al.\ 2016) and on lower luminosity $z\sim2$ galaxies to $-$13 mag
(Alavi et al.\ 2014, 2016).  Later work on the HFF clusters identified
plausible $z\sim6$-7 sources to $-$13 mag (Kawamata et al.\ 2015;
Castellano et al.\ 2016; Livermore et al.\ 2017; Yue et al.\ 2018;
Atek et al.\ 2018; Kawamata et al.\ 2018), while also deriving
constraints on the faint-end slope $\alpha$ (Atek et al.\ 2015a,
2015b, 2018; Ishigaki et al.\ 2015, 2018; Livermore et al.\ 2017;
Bouwens et al.\ 2017b; Bhatawdekar et al.\ 2019) as well as a possible
cut-off at very faint magnitudes (Castellano et al.\ 2016; Livermore
et al.\ 2017; Bouwens et al.\ 2017b; Yue et al.\ 2018; Atek et
al.\ 2018).

Despite the great potential of the HFFs for characterizing the
faintest observable galaxies, the actual process of using lensing
magnification to characterize the faint end ($>$$-$16 mag) of the $UV$
LF is challenging, due to the impact of systematic errors on the
faint-end form of the $UV$ LF.  One of these sources of error is the
size (or surface brightness) distribution assumed for the faintest
high-redshift sources (Bouwens et al.\ 2017a; Atek et al.\ 2018).
This issue is important for quantifying the prevalence of faint
galaxies due to the impact of source size on their detectability
(Grazian et al.\ 2011; Bouwens et al.\ 2017a).  The issue was
appreciated to be especially important at the faint end due to the
surface brightness of star-forming galaxies scaling as the square root
of the luminosity for standard size-luminosity relations (Huang et
al.\ 2013; Shibuya et al.\ 2015), such that 0.001 $L^*$ galaxies would
have surface brightnesses $\sim$30$\times$ lower than for $L^*$
galaxies (Bouwens et al.\ 2017a, 2017b).

A second major source of error are uncertainties in the lensing models
themselves and the impact this has on $UV$ LF results (Bouwens et
al.\ 2017b; Atek et al.\ 2018).  Comparisons of different lensing
models can show a wide range in the predicted magnifications for
individual sources ($\sim$0.3-0.5 dex scatter), with the position of
high magnification critical lines varying by $\lesssim1''$ from one
model to another (Meneghetti et al.\ 2017; Sebesta et al.\ 2016;
Bouwens et al.\ 2017b, 2022b).  The magnification factors for sources
in the highest-magnification regions are accordingly the most
uncertain and can have a particularly large impact on the recovered
$UV$ LF.  The uncertainties are sufficient to completely wash out a 
turn-over at the faint-end of the LF (Bouwens et al.\ 2017b).

Fortunately, significant progress has been made over the last few
years, allowing us to largely overcome the aforementioned challenges.
Detailed quantitative analyses of the rest-$UV$ sizes of the faintest
and highest magnification sources (Bouwens et al.\ 2017a,c, 2022a;
Kawamata et al.\ 2018; Yang et al.\ 2022) indicate that the rest-$UV$
sizes of galaxies are much smaller than one would expect based on
extrapolation from standard size luminosity relations (Huang et
al.\ 2013; Shibuya et al.\ 2015).  The small sizes of faint sources
result in substantially smaller completeness correction than if these
sources were more extended (Bouwens et al.\ 2017b, 2022a; Kawamata et
al.\ 2018).  Similarly, use of the median magnification from multiple
public lensing models (Livermore et al.\ 2017; Bouwens et al.\ 2017b;
Bhatawdekar et al.\ 2019; Bouwens et al.\ 2022b) and forward modeling
(Bouwens et al.\ 2017b; Atek et al.\ 2018) provide us with a very
effective way of accounting for the uncertainties in lensing models
for specific clusters.  By creating mock data sets on the basis of
candidate LFs and specific lensing models and interpreting the
observations using a median of the other magnification maps, one can
replicate the LF recovery process and arrive at realistic
uncertainties on the overall shape of the recovered LF.  This is
illustrated both by Bouwens et al.\ (2017b) and by Atek et
al.\ (2018).

Making use of these advances, Bouwens et al.\ (2017b) were able to
leverage the HFF data and derive faint-end slopes to the $UV$ LF at
$z\sim6$ which were completely consistent with blank-field LF results
(e.g., Bouwens et al.\ 2015a).  This was an important result, given
significant long standing concerns about the impact of systematic
uncertainties on such measurements (e.g., Brada{\v{c}} et al.\ 2009;
Bouwens et al.\ 2009b; Maizy et al.\ 2010).

\begin{deluxetable*}{ccccccccccc}
\tablewidth{0cm}
\tablecolumns{11}
\tabletypesize{\footnotesize}
\tablecaption{Samples of $z=2$-10 Galaxies found over the six HFF cluster fields (including the Oesch et al.\ 2018a $z\sim 10$ selection)\tablenotemark{*}\label{tab:sampnumbers}}
\tablehead{\colhead{Cluster} & \colhead{Area [arcmin$^2$]} & \colhead{$z\sim 2$} & \colhead{$z\sim 3$} & \colhead{$z\sim 4$} & \colhead{$z\sim 5$} & \colhead{$z\sim 6$} & \colhead{$z\sim 7$} & \colhead{$z\sim 8$} & \colhead{$z\sim 9$} & \colhead{$z\sim 10$\tablenotemark{b}}}
\startdata
Abell 2744 & 4.9  & 157 & 233 & ---\tablenotemark{a} & 27 & 49 & 25 & 15 & 4 & 2\tablenotemark{b}\\
MACS0416 & 4.9  & 215 & 233 & ---\tablenotemark{a} & 7 & 50 & 26 & 10 & 6 & 0\\
MACS0717 & 4.9  & 81 & 160 & 32 & ---\tablenotemark{a} & 26 & 14 & 9 & 0 & 0\\
MACS1149 & 4.9  & 134 & 195 & 36 & ---\tablenotemark{a} & 52 & 21 & 5 & 2 & 0\\
Abell S1063 & 4.9  & 96 & 203 & ---\tablenotemark{a} & 11 & 62 & 28 & 6 & 3 & 0\\
Abell 370 & 4.9  & 82 & 152 & ---\tablenotemark{a} & 14 & 35 & 11 & 6 & 1 & 0\\
Total & 29.4 & 765 & 1176 & 68 & 59 & 274 & 125 & 51 & 16 & 2
\enddata
\tablenotetext{*}{Selection is described in the companion paper to this one (Bouwens et al.\ 2022b)}
\tablenotetext{a}{Sources are not selected at this redshift in the
  indicated cluster field, due to concerns about contamination from
  foreground galaxies from the cluster due to the similar position of
  the Lyman break and the 4000\AA break (see Figure 3 from Bouwens et
  al.\ 2022b).}
\tablenotetext{b}{Oesch et al.\ (2018a).  See also Zitrin et al.\ (2014).}
\end{deluxetable*}

With a demonstration of the effectiveness of the approach we pioneered
in Bouwens et al.\ (2017b) for characterizing the faint end of the
$UV$ LF at $z\sim6$, the next step is to apply this metholodogy to the
galaxies over a wider redshift range to derive the relevant LFs.  It
is the purpose of this study to derive such a set of LFs and do it
over the redshift range $z=2$-9 where star-forming galaxies can be
readily identified in the distant universe.  Additionally, we will
characterize the evolution of the faint-end slope $\alpha$ as well as
any potential turn-over at the extreme faint end of each LF.  Mapping
out the extreme faint end of the UV LF is valuable for providing
insight into the efficiency of star formation in lower mass galaxies
and quantifying the total budget of ionizing photons available at
$z\geq 6$ to drive cosmic reionization.  For this effort, we make use
of the extremely large $>$2500-source sample of lensed galaxies
recently identified at $z\sim2$-9 over all six HFF clusters in a
companion paper (Bouwens et al.\ 2022b).

Here we provide a brief outline of our plan for this manuscript.  In
\S2, we begin by reviewing the primary data sets we utilize and
describing the basic properties of our selected high-redshift samples.
\S3 details our procedure for deriving the LF from the lensing
clusters and describes our basic LF results.  In \S4, we compare our
new results with previous work in the literature and consider the
scientific implications of our new results.  Finally, in \S5, we
include a summary.  For consistency with our previous work, results
will frequently be quoted in terms of the luminosity $L_{z=3}^{*}$
Steidel et al.\ (1999) derived at $z\sim3$, i.e.,
$M_{1700,AB}=-21.07$.  The {\it HST} F275W, F336W, F435W, F606W,
F814W, F105W, F125W, F140W, and F160W bands are referred to as
$UV_{275}$, $U_{336}$, $B_{435}$, $V_{606}$, $I_{814}$, $Y_{105}$,
$J_{125}$, $JH_{140}$, and $H_{160}$, respectively, for simplicity.
Where necessary, $\Omega_0 = 0.3$, $\Omega_{\Lambda} = 0.7$, and $H_0
= 70\,\textrm{km/s/Mpc}$ is assumed.  All magnitudes are in the AB
system (Oke \& Gunn 1983).

\section{Data Sets and High-Redshift Samples}

\subsection{Data Sets}

We will base the present deep LF results primarily on the sensitive
near-UV, optical, and near-IR observations obtained by the HFF program
(Coe et al.\ 2015; Lotz et al.\ 2017) and a follow-up GO campaign of
the HFF clusters with WFC3/UVIS (Alavi et al.\ 2014, 2016; Siana 2013,
2015).  Over each of the six clusters in the HFF program, at least 16
orbits of WFC3/UVIS time (8 and 8 in the $UV_{275}$ and $U_{336}$
bands), 70 orbits of optical ACS time (18, 10, and 42 in the
$B_{435}$, $V_{606}$, and $I_{814}$ bands), and 70 orbits of WFC3/IR
time (24, 12, 10, and 26 in the $Y_{105}$, $J_{125}$, $JH_{140}$, and
$H_{160}$ bands) was invested into observations of each cluster, with
additional observations coming from the CLASH (Postman et al.\ 2012)
and GLASS (Schmidt et al.\ 2014) programs.  We made use of the v1.0
reductions of these observations made publicly available by the HFF
team (Koekemoer et al.\ 2014).  For the WFC3/UVIS observations, we
made our own reductions, following similar procedures to what we
utilized in reducing the UVIS observations obtained as part of the
HDUV program (Oesch et al.\ 2018b).

In addition, we will make use of the $z=2$-9 LF constraints available
from the comprehensive set of blank-field HST observations recently
utilized by Bouwens et al.\ (2021a).  For that analysis, Bouwens et
al.\ (2021a) not only made use of the extremely sensitive optical,
near-IR, and rest-UV observations obtained over the Hubble Ultra Deep
Field (Beckwith et al.\ 2006; Koekemoer et al.\ 2013; Illingworth et
al.\ 2013; Teplitz et al.\ 2013), but also made use of the sensitive
ultraviolet, optical, and near-IR data over the five CANDELS fields
and ERS field (Grogin et al.\ 2011; Koekemoer et al.\ 2011; Oesch et
al.\ 2018b), Hubble Frontier Field parallels (Coe et al.\ 2015; Lotz et
al.\ 2016), and pure parallel search fields (Trenti et al.\ 2011; Yan
et al.\ 2011).

\begin{figure*}
\epsscale{0.74}
\plotone{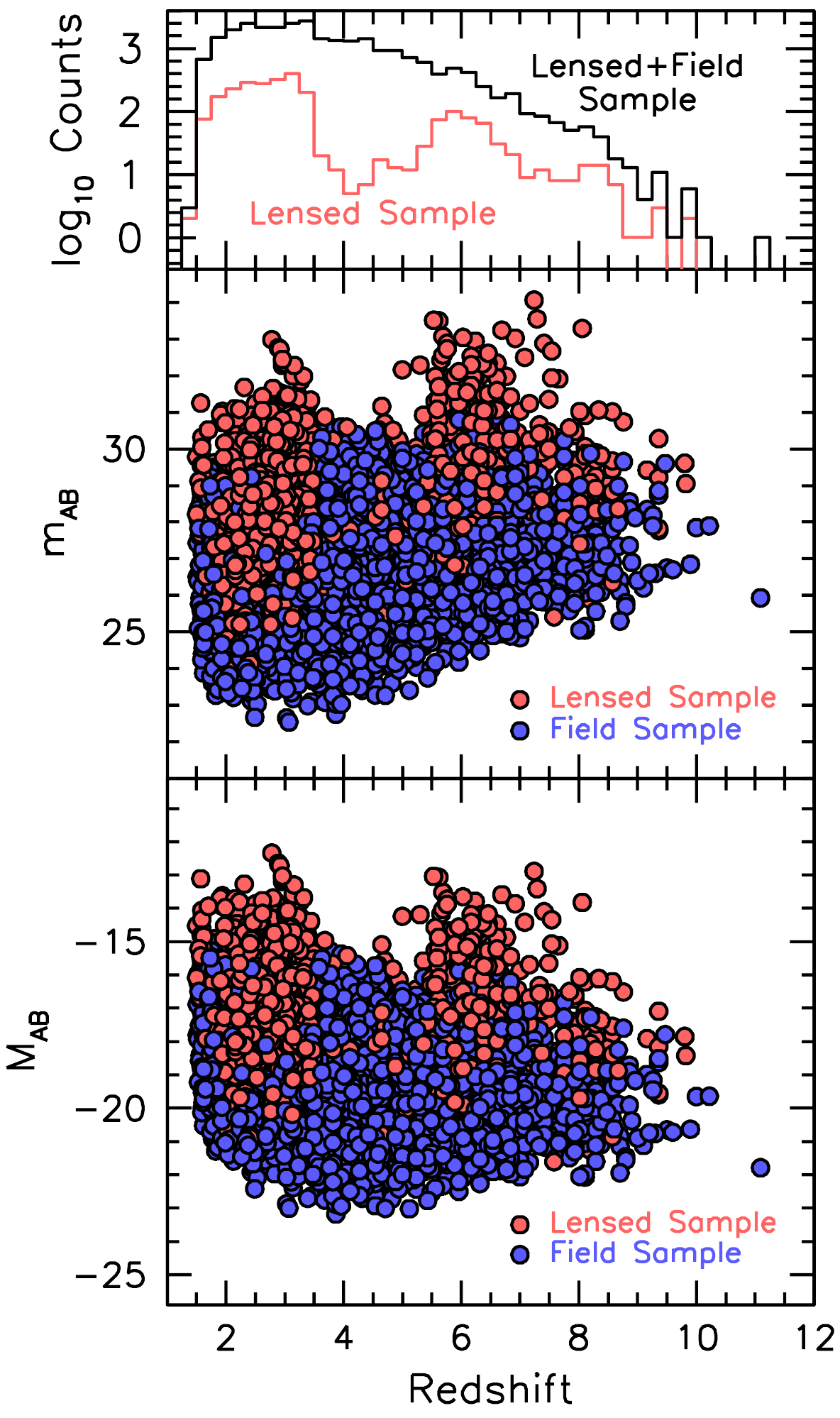}
\caption{(\textit{upper}) The number of sources per unit redshift
  (shown in logarithmic units) in our lensed sample constructed from
  the HFF 'cluster' fields (\textit{light red histogram}) and this
  sample combined with the Bouwens et al.\ (2021a) blank-field sample
  (\textit{black histogram}).  (\textit{middle panel}) The
  distribution of sources in apparent magnitude and redshift from our
  HFF 'cluster' fields (\textit{red circles}: Bouwens et al.\ 2022b)
  and from an HST blank-field sample with $>$24000 sources
  (\textit{blue circles}: Bouwens et al.\ 2021a).  (\textit{lower})
  The distribution of sources in $UV$ luminosity and redshift from our
  HFF 'cluster' fields (\textit{red circles}) and from an HST
  blank-field sample (\textit{blue circles}).  It is clear that
  selections using lensing magnification can reach up to
  $\sim$10-40$\times$ fainter in luminosity than achieved using
  similar selections without the aide of lensing magnification.  The
  relative paucity of $z\sim4$-5 sources in our HFF 'cluster field'
  selections relative to blank-field selections is a direct result of
  the very conservative selection criteria we impose at those two
  redshifts to minimize contamination from galaxies in the foreground
  clusters (see Figure 3 from Bouwens et
  al.\ 2022b).\label{fig:muv_vs_red}}
\end{figure*}

\subsection{Lensed Galaxy Samples at $z=2$-9}

We consider the systematic application of $z=2$-9 Lyman-break galaxy
selection criteria to the HFF near-UV, optical, and near-IR
observations available over the six clusters that make up the HFF
program.  We describe our application of these criteria to the HFF
observations in a companion paper (Bouwens et al.\ 2022b), and it is
very similar to what we previously performed in Bouwens et
al.\ (2015a, 2016a, and 2017b).

In total, we have identified 765, 1176, 68, 59, 274, 125, 51, and 16
sources as part of our $z\sim2$, $z\sim3$, $z\sim4$, $z\sim5$,
$z\sim6$, $z\sim7$, $z\sim8$, and $z\sim9$ selections.  The results
are summarized in Table~\ref{tab:sampnumbers}.  $z\sim4$ galaxies are
exclusively selected around the two highest redshift clusters that
make up the HFF program, i.e., MACS0717 and MACS1149.  Meanwhile,
$z\sim5$ galaxies are selected behind the four lowest redshift clustes
that make up the HFF program, i.e., Abell 2744, MACS0416, Abell S1063,
and Abell 370.  $z\sim4$ and $z\sim5$ selections are exclusively made
behind these specific clusters to minimize contamination from
foreground sources in clusters having spectral breaks at similar
wavelengths to the breaks used to select Lyman-break galaxies.  See
Figures 2-4 from Bouwens et al.\ (2022b).

An illustration of the distribution of the present lensed sample in
redshift and $UV$ magnitude is provided in Figure~\ref{fig:muv_vs_red}
along with the $>$24,000 source sample we utilize from the Bouwens et
al.\ (2021a) blank-field selection.  As should be apparent the lensed
sample reaches $\sim$10-40$\times$ fainter in UV luminosity than does
the blank-field sample, providing with much greater leverage for
probing the faint-end form of the $z\geq2$ LFs.

\section{Luminosity Function Determinations}

\subsection{Basic Procedure}

Here we describe our basic procedure for constraining the shape of the
$UV$ LF for each of our intermediate to high redshift samples
$z\sim2$, $z\sim3$, $z\sim4$, $z\sim5$, $z\sim6$, $z\sim7$, $z\sim8$,
and $z\sim9$.

One significant challenge, in the derivation of the $UV$ LF results at
high redshift from lensed samples, is the impact of errors in the
estimated magnification factors for individual sources.  As
demonstrated in Bouwens et al.\ (2017b), errors in the magnification
factors can effectively wash out faint-end ($>$$-$15 mag) turn-overs
in the $UV$ LF, making it difficult to observationally test for the
presence of such a turn-over in real-life observations.

We have already demonstrated in Bouwens et al. (2017b) how we can
overcome the impact of potential errors in the lensing maps using a
forward-modeling procedure (see Atek et al.\ 2018 for a separate
approach using forward modeling).  The basic idea is to leverage the
availability of the many independent lensing models for each cluster
to estimate the uncertainties.  Model LFs are evaluated by treating
one of the lensing maps as the truth and thus constructing a full
catalog of observables with that map and then interpreting the
observations using another lensing model.  In this way, the expected
number of sources per unit luminosity for a model LF could be
realistically estimated.  Figure 6 of Bouwens et al. (2017b)
illustrates the basic procedure.

Here we will follow the same forward-modeling procedure as we
introduced in Bouwens et al.\ (2017b).  In evaluating model LFs, we
treat one flavor of lensing model as the truth and use it to construct
a complete catalog of background sources behind each cluster.  Sources
are added to a search field in proportion to the product of the model
volume density, selection efficiency $S(m)$, and the cosmic volume
element -- which we take to be the cosmic volume element divided by
the magnification factor.  The selection efficiency $S$ is, in
general, a function of both the apparent magnitude $m$ and the
magnification factor $\mu$, but we can largely ignore the impact of
magnification in the limit that faint sources have very small sizes.
Justification for this is provided in the published results of Bouwens
et al.\ (2017a), Kawamata et al.\ (2018), Bouwens et al.\ (2022a), and
Bouwens et al.\ (2022b).  

We then use a different lensing model to estimate the magnification
factors and $UV$ luminosities for individual sources behind the
cluster.  To maximize the reliability of our results, we made use of
the median magnification from the latest parametric lensing models for
our fiducial LF determinations.  Not only has use of the median
magnification model been shown to provide robust estimates of the
magnification to magnifications of $\gtrsim$50 (Livermore et
al.\ 2017; Bouwens et al.\ 2017b, 2022a), but the parametric models
were shown to best reproduce the input magnification models from the
HFF comparison project (Meneghetti et al.\ 2017), with median
differences ($<$0.1 dex differences) to magnification factors of 30.
The parametric lensing models available for the HFF clusters and
utilizing most of the public multiple image constraints, i.e., v3/v4,
include the CATS (Jullo \& Kneib 2009; Richard et al.\ 2014; Jauzac et
al.\ 2015a,b; Lagattuta et al.\ 2017), Sharon (Johnson et al.\ 2014),
\textsc{GLAFIC} (Oguri 2010; Ishigaki et al.\ 2015; Kawamata et
al.\ 2016), Zitrin-NFW (Zitrin et al.\ 2013, 2015), and Keeton (2010)
results.  When computing the median magnification map used to estimate
the magnification factors for our forward-modeling approach, any model
we treat as the truth is naturally excluded.

While we base our fiducial LF results on the parametric lensing models
available for the HFF clusters, we also derive LF results using the
non-parametric lensing models available for the HFF clusters.  The
non-parametric lensing models have been shown to be a good match
($<$0.1 dex) to the input models from the HFF comparison project
(Meneghetti et al.\ 2017) to magnification factors of $\sim$10-20.
Given the greater flexibility of the non-parametric models relative to
the parametric models, results derived from these models allow us to
assess the impact lensing models have on our LF results.

We evaluate the likelihood of a LF model by comparing the observed
number of sources in various absolute magnitude bins (0.5-mag width)
with the expected number of sources assuming that galaxies are
Poisson-distributed:
\begin{displaymath}
{\cal L} = \Pi_{i} P_i
\end{displaymath}
where 
\begin{equation}
P_i = e^{-N_{exp,i}} \frac{(N_{exp,i})^{N_{obs,i}}}{(N_{obs,i})!}
\end{equation}
$N_{obs,i}$ and $N_{exp,i}$ are the observed and expected number of
sources in magnitude interval $i$.  To reduce the impact of sources
with complex multi-component or morphological structure on our
analysis -- which become common at the bright end of the LF -- we only
consider luminosity bins fainter than $-$19 mag.  As a result of this
choice, this analysis relies entirely on blank-field LF results for
constraints brightward of $-$19 mag.  Additionally, our
forward-modeling simulations typically include $\sim$200$\times$ as
many sources as are present in the actual observations (e.g., see
Figure 6 from Bouwens et al.\ 2017b) to guarantee an accurate
calculation of $N_{exp,i}$ for our likelihood estimates.

\begin{figure}
\epsscale{1.17}
\plotone{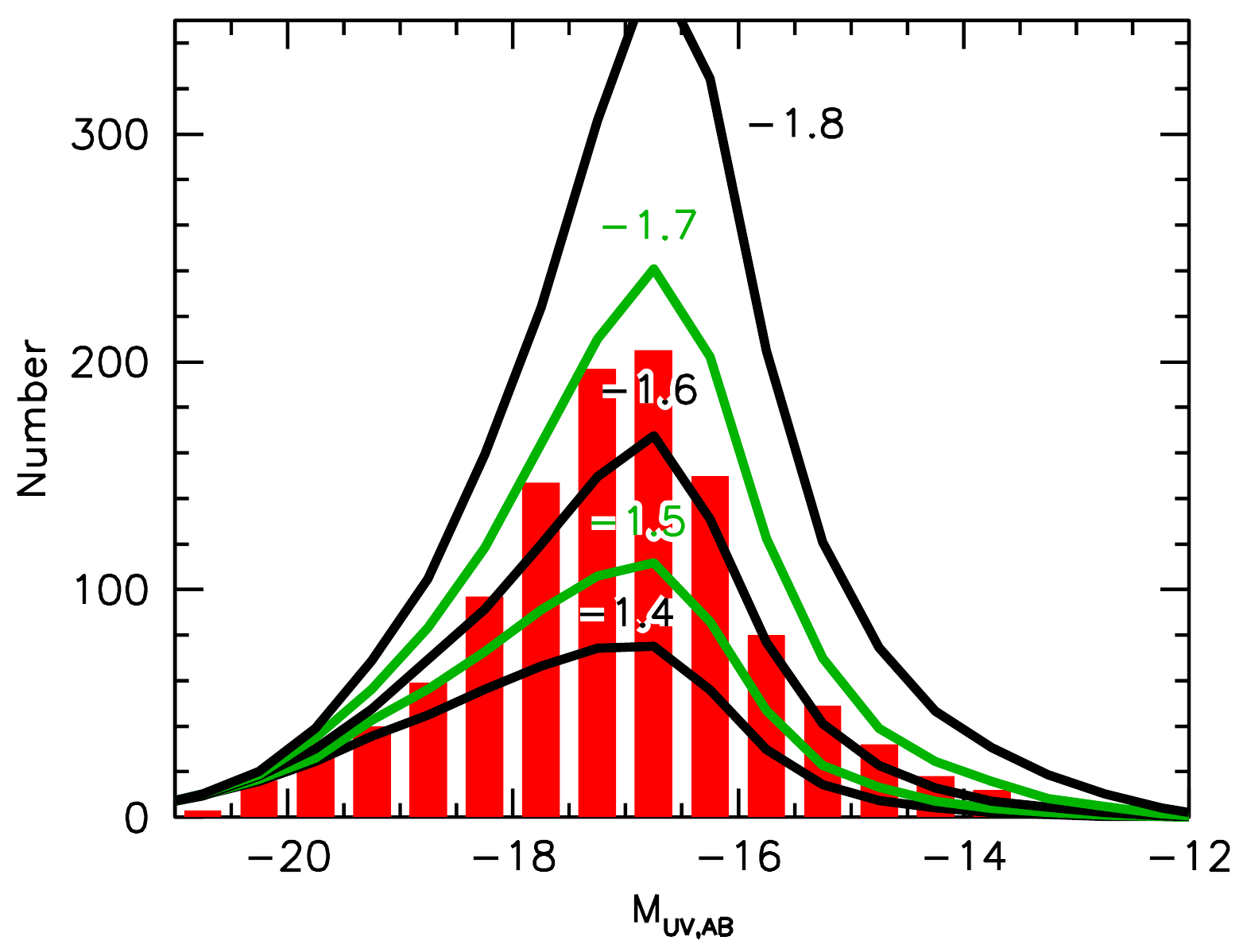}
\caption{Use of the lensed $z\sim3$ samples from the HFF program
  (\textit{red histogram}) to illustrate the leverage available to
  constrain the faint-end slope $\alpha$.  The green and black lines
  show the predicted number of sources per bin for different values of
  the faint-end slope.  The faint-end slope $\alpha$ appears to be
  $\sim$$-$1.65.  Interestingly enough, we can use this figure to
  assess the faint-end slopes $\alpha$ derived earlier by Parsa et
  al.\ (2016) and Alavi et al.\ (2016), i.e., $-1.31\pm0.04$ and
  $-1.94\pm0.06$.  Those results are strictly shallower and steeper,
  respectively, than the faint-end slopes for any of the models shown
  on this figure and thus appear to be inconsistent with our new
  results.  Clearly, there is enormous leverage available in lensed
  HFF samples to constrain the faint-end slope very precisely (as also
  illustrated by the $\sim$2-3\% uncertainties in $\alpha$ presented
  in Table~\ref{tab:lfparmsum}).\label{fig:leverage}}
\end{figure}

To evaluate the likelihood of various LF parameters for our $z\sim6$-9
samples, we must account for our combining five separate selections of
$z\sim6$, $z\sim7$, $z\sim8$, and $z\sim9$ galaxies in creating our
composite sample of $z\sim6$-9 galaxies.  This was done to maximize
the utility of our results to constrain the faint end of the
$z\sim6$-9 LFs.  Selections were constructed by leveraging a separate
source catalog, each made with different parameters to handle the
background parameters and thresholding (Bouwens et al.\ 2022b).  While
this results in a larger number of sources in each bin, many sources
occur in our final catalog multiple times and are not independent.  We
can account for the impact of this by quantifying the fraction of
sources $f_{i,j}$ in each bin $i$ that are counted $1\times$,
$2\times$, $3\times$, $4\times$, and $5\times$ and assuming similar
multiplicity fraction in the modeled statistics.  As such, the
probability $P_i$ for measuring a specific number of counts in bin $i$
is then the following:
\begin{equation}
P_i = \Sigma_{N_{obs,i,j}} \Pi_{j=1,5} e^{-f_{i,j} N_{exp,i}} \frac{(f_{i,j} N_{exp,i})^{N_{obs,i,j}}}{(N_{obs,i,j})!}
\label{eq:pois}
\end{equation}
where the summation $\Sigma_{N_{obj,i,j}}$ runs over all $N_{obj,i,j}$
where the sum $\Sigma_{j=1,5} j N_{obj,i,j}$ is equal to $N_{obj,i}$.
We have verified that Eq.~\ref{eq:pois} reduces to the appropriate
Poissonian likelihood distribution in the limit that all sources are
present in our catalogs with a fixed multiplicity (e.g., once or five
times).

Due to the relatively modest depth of the $UV_{275}$ and $U_{336}$
data available to select our $z\sim2$-3 samples, we only include
sources brightward of 28.0 mag and 28.5 mag ($V_{606}$ and $I_{814}$
bands, respectively) to mitigate the impact of the uncertain
completeness corrections faintward of these limits.  For our $z\sim4$
and $z\sim 5$ selections, where contamination from evolved galaxies at
the redshift of the cluster is a particular concern, we restrict
ourselves to using sources brightward of 27.3 and 27.5 mag,
respectively.

We used extensive source recovery simulations to estimate selection
efficiencies $S(m)$ for each of our intermediate to high-redshift
samples.  For each of the simulations, we first constructed mock
catalogs of sources over the general redshift ranges spanned by each
of our $z\sim2$-9 samples, i.e., $z=1-4$, $z=1$-4, $z=2.5$-5.0,
$z=4$-6, $z=5$-7, $z=5.5$-8, $z=6.5$-9.5, and $z=7$-10 for our
$z\sim2$, 3, 4, 5, 6, 7, 8, and 9 selections, respectively.  We then
created artificial two-dimensional images for each of the sources in
these catalogs in all HST bands used for the selection and detection
of the sources and then added these images to the real observations.
We then repeated both our catalog creation procedure and source
selection procedure in the same way as on the real observations
(Bouwens et al.\ 2017a,b).

Motivated by our earlier findings regarding source sizes for faint
$z\sim2$-8 samples from the HFFs (Bouwens et al.\ 2017a; see also
Kawamata et al.\ 2018; Bouwens et al.\ 2017c, 2022a; Yang et
al.\ 2022), we adopted a point-source size distribution in modeling
the completeness of sources over the HFFs for our fiducial LF results.
Additionally, we took the $UV$-continuum slope $\beta$ distribution to
a median value of $-2$ at $z\sim2$-3, $-2.1$ at $z\sim4$-5, and $-2.3$
at $z\sim6$-8 consistent with the Kurczynski et al.\ (2014) and
Bouwens et al.\ (2014) $UV$-continuum slope measurements.  We have
verified that for unlensed sources at the faint end of our HFF
selections, we estimate almost identical selection volumes to what we
estimate by computing the selection volumes using randomly-selected
$z\sim2$-4 galaxies as templates in our image simulations.  We expect
that this is due to the combination of the high surface brightness
sensitivity of the HFF observations and the faintest sources in our
fields having small sizes.

To quantify the possible systematic uncertainties that could result in
our LF determinations from our size modeling, we have also repeated
these completeness estimates using the size-luminosity relations of
both Shibuya et al.\ (2015) and Bouwens et al.\ (2022a) to illustrate
how large the systematic uncertainties could be, similar to the
exercise we performed in \S5.4 of Bouwens et al.\ (2017b).  While we
include these estimates as a possible systematic error on the derived
faint-end LF results, we emphasize that this is a worst case scenario,
as the results from Bouwens et al.\ (2017a), Kawamata et al.\ (2018),
and Yang et al.\ (2022) results all point towards substantially
smaller source sizes.

As an illustration of the substantial leverage available from the HFF
samples to constrain the faint-end slope, we show in
Figure~\ref{fig:leverage} the number of sources behind the HFF
clusters as a function of $UV$ luminosity and compare that to the
predicted numbers for specific values for the faint-end slope of the
$UV$ LF at $z\sim3$.  From this figure and substantial numbers of
sources faintward of $-19$ mag, it is clear that the faint-end slope
of the $z\sim3$ LF must be fairly close to $-$1.63 and faint-end
slopes of $-1.43$ and $-1.83$ can both be excluded at high confidence
on the basis of the HFF results.

\subsection{Functional Form and Optimization Procedure\label{sec:fform}}

As in Bouwens et al.\ (2017b), we adopt a standard Schechter
functional form for our LF
\begin{displaymath}
\phi^* (\ln(10)/2.5) 10^{-0.4(M-M^{*})(\alpha+1)} e^{-10^{-0.4(M-M^{*})}}
\end{displaymath}
but modified to allow for curvature in the faint-end slope $\alpha$ at
the faint end.  We implement this using a new curvature parameter
$\delta$ and multiply the standard Schechter form with the following
expression faintward of $-$16 mag:
\begin{displaymath}
10^{-0.4\delta (M+16)^2}
\end{displaymath}
As we demonstrate in Bouwens et al.\ (2017b), positive values of
$\delta$ result in a turn-over in the LF at
\begin{equation}
M_T = -16 - \frac{\alpha+1}{2\delta}
\label{eq:mt}
\end{equation}
while negative values for $\delta$ result in the LF turning concave
upwards.

Given our use of four separate parameters to characterize the overall
shape of the $UV$ LF, we utilize a Markov-Chain Monte Carlo procedure
both to determine the maximum likelihood value and to characterize the
observational uncertainties.  We begin the MCMC optimization process
using the best-fit blank-field LF parameters from Bouwens et
al.\ (2021a) and run $\sim$1000 iterations to find the best-fit LF
parameters and map out the likelihood space.

\subsection{LF Results at $z=2$-9}

\subsubsection{Leveraging Only Lensed Sources from the HFF Program\label{sec:hffonly}}

As in Bouwens et al.\ (2017b), we commence our analysis by exclusively
making use of the lensed HFF samples to derive our initial $UV$ LF
results.  The value in doing this first is that it allows us to test
for the presence of any systematic errors in lensing-cluster LF
determinations vis-a-vis blank-field determinations.  The two
approaches have their own strengths and are subject to different
sources of systematic error, so it is valuable to first look into this
issue before combining the results to arrive the best constraints on
the overall shape of the $UV$ LF.

In Bouwens et al.\ (2017b), we demonstrated we could obtain accurate
constraints on both the faint-end slope $\alpha$ and $\phi^*$ at
$z\sim6$ relying only on the lensed $z\sim6$ sources from the HFF
lensing clusters.  There, a faint-end slope $\alpha$ of $-1.92\pm0.04$
and a $\phi^*$ of 0.66$\pm$0.06 $\times$ 10$^{-3}$ Mpc$^{-3}$ were
found, consistent ($\lesssim$1$\sigma$) with the $-1.87\pm0.10$ slope
and 0.51$_{-0.10}^{+0.12}$ $\times$ 10$^{-3}$ Mpc$^{-3}$ normalization
found from blank-field observations (Bouwens et al.\ 2015a).  For
these determinations, the characteristic luminosity $M^*$ was fixed to
$-$20.94 mag, the value obtained from wide-area blank-field studies,
due to their being insufficient volume behind the HFF clusters to
achieve strong constraints on this luminosity.

We adopt a similar approach here.  We begin by fixing the
characteristic luminosities $M^*$ for the $z\sim2$-9 LFs to the values
obtained by the Bouwens et al.\ (2021a) blank-field analysis and then
use the described MCMC approach to identify the values of $\phi^*$,
$\alpha$, and $\delta$ that maximizes the likelihood of recovering our
$z\sim2$-9 samples from the full HFF program.  Uncertainties on the
individual LF parameters can be calculated based on fits to the
multi-dimensional likelihood surface derived from our MCMC
simulations.

\begin{figure}
\epsscale{1.17}
\plotone{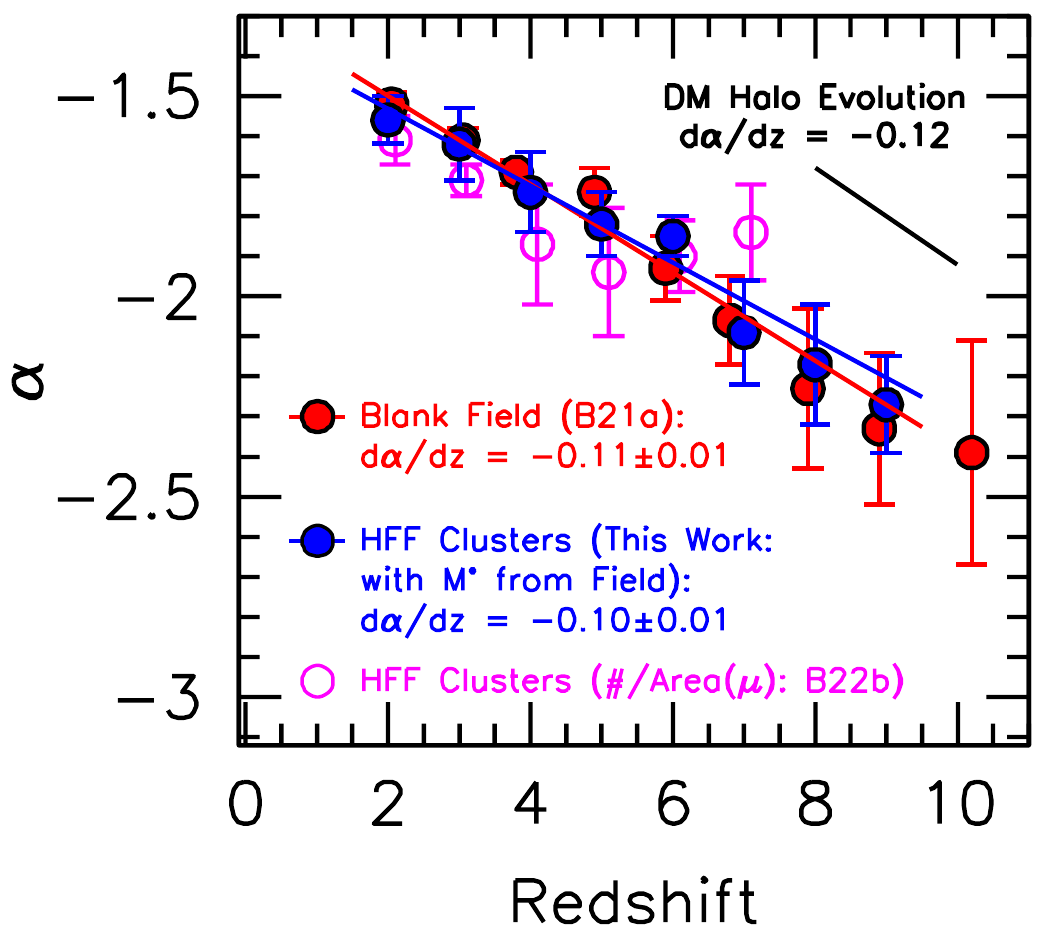}
\caption{Comparison of faint-end slope $\alpha$ determinations from
  lensed HFF galaxy samples (\textit{blue solid circles}) with similar
  blank-field determinations of these slopes $\alpha$ (\textit{red
    solid circles}: Bouwens et al.\ 2021a).  The faint-end slope
  determinations shown here from the lensed HFF samples do not make
  use of any information from the blank-field search results, except
  for the value of characteristic luminosity $M^*$ (\S3.3.1: but see
  Figure~\ref{fig:schevol}, Table~\ref{tab:lfparmsum}, and
  \S\ref{sec:lfresults} for $\alpha$ determinations using both the
  lensed HFF + blank-field constraints).  The purple open circles are
  based on the surface density $\Sigma$ vs. magnification $\mu$ trend
  found in the companion paper (Bouwens et al.\ 2022b) for galaxies at
  $z\sim2$, 3, 4, 5, 6, and 7.  The blue and red lines show the
  essentially identical, and remarkable, best-fit evolution in the
  faint-end slope $\alpha$ from the lensing cluster and blank-field
  search results, respectively.  This strongly suggests that there are
  minimal systematic errors in either determination, and that we can
  combine both probes to dramatically improve the statistical
  constraints on the evolution of the $UV$ LF and faint-end slope
  $\alpha$.  \textit{This is the first time such agreement in the
    faint-end slope $\alpha$ evolution has been demonstrated over such
    a large range in redshift.}\label{fig:alpha}}
\end{figure}

We present the LF results we derive using our lensed HFF samples and
fixed values for the characteristic luminosity in
Table~\ref{tab:lfparmsum}.  In addition, in Figure~\ref{fig:alpha}, we
present our determinations of the faint-end slope $\alpha$ of the LF
vs. redshift.  A simple linear fit to the faint-end slope $\alpha$
results we derive vs. redshift yields the relation
\begin{equation}
\alpha = (-1.92\pm0.03) + (-0.10\pm0.01) (z-6)
\end{equation}
and presented on Figure~\ref{fig:alpha} as a blue line.  For context,
figure~\ref{fig:alpha} also shows a recent determination of the
faint-end slope $\alpha$ evolution based on blank-field observations
alone (Bouwens et al.\ 2021a: \textit{red circles}), with the observed
trend with redshift (\textit{red line}):
\begin{equation}
\alpha = (-1.92\pm0.03) + (-0.11\pm0.01) (z-6)
\end{equation}
Encouragingly enough, the new faint-end slope $\alpha$ results we
derive from lensing cluster observations seem very consistent (within
the $1\sigma$ errors) both in terms of its slope and intercept with
blank-field results over the entire redshift range we are examining
$z\sim2$ to $z\sim9$.

\begin{deluxetable*}{cccccc}
  \renewcommand{\arraystretch}{0.75}
\tablewidth{16.2cm}
\tablecolumns{6}
\tabletypesize{\footnotesize}
\tablecaption{Summary of our Final Fiducial Constraints on the $z\sim2$-9 $UV$ LFs\label{tab:lfparmsum}}
\tablehead{
\colhead{Data Sets + Method} & \colhead{$M_{UV} ^{*}$} & \colhead{$\phi^*$ $(10^{-3}$ Mpc$^{-3}$)} & \colhead{$\alpha$} & \colhead{$\delta$\tablenotemark{a}} & \colhead{$M_{T}$\tablenotemark{$\dagger$}}}
\startdata
\multicolumn{5}{c}{$z\sim2$}\\
HFF + Blank-Field (Parametric, Fiducial) & $-$20.30$\pm$0.08 & 3.4$\pm$0.4 & $-$1.53$\pm$0.03 & 0.09$\pm$0.11 & $>$$-$15.2\\
HFF + Blank-Field (Non-Parametric)\tablenotemark{b} & $-$20.31$\pm$0.09 & 4.0$\pm$0.7 & $-$1.53$\pm$0.03 & 0.20$\pm$0.13 & $>$$-$15.3\\
HFF (Parametric) + $M^*$ Fixed & $-$20.28 (fixed) & 3.0$\pm$0.6 & $-$1.56$\pm$0.06 & 0.15$\pm$0.13 \\
HFF (Parametric) + $M^*+\phi^*$ Fixed & $-$20.28 (fixed) & 4.0 (fixed) & $-$1.49$\pm$0.03 & 0.05$\pm$0.11 \\
$\Sigma(\mu)$ Fit\tablenotemark{c} & & & $\leq-1.61\pm0.06$ \\
Blank-Field (B21a) & $-$20.28$\pm$0.09 & 4.0$_{-0.4}^{+0.5}$ & $-$1.52$\pm$0.03 & --- \\
\\
\multicolumn{5}{c}{$z\sim3$}\\
HFF + Blank-Field (Parametric, Fiducial) & $-$20.84$\pm$0.07 & 2.3$\pm$0.4 & $-$1.60$\pm$0.03 & $-$0.06$\pm$0.05 & $>$$-$13.1\\
HFF + Blank-Field (Non-Parametric)\tablenotemark{b} & $-$20.96$\pm$0.09 & 1.9$\pm$0.8 & $-$1.65$\pm$0.04 & 0.05$\pm$0.07 & $>$$-$12.3\\
HFF (Parametric) + $M^*$ Fixed & $-$20.87 (fixed) & 2.0$\pm$1.5 & $-$1.62$\pm$0.09 & $-$0.03$\pm$0.06 \\
HFF (Parametric) + $M^*+\phi^*$ Fixed & $-$20.87 (fixed) & 2.1 (fixed) & $-$1.60$\pm$0.03 & $-$0.06$\pm$0.04 \\
$\Sigma(\mu)$ Fit\tablenotemark{c} & & & $\leq-1.71\pm0.04$ \\
Blank-Field (B21a) & $-$20.87$\pm$0.09 & 2.1$_{-0.3}^{+0.3}$ & $-$1.61$\pm$0.03 & --- \\
\\
\multicolumn{5}{c}{$z\sim4$}\\
HFF + Blank-Field (Parametric, Fiducial) & $-$20.93$\pm$0.07 & 1.6$\pm$0.3 & $-$1.69$\pm$0.03 & $-$0.18$\pm$0.18 & $>$$-$15.3\\
HFF + Blank-Field (Non-Parametric)\tablenotemark{b} & $-$20.95$\pm$0.07 & 1.8$\pm$0.3 & $-$1.71$\pm$0.03 & $-$0.06$\pm$0.12 & $>$$-$15.6\\
HFF (Parametric) + $M^*$ Fixed & $-$20.93 (fixed) & 1.5$\pm$0.6 & $-$1.71$\pm$0.09 & $-$0.12$\pm$0.22 \\
HFF (Parametric) + $M^*+\phi^*$ Fixed & $-$20.93 (fixed) & 1.69 (fixed) & $-$1.68$\pm$0.05 & $-$0.19$\pm$0.17 \\
$\Sigma(\mu)$ Fit\tablenotemark{c} & & & $\leq-1.87\pm0.15$ \\
Blank-Field (B21a) & $-$20.93$\pm$0.08 & 1.69$_{-0.20}^{+0.22}$ & $-$1.69$\pm$0.03 & --- \\
Bouwens et al.\ (2007) & $-$20.98$\pm$0.10 & 1.3$\pm$0.2 & $-$1.73$\pm$0.05 & --- \\
Bouwens et al.\ (2015a) & $-$20.88$\pm$0.08 & 1.97$_{-0.29}^{+0.34}$ & $-$1.64$\pm$0.04\\
\\
\multicolumn{5}{c}{$z\sim5$}\\
HFF + Blank-Field (Parametric, Fiducial) & $-$21.13$\pm$0.09 & 0.73$\pm$0.13 & $-$1.78$\pm$0.04 & $-$0.07$\pm$0.20 & $>$$-$14.7\\
HFF + Blank-Field (Non-Parametric)\tablenotemark{b} & $-$21.15$\pm$0.10 & 0.74$\pm$0.14 & $-$1.80$\pm$0.05 & $-$0.02$\pm$0.24 & $>$$-$14.8\\
HFF (Parametric) + $M^*$ Fixed & $-$21.10 (fixed) & 0.66$\pm$0.15 & $-$1.82$\pm$0.08 & $-$0.03$\pm$0.21 \\
HFF (Parametric) + $M^*+\phi^*$ Fixed & $-$21.10 (fixed) & 0.79 (fixed) & $-$1.77$\pm$0.06 & $-$0.07$\pm$0.20 \\
$\Sigma(\mu)$ Fit\tablenotemark{c} & & & $\leq-1.94\pm0.16$ \\
Blank-Field (B21a) & $-$21.10$\pm$0.11 & 0.79$_{-0.13}^{+0.16}$ & $-$1.74$\pm$0.06 & --- \\
Bouwens et al.\ (2015a) & $-$21.17$\pm$0.12 & 0.74$_{-0.14}^{+0.18}$ & $-$1.76$\pm$0.05\\
\\
\multicolumn{5}{c}{$z\sim6$}\\
HFF + Blank-Field (Parametric, Fiducial) & $-$20.87$\pm$0.07 & 0.57$\pm$0.11 & $-$1.87$\pm$0.04 & 0.05$\pm$0.10 & $>$$-$14.3\\
HFF + Blank-Field (Non-Parametric)\tablenotemark{b} & $-$20.98$\pm$0.08 & 0.45$\pm$0.10 & $-$1.98$\pm$0.06 & 0.36$\pm$0.18 & $>$$-$15.2\\
HFF (Parametric) + $M^*$ Fixed & $-$20.93 (fixed) & 0.61$\pm$0.29 & $-$1.85$\pm$0.05 & 0.00$\pm$0.14 \\
HFF (Parametric) + $M^*+\phi^*$ Fixed & $-$20.93 (fixed) & 0.51 (fixed) & $-$1.87$\pm$0.02 & 0.09$\pm$0.09 \\
$\Sigma(\mu)$ Fit\tablenotemark{c} & & & $\leq-1.90\pm0.09$ \\
Blank-Field (B21a) & $-$20.93$\pm$0.09 & 0.51$_{-0.10}^{+0.12}$ & $-$1.93$\pm$0.08 & --- \\
Bouwens et al.\ (2015a) & $-$20.94$\pm$0.20 & 0.50$_{-0.16}^{+0.22}$ & $-$1.87$\pm$0.10\\
\\
\multicolumn{5}{c}{$z\sim7$}\\
HFF + Blank-Field (Parametric, Fiducial) & $-$21.13$\pm$0.08 & 0.20$\pm$0.05 & $-$2.05$\pm$0.06 & 0.24$\pm$0.20 & $>$$-$15.2\\
HFF + Blank-Field (Non-Parametric)\tablenotemark{b} & $-$21.21$\pm$0.09 & 0.17$\pm$0.05 & $-$2.13$\pm$0.07 & 0.56$\pm$0.26 & $>$$-$15.4\\
HFF (Parametric) + $M^*$ Fixed & $-$21.15 (fixed) & 0.17$\pm$0.12 & $-$2.09$\pm$0.13 & 0.36$\pm$0.28 \\
HFF (Parametric) + $M^*+\phi^*$ Fixed & $-$21.15 (fixed) & 0.19 (fixed) & $-$2.06$\pm$0.04 & 0.25$\pm$0.16 \\
$\Sigma(\mu)$ Fit\tablenotemark{c} & & & $\leq-1.84\pm0.12$ \\
Blank-Field (B21a) & $-$21.15$\pm$0.13 & 0.19$_{-0.06}^{+0.08}$ & $-$2.06$\pm$0.11 & --- \\
Bouwens et al.\ (2015a) & $-$20.87$\pm$0.26 & 0.29$_{-0.12}^{+0.21}$ & $-$2.06$\pm$0.13\\
\\
\multicolumn{5}{c}{$z\sim8$}\\
HFF + Blank-Field (Parametric, Fiducial) & $-$20.90$\pm$0.19 & 0.096$\pm$0.065 & $-$2.20$\pm$0.09 & 0.23$\pm$0.28 & $>$$-$15.2\\
HFF + Blank-Field (Non-Parametric)\tablenotemark{b} & $-$20.96$\pm$0.19 & 0.094$\pm$0.051 & $-$2.27$\pm$0.09 & 0.50$\pm$0.30 & $>$$-$15.3\\
HFF (Parametric) + $M^*$ Fixed & $-$20.93 (fixed) & 0.11$\pm$0.09 & $-$2.17$\pm$0.15 & 0.15$\pm$0.31 \\
HFF (Parametric) + $M^*+\phi^*$ Fixed & $-$20.93 (fixed) & 0.09 (fixed) & $-$2.21$\pm$0.05 & 0.32$\pm$0.17 \\
Blank-Field (B21a) & $-$20.93$\pm$0.28 & 0.09$_{-0.05}^{+0.09}$ & $-$2.23$\pm$0.20 & --- \\
Bouwens et al.\ (2015a) & $-$20.63$\pm$0.36 & 0.21$_{-0.11}^{+0.23}$ & $-$2.02$\pm$0.23\\
\\
\multicolumn{5}{c}{$z\sim9$}\\
HFF + Blank-Field (Parametric, Fiducial) & $-$21.15$\pm$0.12 & 0.018$\pm$0.009 & $-$2.28$\pm$0.10 & 0.53$\pm$0.41 & $>$$-$15.6\\
HFF + Blank-Field (Non-Parametric)\tablenotemark{b} & $-$21.15$\pm$0.11 & 0.021$\pm$0.011 & $-$2.34$\pm$0.11 & 0.52$\pm$0.41 & $>$$-$15.6\\
HFF (Parametric) + $M^*$ Fixed & $-$21.15 (fixed) & 0.019$\pm$0.012 & $-$2.27$\pm$0.12 & 0.53$\pm$0.29 \\
HFF (Parametric) + $M^*+\phi^*$ Fixed & $-$21.15 (fixed) & 0.021 (fixed) & $-$2.21$\pm$0.05 & 0.33$\pm$0.19 \\
Blank-Field (B21a) & $-$21.15 (fixed) & 0.021$_{-0.009}^{+0.014}$ & $-$2.33$\pm$0.19 & --- 
\enddata
\tablenotetext{a}{Best-fit curvature in the shape of the $UV$ LF faintward of $-16$ mag (\S\ref{sec:fform}).}
\tablenotetext{b}{The non-parametric LF results presented in this table represent a mean of our derived LFs treating the non-parametric lensing models, i.e., grale and Diego, as the truth.}
\tablenotetext{c}{Constraints based on the observed trend in source surface density vs. model magnification factor $\mu$ (Bouwens et al.\ 2022b)}
\tablenotetext{$\dagger$}{Brightest luminosity at which the current constraints from the HFF permit a turn-over in the $UV$ LF (95\% confidence).}
\end{deluxetable*}

\begin{deluxetable*}{lclclclc}
\tablewidth{0pt}
\tabletypesize{\footnotesize}
\tablecaption{68\% Likelihood Intervals Derived for the $z\sim2$-9 $UV$ LFs Using Forward Modeling (\S\ref{sec:lfresults})\tablenotemark{a}\label{tab:likelihood}}
\tablehead{
\colhead{$M_{1600,AB}$\tablenotemark{b}} & \colhead{$\phi_k$} & \colhead{$M_{1600,AB}$\tablenotemark{b}} & \colhead{$\phi_k$} & \colhead{$M_{1600,AB}$\tablenotemark{b}} & \colhead{$\phi_k$} & \colhead{$M_{1600,AB}$\tablenotemark{b}} & \colhead{$\phi_k$} \\
\colhead{} & \colhead{(Mpc$^{-3}$ mag$^{-1}$)} & \colhead{} & \colhead{(Mpc$^{-3}$ mag$^{-1}$)} & \colhead{} & \colhead{(Mpc$^{-3}$ mag$^{-1}$)} & \colhead{} & \colhead{(Mpc$^{-3}$ mag$^{-1}$)} }
\startdata
\multicolumn{2}{c}{$z\sim2$ galaxies} & \multicolumn{2}{c}{$z\sim4$ galaxies} & \multicolumn{2}{c}{$z\sim6$ galaxies} & \multicolumn{2}{c}{$z\sim8$ galaxies}\\
$-$18.75 & 0.0051$_{-0.0003}^{+0.0003}$ & $-$18.75 & 0.0053$_{-0.0005}^{+0.0006}$ & $-$18.75 & 0.0025$_{-0.0003}^{+0.0003}$ & $-$18.75 & 0.0008$_{-0.0001}^{+0.0001}$\\
$-$18.25 & 0.0072$_{-0.0004}^{+0.0005}$ & $-$18.25 & 0.0077$_{-0.0008}^{+0.0009}$ & $-$18.25 & 0.0040$_{-0.0004}^{+0.0004}$ & $-$18.25 & 0.0015$_{-0.0002}^{+0.0002}$\\
$-$17.75 & 0.0096$_{-0.0006}^{+0.0006}$ & $-$17.75 & 0.011$_{-0.001}^{+0.001}$ & $-$17.75 & 0.0063$_{-0.0006}^{+0.0007}$ & $-$17.75 & 0.0027$_{-0.0003}^{+0.0003}$\\
$-$17.25 & 0.013$_{-0.001}^{+0.001}$ & $-$17.25 & 0.015$_{-0.002}^{+0.002}$ & $-$17.25 & 0.0097$_{-0.0010}^{+0.0015}$ & $-$17.25 & 0.0048$_{-0.0006}^{+0.0008}$\\
$-$16.75 & 0.017$_{-0.001}^{+0.003}$ & $-$16.75 & 0.021$_{-0.002}^{+0.005}$ & $-$16.75 & 0.015$_{-0.002}^{+0.003}$ & $-$16.75 & 0.0086$_{-0.0011}^{+0.0022}$\\
$-$16.25 & 0.021$_{-0.001}^{+0.005}$ & $-$16.25 & 0.029$_{-0.003}^{+0.009}$ & $-$16.25 & 0.023$_{-0.003}^{+0.007}$ & $-$16.25 & 0.015$_{-0.002}^{+0.005}$\\
$-$15.75 & 0.027$_{-0.002}^{+0.009}$ & $-$15.75 & 0.041$_{-0.005}^{+0.017}$ & $-$15.75 & 0.034$_{-0.004}^{+0.014}$ & $-$15.75 & 0.026$_{-0.004}^{+0.012}$\\
$-$15.25 & 0.034$_{-0.003}^{+0.016}$ & $-$15.25 & 0.062$_{-0.010}^{+0.035}$ & $-$15.25 & 0.050$_{-0.007}^{+0.027}$ & $-$15.25 & 0.040$_{-0.007}^{+0.024}$\\
$-$14.75 & 0.040$_{-0.006}^{+0.026}$ & $-$14.75 & 0.10$_{-0.03}^{+0.08}$ & $-$14.75 & 0.071$_{-0.014}^{+0.050}$ & $-$14.75 & 0.054$_{-0.016}^{+0.047}$\\
$-$14.25 & 0.046$_{-0.012}^{+0.042}$ & $-$14.25 & 0.18$_{-0.08}^{+0.23}$ & $-$14.25 & 0.097$_{-0.027}^{+0.094}$ & $-$14.25 & 0.065$_{-0.030}^{+0.092}$\\
$-$13.75 & 0.050$_{-0.019}^{+0.065}$ & $-$13.75 & 0.35$_{-0.20}^{+0.73}$ & $-$13.75 & 0.13$_{-0.05}^{+0.17}$ & $-$13.75 & 0.066$_{-0.042}^{+0.161}$\\
$-$13.25 & 0.052$_{-0.027}^{+0.097}$ & $-$13.25 & 0.74$_{-0.53}^{+2.50}$ & $-$13.25 & 0.17$_{-0.08}^{+0.31}$ & $-$13.25 & 0.058$_{-0.045}^{+0.250}$\\
$-$12.75 & 0.052$_{-0.033}^{+0.136}$ & $-$12.8 & 1.7$_{-1.4}^{+9.7}$ & $-$12.75 & 0.21$_{-0.13}^{+0.53}$ & $-$12.75 & 0.047$_{-0.041}^{+0.377}$\\
$-$12.25 & 0.050$_{-0.036}^{+0.186}$ & $-$12.2 & 4.2$_{-3.8}^{+43.0}$ & $-$12.25 & 0.25$_{-0.18}^{+0.91}$ & $-$12.25 & 0.033$_{-0.031}^{+0.550}$\\
\multicolumn{2}{c}{$z\sim3$ galaxies} & \multicolumn{2}{c}{$z\sim5$ galaxies} & \multicolumn{2}{c}{$z\sim7$ galaxies} & \multicolumn{2}{c}{$z\sim9$ galaxies}\\
$-$18.75 & 0.0060$_{-0.0008}^{+0.0010}$ & $-$18.75 & 0.0033$_{-0.0003}^{+0.0003}$ & $-$18.75 & 0.0017$_{-0.0001}^{+0.0001}$ & $-$18.75 & 0.0002$_{-0.0001}^{+0.0001}$\\
$-$18.25 & 0.0083$_{-0.0011}^{+0.0013}$ & $-$18.25 & 0.0049$_{-0.0004}^{+0.0004}$ & $-$18.25 & 0.0028$_{-0.0002}^{+0.0002}$ & $-$18.25 & 0.0005$_{-0.0001}^{+0.0001}$\\
$-$17.75 & 0.011$_{-0.001}^{+0.002}$ & $-$17.75 & 0.0073$_{-0.0006}^{+0.0007}$ & $-$17.75 & 0.0047$_{-0.0004}^{+0.0004}$ & $-$17.75 & 0.0009$_{-0.0001}^{+0.0002}$\\
$-$17.25 & 0.015$_{-0.002}^{+0.003}$ & $-$17.25 & 0.011$_{-0.001}^{+0.001}$ & $-$17.25 & 0.0078$_{-0.0007}^{+0.0010}$ & $-$17.25 & 0.0016$_{-0.0002}^{+0.0003}$\\
$-$16.75 & 0.020$_{-0.003}^{+0.005}$ & $-$16.75 & 0.015$_{-0.002}^{+0.003}$ & $-$16.75 & 0.013$_{-0.001}^{+0.003}$ & $-$16.75 & 0.0029$_{-0.0004}^{+0.0008}$\\
$-$16.25 & 0.027$_{-0.003}^{+0.009}$ & $-$16.25 & 0.022$_{-0.002}^{+0.007}$ & $-$16.25 & 0.021$_{-0.002}^{+0.006}$ & $-$16.25 & 0.0052$_{-0.0008}^{+0.0020}$\\
$-$15.75 & 0.036$_{-0.004}^{+0.015}$ & $-$15.75 & 0.032$_{-0.004}^{+0.013}$ & $-$15.75 & 0.033$_{-0.004}^{+0.014}$ & $-$15.75 & 0.0092$_{-0.0016}^{+0.0045}$\\
$-$15.25 & 0.048$_{-0.006}^{+0.025}$ & $-$15.25 & 0.047$_{-0.008}^{+0.027}$ & $-$15.25 & 0.049$_{-0.008}^{+0.027}$ & $-$15.25 & 0.013$_{-0.003}^{+0.008}$\\
$-$14.75 & 0.068$_{-0.010}^{+0.044}$ & $-$14.75 & 0.072$_{-0.020}^{+0.062}$ & $-$14.75 & 0.065$_{-0.016}^{+0.051}$ & $-$14.75 & 0.014$_{-0.005}^{+0.014}$\\
$-$14.25 & 0.097$_{-0.019}^{+0.079}$ & $-$14.25 & 0.11$_{-0.05}^{+0.16}$ & $-$14.25 & 0.076$_{-0.029}^{+0.090}$ & $-$14.25 & 0.012$_{-0.007}^{+0.022}$\\
$-$13.75 & 0.14$_{-0.04}^{+0.15}$ & $-$13.75 & 0.19$_{-0.12}^{+0.43}$ & $-$13.75 & 0.080$_{-0.042}^{+0.145}$ & $-$13.75 & 0.0084$_{-0.0063}^{+0.0304}$\\
$-$13.25 & 0.22$_{-0.07}^{+0.28}$ & $-$13.25 & 0.32$_{-0.24}^{+1.25}$ & $-$13.25 & 0.077$_{-0.051}^{+0.217}$ & $-$13.25 & 0.0044$_{-0.0039}^{+0.0362}$\\
$-$12.75 & 0.34$_{-0.13}^{+0.54}$ & $-$12.75 & 0.55$_{-0.47}^{+3.92}$ & $-$12.75 & 0.067$_{-0.053}^{+0.308}$ & $-$12.75 & 0.0019$_{-0.0018}^{+0.0383}$\\
$-$12.25 & 0.55$_{-0.25}^{+1.07}$ & $-$12.2 & 1.0$_{-0.9}^{+13.7}$ & $-$12.25 & 0.052$_{-0.045}^{+0.409}$ & $-$12.25 & 0.0006$_{-0.0006}^{+0.0371}$
\enddata
\tablenotetext{a}{68\% confidence intervals on the $z\sim6$ $UV$ LF we
  achieve (\S\ref{sec:lfresults}) by applying our forward modeling formalism to
  observations of all six HFF clusters in \S\ref{sec:lfresults}.  The
  quoted constraints give the geometric mean of the forward-modeling
  results using the \textsc{GLAFIC}, CATS, Sharon/Johnson, and Keeton
  parametric models as inputs.  Examples of these constraints for
  individual lensing models are provided in figures \ref{fig:lf2} and
  \ref{fig:lf6}.  The $1\sigma$ upper limits indicate the upper limits
  if one adopts the Bouwens et al.\ (2022a) size-luminosity scalings
  (implemented in a similar manner to \S5.4 of Bouwens et al.\ 2017b).
  As the intervals provided here are based on our parameterized
  modeling, the error bars on individual bins are not independent.}
\tablenotetext{b}{Derived at a rest-frame wavelength of 1600\AA.}
\end{deluxetable*}

\begin{deluxetable*}{lclclclc}
\tablewidth{0pt}
\tabletypesize{\footnotesize}
\tablecaption{Binned Determinations of the rest-frame $UV$ LF at $z\sim2$-9 (\S\ref{sec:lfresults})\tablenotemark{a}\label{tab:swlf}}
\tablehead{
\colhead{$M_{1600,AB}$\tablenotemark{b}} & \colhead{$\phi_k$\tablenotemark{c}} & \colhead{$M_{1600,AB}$\tablenotemark{b}} & \colhead{$\phi_k$\tablenotemark{c}} & \colhead{$M_{1600,AB}$\tablenotemark{b}} & \colhead{$\phi_k$\tablenotemark{c}} & \colhead{$M_{1600,AB}$\tablenotemark{b}} & \colhead{$\phi_k$\tablenotemark{c}} \\
\colhead{} & \colhead{(Mpc$^{-3}$ mag$^{-1}$)} & \colhead{} & \colhead{(Mpc$^{-3}$ mag$^{-1}$)} & \colhead{} & \colhead{(Mpc$^{-3}$ mag$^{-1}$)} & \colhead{} & \colhead{(Mpc$^{-3}$ mag$^{-1}$)} }
\startdata
\multicolumn{2}{c}{$z\sim2$ galaxies} & \multicolumn{2}{c}{$z\sim4$ galaxies} & \multicolumn{2}{c}{$z\sim6$ galaxies} & \multicolumn{2}{c}{$z\sim8$ galaxies}\\
$-$18.75 & 0.0046$_{-0.0006}^{+0.0006}$ & $-$18.75 & 0.0050$_{-0.0017}^{+0.0017}$ & $-$18.75 & 0.0029$_{-0.0007}^{+0.0007}$ & $-$18.75 & 0.0018$_{-0.0006}^{+0.0006}$\\
$-$18.25 & 0.0074$_{-0.0008}^{+0.0008}$ & $-$18.25 & 0.0075$_{-0.0021}^{+0.0021}$ & $-$18.25 & 0.0051$_{-0.0010}^{+0.0010}$ & $-$18.25 & 0.0016$_{-0.0007}^{+0.0007}$\\
$-$17.75 & 0.0067$_{-0.0008}^{+0.0008}$ & $-$17.75 & 0.011$_{-0.003}^{+0.003}$ & $-$17.75 & 0.0096$_{-0.0015}^{+0.0015}$ & $-$17.75 & 0.0037$_{-0.0012}^{+0.0012}$\\
$-$17.25 & 0.012$_{-0.001}^{+0.002}$ & $-$17.25 & 0.025$_{-0.005}^{+0.006}$ & $-$17.25 & 0.0092$_{-0.0017}^{+0.0021}$ & $-$17.25 & 0.0058$_{-0.0020}^{+0.0023}$\\
$-$16.75 & 0.017$_{-0.001}^{+0.004}$ & $-$16.75 & 0.0094$_{-0.0047}^{+0.0062}$ & $-$16.75 & 0.014$_{-0.003}^{+0.004}$ & $-$16.75 & 0.0063$_{-0.0030}^{+0.0040}$\\
$-$16.25 & 0.017$_{-0.002}^{+0.005}$ & $-$16.25 & 0.030$_{-0.012}^{+0.020}$ & $-$16.25 & 0.014$_{-0.004}^{+0.007}$ & $-$16.25 & 0.019$_{-0.008}^{+0.013}$\\
$-$15.75 & 0.018$_{-0.002}^{+0.008}$ & $-$15.75 & 0.063$_{-0.026}^{+0.051}$ & $-$15.75 & 0.019$_{-0.006}^{+0.013}$ & $-$15.75 & 0.016$_{-0.011}^{+0.018}$\\
$-$15.25 & 0.027$_{-0.004}^{+0.016}$ & $-$15.25 & 0.090$_{-0.045}^{+0.095}$ & $-$15.25 & 0.059$_{-0.017}^{+0.045}$ & $-$15.25 & 0.014$_{-0.011}^{+0.026}$\\
$-$14.75 & 0.026$_{-0.006}^{+0.022}$ & $-$14.75 & $<$0.088 & $-$14.75 & 0.087$_{-0.032}^{+0.087}$ & $-$14.75 & $<$0.068\\
$-$14.25 & 0.031$_{-0.011}^{+0.035}$ & $-$14.25 & $<$0.35 & $-$14.25 & 0.085$_{-0.048}^{+0.124}$ & $-$14.25 & 0.11$_{-0.09}^{+0.23}$\\
$-$13.75 & 0.054$_{-0.022}^{+0.075}$ & $-$13.75 & 0.93$_{-0.74}^{+2.20}$ & $-$13.75 & 0.13$_{-0.09}^{+0.24}$ & $-$13.75 & 0.53$_{-0.39}^{+1.03}$\\
$-$13.25 & 0.042$_{-0.030}^{+0.087}$ &  &  & $-$13.25 & 0.32$_{-0.22}^{+0.66}$ & $-$13.25 & 0.20$_{-0.16}^{+0.99}$\\
$-$12.75 & $<$0.11 &  &  & $-$12.75 & $<$0.97 &  & \\
$-$12.25 & 0.13$_{-0.10}^{+0.41}$ &  &  &  &  &  & \\
\multicolumn{2}{c}{$z\sim3$ galaxies} & \multicolumn{2}{c}{$z\sim5$ galaxies} & \multicolumn{2}{c}{$z\sim7$ galaxies} & \multicolumn{2}{c}{$z\sim9$ galaxies}\\
$-$18.75 & 0.0046$_{-0.0006}^{+0.0006}$ & $-$18.75 & 0.0016$_{-0.0005}^{+0.0005}$ & $-$18.75 & 0.0014$_{-0.0005}^{+0.0005}$ & $-$18.75 & 0.0003$_{-0.0002}^{+0.0003}$\\
$-$18.25 & 0.0076$_{-0.0008}^{+0.0008}$ & $-$18.25 & 0.0032$_{-0.0009}^{+0.0009}$ & $-$18.25 & 0.0019$_{-0.0006}^{+0.0006}$ & $-$18.25 & 0.0007$_{-0.0005}^{+0.0005}$\\
$-$17.75 & 0.012$_{-0.001}^{+0.001}$ & $-$17.75 & 0.0034$_{-0.0013}^{+0.0013}$ & $-$17.75 & 0.0027$_{-0.0008}^{+0.0008}$ & $-$17.75 & 0.0006$_{-0.0005}^{+0.0005}$\\
$-$17.25 & 0.017$_{-0.001}^{+0.002}$ & $-$17.25 & 0.0090$_{-0.0032}^{+0.0036}$ & $-$17.25 & 0.0090$_{-0.0018}^{+0.0021}$ & $-$17.25 & 0.0009$_{-0.0007}^{+0.0010}$\\
$-$16.75 & 0.021$_{-0.001}^{+0.004}$ & $-$16.75 & 0.014$_{-0.006}^{+0.008}$ & $-$16.75 & 0.013$_{-0.003}^{+0.005}$ & $-$16.75 & 0.0016$_{-0.0013}^{+0.0022}$\\
$-$16.25 & 0.022$_{-0.002}^{+0.006}$ & $-$16.25 & 0.013$_{-0.009}^{+0.013}$ & $-$16.25 & 0.013$_{-0.004}^{+0.007}$ & $-$16.25 & 0.0091$_{-0.0065}^{+0.0094}$\\
$-$15.75 & 0.021$_{-0.002}^{+0.009}$ & $-$15.75 & 0.029$_{-0.021}^{+0.035}$ & $-$15.75 & 0.025$_{-0.009}^{+0.019}$ & $-$15.75 & 0.0021$_{-0.0017}^{+0.0066}$\\
$-$15.25 & 0.028$_{-0.004}^{+0.016}$ & $-$15.25 & $<$0.051 & $-$15.25 & 0.030$_{-0.015}^{+0.032}$ & $-$15.25 & $<$0.037\\
$-$14.75 & 0.040$_{-0.007}^{+0.029}$ & $-$14.75 & 0.16$_{-0.11}^{+0.24}$ & $-$14.75 & 0.037$_{-0.025}^{+0.054}$ & $-$14.75 & $<$0.097\\
$-$14.25 & 0.057$_{-0.013}^{+0.053}$ & $-$14.25 & 0.19$_{-0.15}^{+0.41}$ & $-$14.25 & 0.098$_{-0.063}^{+0.156}$ & $-$14.25 & $<$0.27\\
$-$13.75 & 0.086$_{-0.024}^{+0.100}$ & $-$13.75 & $<$3.0 & $-$13.75 & 0.082$_{-0.066}^{+0.212}$ & $-$13.75 & $<$0.98\\
$-$13.25 & 0.077$_{-0.035}^{+0.126}$ &  &  & $-$13.25 & $<$0.56 & $-$13.25 & $<$5.8\\
$-$12.75 & 0.18$_{-0.08}^{+0.33}$ &  &  & $-$12.75 & $<$3.7 &  & \\
$-$12.25 & 0.097$_{-0.078}^{+0.308}$ &  &  &  &  &  & 
\enddata

\tablenotetext{a}{The LF results presented in this table are derived
  using Eq.~\ref{eq:phi} from \S\ref{sec:lfresults}.  One advantage of
  the results presented in this table is that each binned value is
  independent of the other bins.  On the other hand, no account is
  made for uncertainties in the magnification map in deriving
  constraints on the overall shape of the $UV$ LF, so this is one
  disadvantage.}
\tablenotetext{b}{Derived at a rest-frame wavelength of 1600\AA.}
\tablenotetext{c}{The $1\sigma$ upper limits indicate the upper limits
  if one adopts the Bouwens et al.\ (2022a) size-luminosity scalings (implemented
  in a similar manner to \S5.4 of Bouwens et al.\ 2017b).}
\end{deluxetable*}

\begin{figure*}
\epsscale{0.9}
\plotone{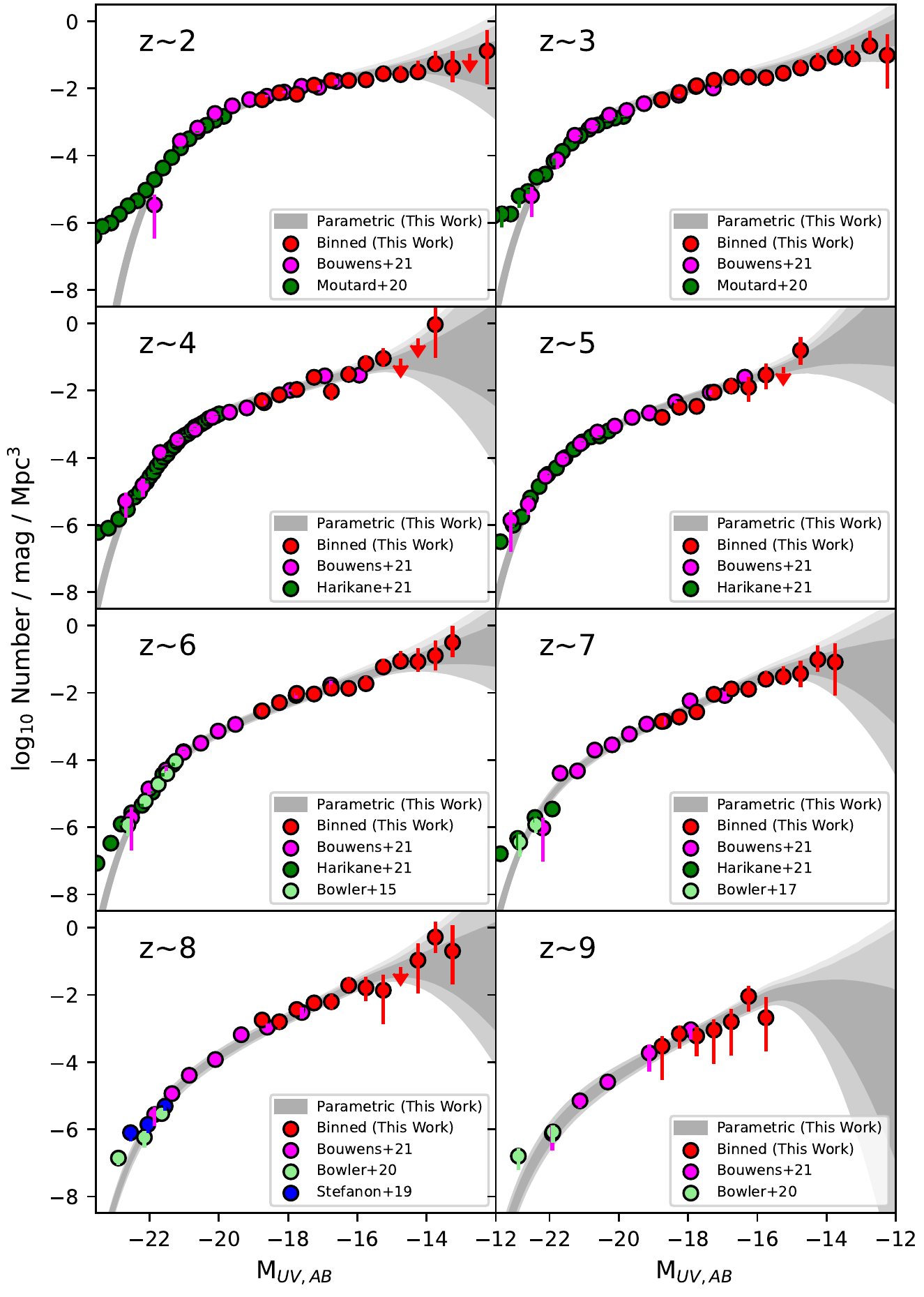}
\caption{The 68\% and 95\% likelihood contours (\textit{dark and light
    grey shaded regions, respectively}) we derive on the shape of the
  $UV$ LFs at $z\sim2$, $z\sim3$, $z\sim4$, $z\sim5$, $z\sim6$,
  $z\sim7$, $z\sim8$, and $z\sim9$ based on our lensed HFF samples and
  the presented blank-field constraints on the LF.  The presented
  contours give equal weight to the contours we derive treating the
  CATS, Sharon/Johnson, \textsc{GLAFIC}, and Keeton models as the
  truth in our forward-modeling procedure (Bouwens et al.\ 2017b) and
  then recover LF results using the median magnification factors from
  the other parametric lensing models.  The red solid circles show the
  binned constraints (with $1\sigma$ error bars) we obtain by dividing
  the number of sources in a luminosity bin $N_m$ by the selection
  volume $V_m$ in that bin (Table~\ref{tab:swlf}).  The red upper
  limits are $1\sigma$.  The light red shaded region ($2\sigma$)
  indicates the LF constraints if we adopt larger sizes in modeling
  the completeness of the faintest sources, as per the Shibuya et
  al.\ (2015) size-luminosity relations.  At $z\sim2$-9, the
  blank-field constraints are from Bouwens et al.\ (2021a), but with
  comparisons to the results from Bouwens et al.\ (2007), Bouwens et
  al.\ (2015), Bowler et al.\ (2015), Bowler et al.\ (2017), Bouwens
  et al.\ (2019), Stefanon et al.\ (2019), and Bowler et al.\ (2020)
  also shown.  \label{fig:lf29p}}
\end{figure*}

The observed agreement between the two results is remarkable given the
enormous differences between the two approaches and their different
challenges.  For example, while blank-field probes provide us with
less leverage in luminosity to constrain the faint-end slope $\alpha$
of the $UV$ LF, they are much less sensitive to assumptions about
source size near the detection limits (since almost all sources at
these limits are unresolved) and sample sufficient volume to obtain
much better constraints on the bright end of the LF.  Meanwhile, for
lensing cluster probes, despite the sensitivity of this approach to an
accurate modeling of the sizes (e.g., Bouwens et al.\ 2017a), the
additional leverage provided by lensing allows us to probe
substantially fainter in luminosity, providing us with much greater
leverage to probe the faint-end slope.  Figure~\ref{fig:leverage}
illustrates this leverage quite well.

We emphasize that our use of very small sizes is critical for
achieving consistent faint-end slope $\alpha$ results to blank-field
studies.  As demonstrated in both Bouwens et al.\ (2017a) and Bouwens
et al.\ (2022a), if we had instead assumed that galaxies followed a
more standard size-luminosity relation (e.g., Huang et al.\ 2013;
Shibuya et al.\ 2015), completeness would be a decreasing function of
the magnification of the source.  This would result in much steeper
faint-end slopes $\alpha$ and significantly higher volume densities of
sources $>$$-$15 mag (Bouwens et al.\ 2017a, 2022a).

This is the first time such consistent faint-end slope results have
been found between blank-field and lensing-cluster analyses over such
an extended range in redshift.  The independence of the two approaches
and their consistency strongly suggests that we can combine the two
approaches to arrive at even more robust determinations of the overall
shape and evolution of the $UV$ LF.

\subsubsection{Leveraging Both Blank-Field and HFF Results\label{sec:lfresults}}

Having demonstrated we can use lensed samples of $z=2$-9 galaxies to
obtain constraints on the faint-end slope $\alpha$ consistent with
blank-field studies, we now leverage both data sets to arrive at our
best overall estimate on the shape of the $UV$ LFs at $z=2$-9.

For our blank-field constraints on the $z\sim2$-9 LF results, we rely
on the likelihood constraints Bouwens et al.\ (2021a) derived from the
comprehensive set of HST fields.  Following the treatment we provided
in Bouwens et al.\ (2017b), we again allow for a 20\% uncertainty in
the relative normalization of the value of blank-field and cluster LF
results at $z\sim2$, 3, 6, 7, 8, and 9.  This 20\% uncertainty
includes a $\sim$15\% contribution from cosmic variance (Robertson et
al.\ 2014) and a $\sim$10\% uncertainty in the selection volume
calculation for both the bright and faint-end of the LFs.  For our
$z\sim4$ and $z\sim5$ results, we allow for a 26\% and 18\% cosmic
variance uncertainty (and 28\% and 22\% relative uncertainty in total)
owing to our consideration of lensed sources behind only 2 and 4
clusters, respectively, due to the challenge in identifying $z\sim4$-5
galaxies behind those clusters without substantial contamination.

Additionally and motivated by the modest uncertainties in the size
distribution of the faintest sources, we allow for modest
uncertainties in the selection efficiency of galaxies in the faintest
magnitude intervals.  In particular, we consider both the impact of
increasing the selection efficiencies in the 27.5-28.0 mag, 28.0-28.5
mag, 28.5-29.0 mag, and $>$29.0 mag intervals by 2\%, 5\%, 20\%, and
30\% and decreasing the selection efficiencies by the same amount and
then marginalize the results over both the regularly calculated
efficiencies and the two others.  These percentage changes indicate
the approximate impact of $\sim$10\% differences between the model and
derived S/N for sources of a given magnitude.  For our derived LF
parameters at $z\sim 2$ and 3, similar adjustments were made to
selection volumes to determine the impact of possible systematics in
the estimated volumes, but 1.2 and 0.7 mag brighter (corresponding to
the shallower $UV_{275}$ and $U_{336}$ coverage over the HFF clusters
and also shallower $z\sim2$ and $z\sim3$ selections).

In Figure~\ref{fig:lf29p}, we show the 68\% and 95\% confidence
intervals on our LF results marginalizing over the results for the
four families of parametric models, i.e., \textsc{GLAFIC}, CATS,
Sharon/Johnson, and Keeton, while comparing against independent
constraints on the bright-end of the $UV$ LFs from various blank-field
probes (Bouwens et al.\ 2007, 2015a, 2019, 2021a; Bowler et al.\ 2015,
2017, 2020; Stefanon et al.\ 2019; Harikane et al.\ 2021).  Along with
the formal confidence intervals, we also show the allowed LF results
(\textit{light shaded region}) if the mean sizes of lower luminosity
galaxies are more consistent with the Shibuya et al.\ (2015) scalings.
We stress that the observational results of Bouwens et al.\ (2017a),
Kawamata et al.\ (2018), Bouwens et al.\ (2022a), and Yang et
al.\ (2022) strongly suggest the true size distribution is
significantly smaller than this, but show these allowed regions to
present the possible systematic error.

For convenience, in Table~\ref{tab:likelihood}, we include a
compilation of our derived 68\% confidence intervals for our $z=2$-9
LF results from $z\sim9$ to $z\sim2$.  The formal uncertainties
provided in Table~\ref{tab:likelihood} include not only the formal
68\% confidence results from our MCMC fitting results using the
parametric models, but include systematic uncertainties on the volume
densities if the true sizes of lower luminosity galaxies are much
larger than implied by the Bouwens et al.\ (2017a, 2017c), Kawamata et
al.\ (2018), and Bouwens et al.\ (2022a) results.

We also include a simple binned version of our LF results in
Figure~\ref{fig:lf29p} using the relation
\begin{equation}
  \phi_m = \frac{N_m}{V_m}
  \label{eq:phi}
  \end{equation}
where $\phi_m$ is the derived volume density of sources in magnitude
bin $m$, $N_m$ is the number of sources in magnitude bin $m$, and
$V_m$ is the estimated selection volume in magnitude bin $m$.  We
calculate $V_m$ as follows:
\begin{equation}
V_m = \int_{A} \int_{dz} C(z,m,\mu) \frac{1}{\mu(A)}
\frac{d^2 V(z)}{dz dA} dz dA
\end{equation}
where $dA$ is a differential area in the image plane, $C(z,m,\mu)$ is
the selection completeness as a function of redshift $z$, apparent
magnitude $m$, an the magnification factor $\mu$, and $\frac{d^2
  V(z)}{dz dA}$ is a differential volume element.  The plotted
uncertainties in Figure~\ref{fig:lf29p} include not only the formal
Poissonian uncertainties but also the systematic uncertainties on the
volume densities if the true sizes of lower luminosity galaxies are as
given by Bouwens et al.\ (2022a) size-luminosity relation instead of
adopting point-source sizes.  No account is made for the impact of
uncertainties in magnification factors on the binned constraints.  For
convenience, we list the binned constraints shown in
Figure~\ref{fig:lf29p} in Table~\ref{tab:swlf}.

Results derived using the individual parametric and non-parametric
magnification models are presented in Table~\ref{tab:lfparm} of
Appendix A.  Figures~\ref{fig:lf2} and \ref{fig:lf6} from Appendix A
show the 68\% and 95\% confidence intervals on the $z\sim2$ and
$z\sim6$ LF results, respectively, using the same set of parametric
and non-parametric models.  In general, very similar LF results are
obtained utilizing different lensing models, with derived parameters
typically differing by much less than the formal uncertainties derived
for a single model.  In general, LFs derived using the non-parametric
lensing models showed higher values for the curvature parameter
$\delta$.  This appears to derive from the larger differences seen
between the magnification factors from these models and those in the
parametric lensing models and the impact this has in washing out
potential turn-overs at the faint-end of the $UV$ LFs.

\begin{figure*}
\epsscale{1.18}
\plotone{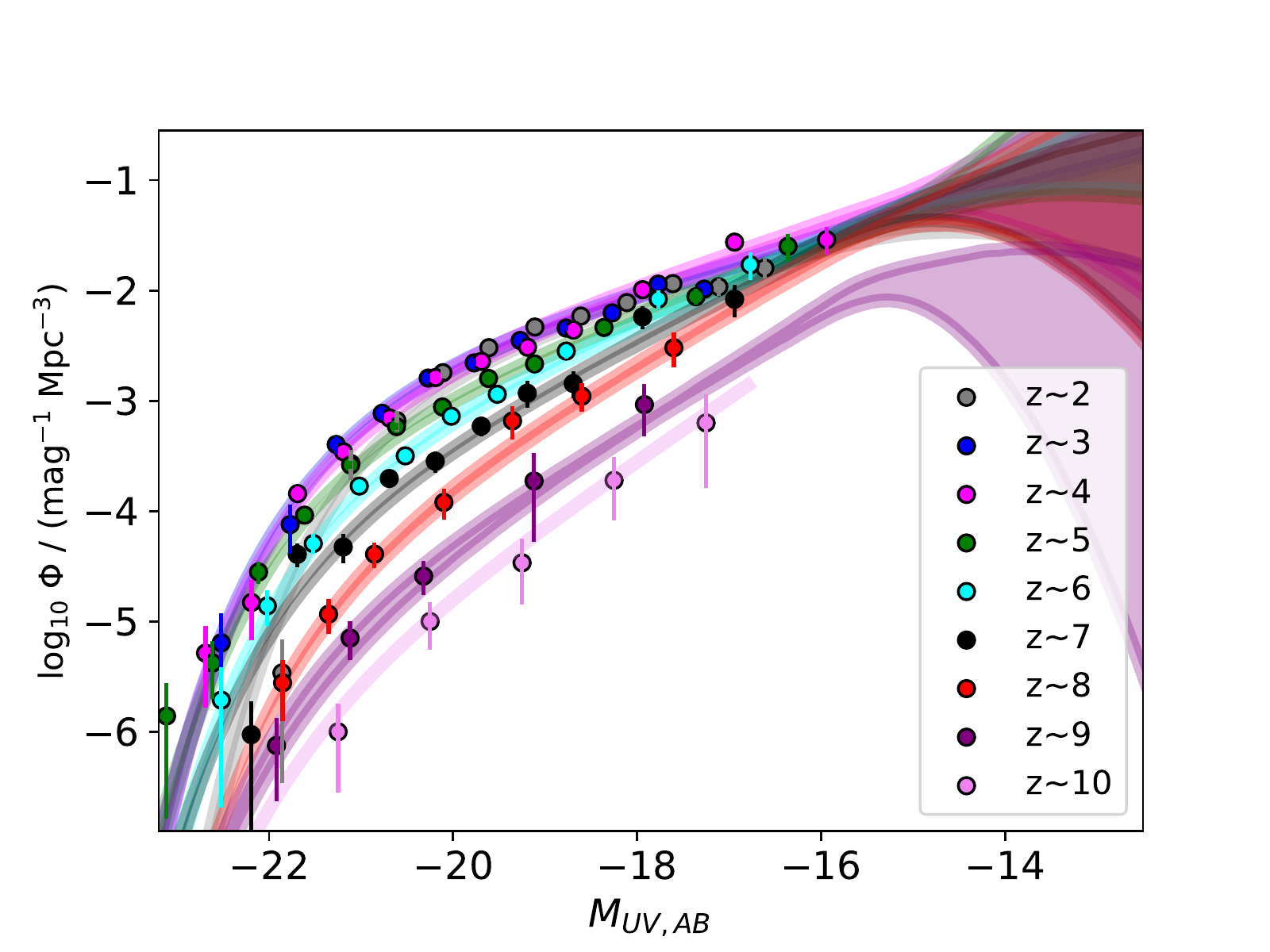}
\caption{The 68\% likelihood contours we derive on the shape of the
  faint end of the $UV$ LFs at $z\sim2$, $z\sim3$, $z\sim4$, $z\sim5$,
  $z\sim6$, $z\sim7$, $z\sim8$, and $z\sim9$ based on our lensed HFF
  samples and the presented constraints on the LF from blank-field
  studies.  The solid circles show our LF constraints from the Bouwens
  et al.\ (2021a) blank-field analysis.  The $z\sim10$ LF results
  shown here are from Oesch et al.\ (2018a) and rely on both
  blank-field and lensing-field results.\label{fig:lfall}}
\end{figure*}

\begin{figure}
\epsscale{1.17}
\plotone{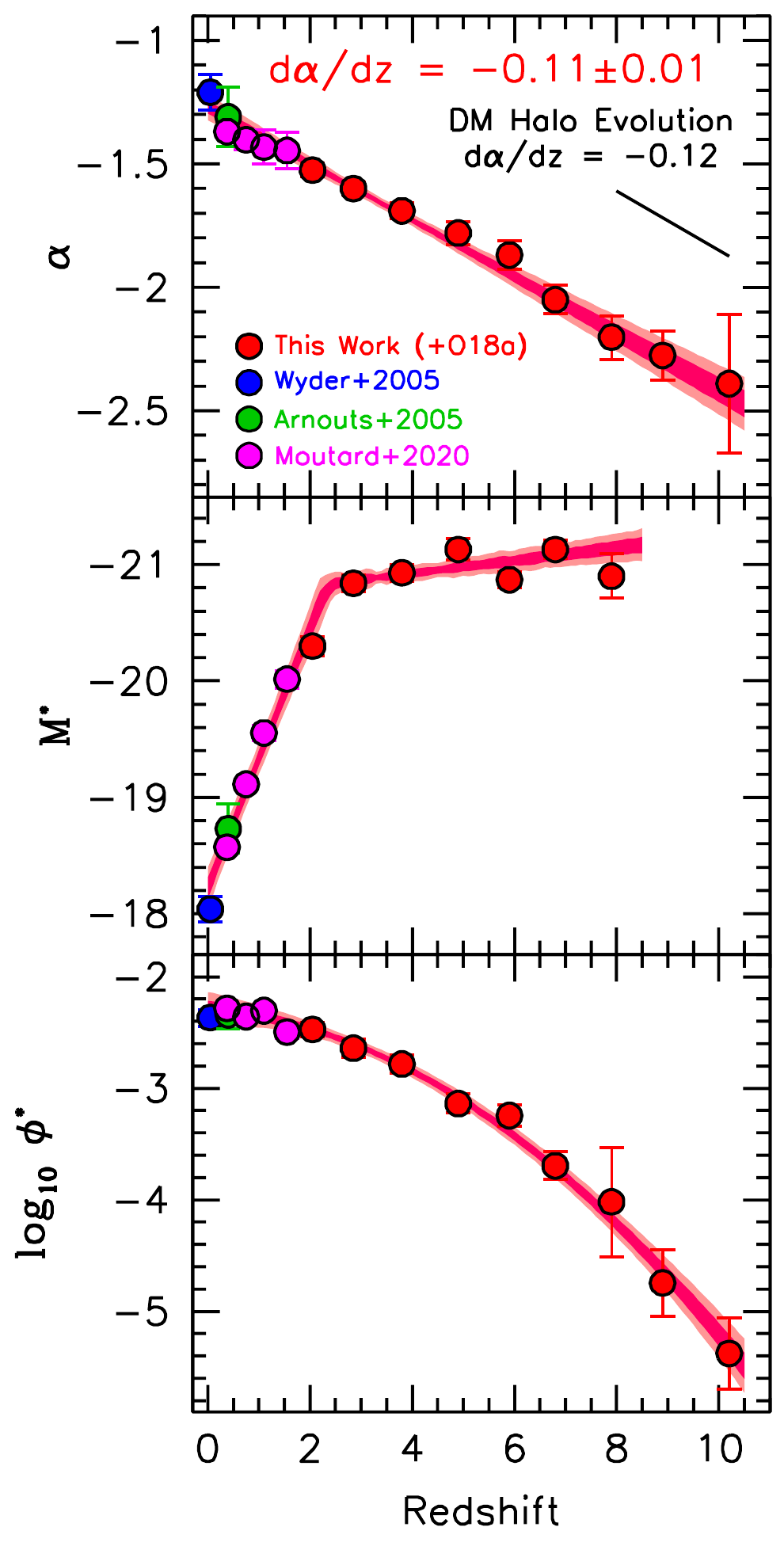}
\caption{Apparent evolution of the faint-end slope $\alpha$
  (\textit{upper panel}), characteristic luminosities $M^*$
  (\textit{center panel}), and normalization $\phi^*$ (\textit{lower
    panel}) of the $UV$ LF with redshift.  The red circles present the
  current LF constraints at $z=2$-9 and those of Oesch et al.\ (2018a)
  at $z\sim10$, while the blue, green, and magenta circles give the
  Wyder et al.\ (2005), Arnouts et al.\ (2005), and Moutard et
  al.\ (2020) constraints at $z\sim0.05$, $z\sim0.4$, and over the
  redshift range $z\sim0.3$-1.5.  The shaded red line illustrates our
  best-fit relation for the evolution (\S\ref{sec:schparm}).  The
  evolution of the faint-end slope $\alpha$ is now extremely well
  determined as a function of redshift and remarkably consistent with
  the evolution expected based on the change in slope of the halo mass
  function (Tacchella et al.\ 2013, 2018; Bouwens et al.\ 2015a,
  2021a; Mason et al.\ 2015; Mashian et al.\ 2016; Park et al.\ 2019).
  Previous work by Bouwens et al.\ (2015), Parsa et al.\ (2016), and
  Finkelstein (2016) arrived at very similar d$\alpha$/dz trends as
  what is found here by fitting to the available $\alpha$
  determinations.\label{fig:schevol}}
\end{figure}

A summary of the mean LF parameters $M^*$, $\phi^*$, $\alpha$, and
$\delta$ we derive on the basis of the four parametric magnification
models (\textsc{GLAFIC}, CATS, Sharon/Johnson, Keeton) and the two
non-parametric models (\textsc{grale} and Diego) is provided in
Table~\ref{tab:lfparmsum} and are indicated by the descriptors
``Parametric'' and ``Non-parametric,'' respectively.  Finally,
figure~\ref{fig:lfall} shows our best-fit $z\sim2$, $z\sim3$,
$z\sim4$, $z\sim5$, $z\sim6$, $z\sim7$, $z\sim8$, and $z\sim9$ LFs,
along with the $z\sim10$ results from Oesch et al.\ (2018a).  The
$z\sim10$ LF results from Oesch et al.\ (2018a) rely on both
blank-field and lensing-field results.  From the plotted constraints,
it is clear that the $UV$-bright galaxies undergo a much more rapid
evolution in volume density than galaxies at the faint end of the LF.
In the next section, we parameterize the evolution of the $UV$ LF in
terms of convenient fitting formulas.

\subsection{Evolution of the Schechter Parameters with Cosmic Time\label{sec:schparm}}

The availability of deep lensed sample of $z=2$-9 galaxies behind the
HFF clusters have made it possible to significantly improve our
constraints on the faint-end slope of the $UV$ LF and therefore the
evolution of the $UV$ LF from $z\sim10$ to $z\sim2$.

Our estimates of the Schechter parameters in Table~\ref{tab:lfparmsum}
provide a good illustration of this.  Comparing the uncertainties on
the faint-end slope $\alpha$ from our most recent blank-field
determinations, i.e., Bouwens et al.\ (2021a), and those combining
blank-field constraints with the lensing cluster constraints, we have
been able to reduce the uncertainties on the faint-end slope $\alpha$
by a factor of $\sim$1.5, 2, 2, 2, and 2 at $z\sim5$, 6, 7, 8, and 9,
respectively.

\begin{figure*}
\epsscale{1.17}
\plotone{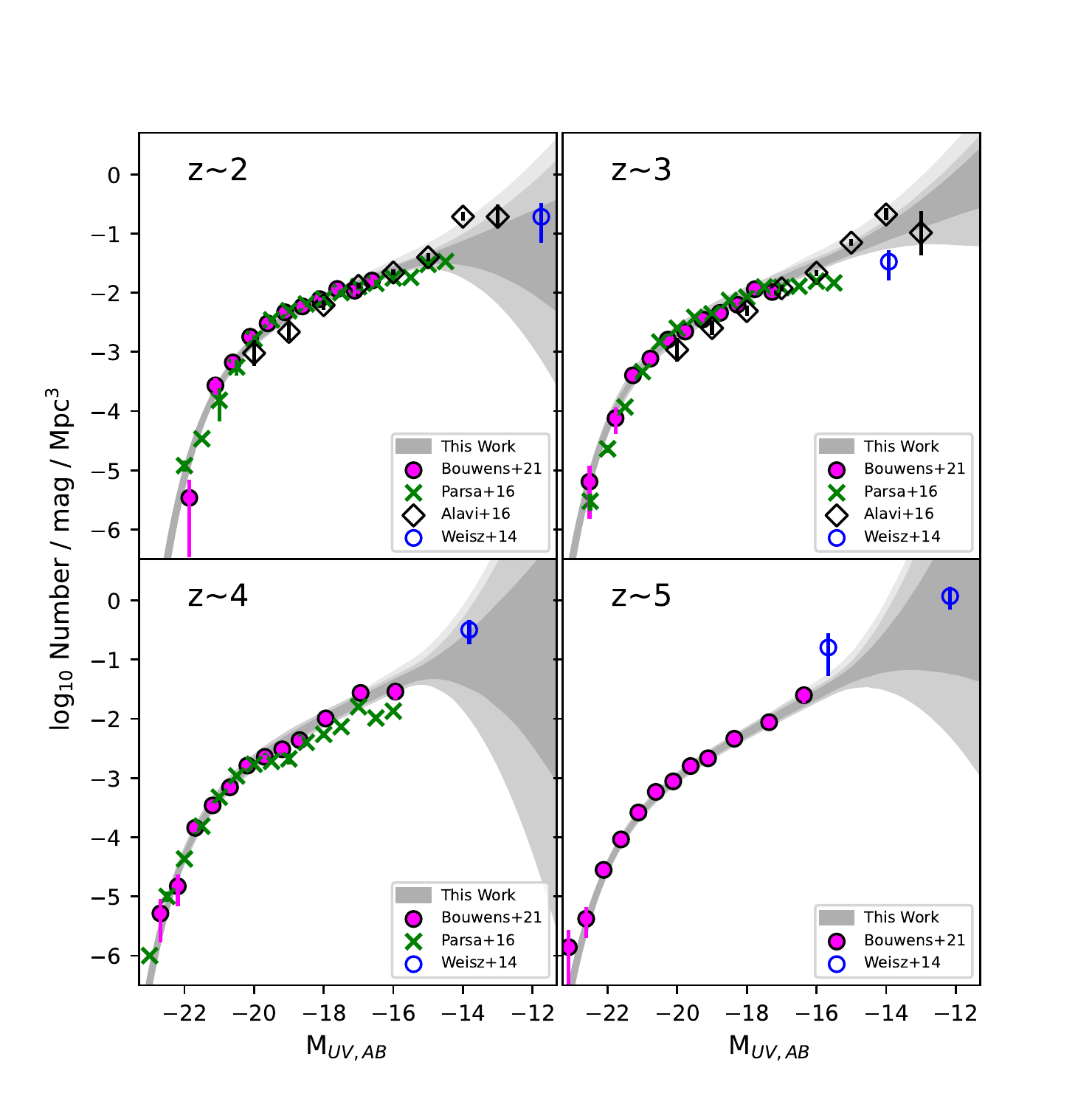}
\caption{Comparison of the new $z=2$-5 $UV$ LFs we have derived from
  the HFFs (\textit{light and dark grey shaded regions}) and the new
  blank-field results of Bouwens et al.\ (2021a: \textit{solid magenta
    circles}) against previous LF results in the literature including
  those from Parsa et al.\ (2016: \textit{green crosses}), Alavi et
  al.\ (2016: \textit{black open diamonds}), and Weisz et al.\ (2014:
  \textit{open blue circles}).  While the comparisons here focus on
  previous results focusing on the faint-end of the $UV$ LF,
  comparisons to brighter $UV$ LF work (e.g., Steidel et al.\ 1999;
  Reddy \& Steidel 2009; Oesch et al.\ 2010; van der Burg et
  al.\ 2010; Finkelstein et al.\ 2015; McLeod et al.\ 2016; Mehta et
  al.\ 2017; Adams et al.\ 2020) can be found in Bouwens et
  al.\ (2021a).  See \S\ref{sec:litcomp} for
  discussion.\label{fig:comp25}}
\end{figure*}

Taking advantage of these new constraints, we are in position to
further refine our characterization of the evolution in each Schechter
parameters.  Following Bouwens et al.\ (2021a), we fit the evolution
in $\alpha$ as a linear function of redshift, $\log_{10}$ $\phi^*$ as
a quadratic function of redshift, and $M^*$ as a linear function of
redshift, with a break at $z\sim2.5$.  Simultaneously fitting the
present LF constraints over the redshift range $z\sim2$-9 as well as
the Oesch et al.\ (2018a) results at $z\sim10$, we find the following
best-fit relation:
\begin{eqnarray*}
M_{UV} ^{*} =& \left\{\begin{array}{ll}
               (-20.87\pm0.08) + (-1.10\pm0.06) (z - z_t), & \textrm{for}~ z < z_t\\
               (-21.04\pm0.04) + (-0.05\pm0.02) (z - 6), & \textrm{for}~ z > z_t\\               
               \end{array} \right. \\
\phi^* =& (0.38\pm0.03)(10^{-3} \textrm{Mpc}^{-3}) 10^{(-0.35\pm0.01)(z-6)+(-0.027\pm0.004)(z-6)^2}\\
\alpha =& (-1.95\pm0.02) + (-0.11\pm0.01)(z-6)
\end{eqnarray*}
where $z_t = 2.42\pm0.09$.  Figure~\ref{fig:schevol} compares the
observed evolution of $\alpha$, $M^*$, and $\phi^*$ with the above
relation.

\begin{figure*}
\epsscale{1.15}
\plotone{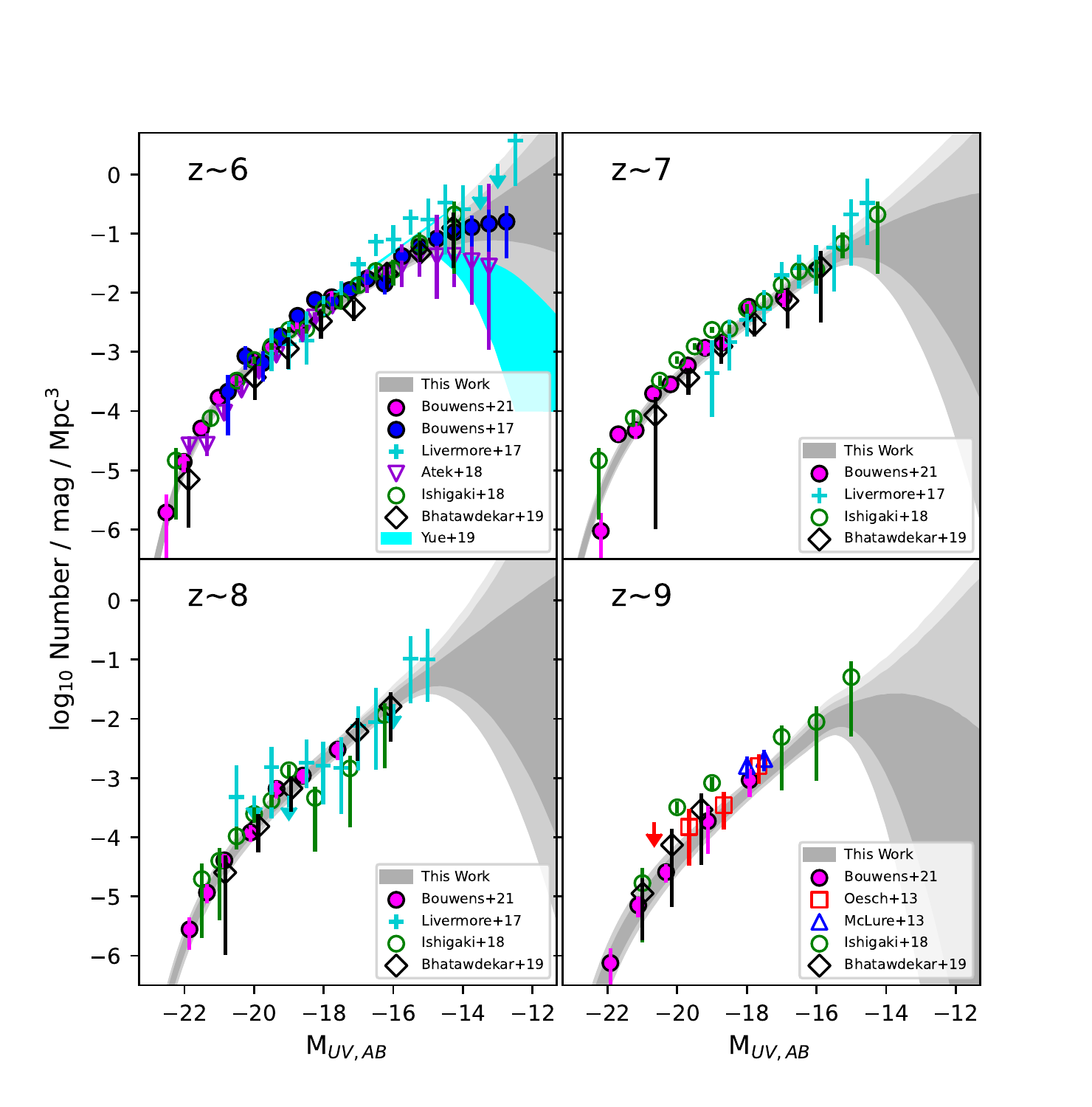}
\caption{Similar to Figure~\ref{fig:comp25} but for our derived LFs at
  $z=6$-9.  Included in these comparisons are the blank-field results
  of Bouwens et al.\ (2021a: \textit{solid magenta circles}), McLure
  et al.\ (2013: \textit{open blue triangles}), and Oesch et
  al.\ (2013: \textit{open red squares}) and the HFF results of
  Livermore et al.\ (2017: \textit{bluish green plus signs}), Bouwens
  et al.\ (2017b: \textit{solid blue circles}), Ishigaki et
  al.\ (2018: \textit{open green circles}), Atek et al.\ (2018:
  \textit{open purple triangles}), Yue et al.\ (2018: \textit{shaded
    cyan region}), and Bhatawdekar et al.\ (2019: \textit{open black
    diamonds}).  See \S\ref{sec:litcomp}.\label{fig:comp69}}
\end{figure*}

As we previously noted (Bouwens et al.\ 2015a, 2021a), the evolution
in the faint-end slope $\alpha$ agrees remarkably well with change in
slope expected based on the evolution of the halo mass function over
the same range in redshift, i.e., $d\alpha/dz \sim -0.12$ (Bouwens et
al.\ 2015a).  This is fairly similar to the faint-end slope evolution
we recovered in Bouwens et al.\ (2015a), i.e.,
$d\alpha/dz\sim-0.10\pm0.03$, the $d\alpha/dz\sim -0.11\pm0.01$ trends
derived by Parsa et al.\ (2016) and Finkelstein (2016) fitting to the
available $z\sim0$-10 and $z\sim4$-10 LF results, respectively, and
$d\alpha/dz\sim-0.09\pm0.05$ trend recently derived by Bowler et
al.\ (2020: but who express the LF evolution they derive using a
double power-law form).

The slow evolution in the characteristic luminosity $M^*$ at $z>z_t$
seems very likely to be a consequence of the fact that $UV$ luminosity
reaches a maximum value of $\sim-21$ to $-23$ mag due to the
increasing importance of dust extinction in the highest stellar mass
and SFR sources (Bouwens et al.\ 2009a; Reddy et al.\ 2010).  Finally,
as Bouwens et al.\ (2015a, 2021a) have argued, the evolution in
$\phi^*$ can readily be explained by evolution in the halo mass
function and no significant evolution in the star formation
efficiency.  As in Bouwens et al.\ (2021a), we find that $\log_{10}$
$\phi^*$ depends on redshift with a clear quadratic dependence (but
here significant at $6\sigma$ instead of $4\sigma$).

\section{Discussion}

\subsection{Comparison with Previous LF Results\label{sec:litcomp}}

Before discussing in more detail the implications of the present LF
determinations, it makes sense to compare our new results with the
many previous determinations of these LFs that exist in the
literature.  Comparing with previous work is very valuable for
improving LF determinations in general, as it allows us to identify
differences in the results and ascertain the best path to improve LF
results in the future.  

To this end, we present our new $z\sim2$, 3, 4, 5, 6, 7, 8, and 9 LF
results in Figures~\ref{fig:comp25} and \ref{fig:comp69} using the
68\% and 95\% confidence intervals we have derived (\textit{light
  grey} and \textit{dark grey} shaded regions, respectively).  For the
comparisons we provide, we will focus on previous LF results that
provided results on the faint-end of the $z=2$-9 LFs.  In the
preparatory work to this using blank-field data (Bouwens et
al.\ 2021a), we already provided a significant discussion of previous
LF results which are more relevant for the bright end of the $UV$ LFs
(their \S4.1).

To help with the discussion of various faint-end LF results, we also
include on Figure~\ref{fig:comp25} and \ref{fig:comp69} the
approximate regions in parameter space allowed if we take galaxy sizes
to follow the Bouwens et al.\ (2022a) size-luminosity relation
(\textit{light shaded regions above the nominal 68\% and 95\%
  contours}).

\begin{deluxetable}{ccccc}
  \renewcommand{\arraystretch}{0.75}
\tabletypesize{\footnotesize}
\tablecaption{68\% and 95\% Confidence Intervals on the $UV$ Luminosity Density
at $z\sim2$-9 to Various Limiting Luminosities\tablenotemark{a,b}\label{tab:lumd}}
\tablehead{
\colhead{} & \multicolumn{4}{c}{$\log_{10} \rho_{UV}$ ($UV$ Luminosity Density)}\\
\colhead{} & \multicolumn{4}{c}{(ergs s$^{-1}$ Hz$^{-1}$ Mpc$^{-3}$)}\\
\colhead{} & \multicolumn{2}{c}{Lower Bound} & \multicolumn{2}{c}{Upper Bound}\\
\colhead{Faint-End Limit} & \colhead{95\%} & \colhead{68\%} & \colhead{68\%} & \colhead{95\%}}
\startdata
& \multicolumn{4}{c}{$z\sim2$}\\
$M_{UV}<-17$ & 26.38 & 26.41 & 26.46 & 26.49\\
$M_{UV}<-15$ & 26.46 & 26.49 & 26.54 & 26.57\\
$M_{UV}<-13$ & 26.48 & 26.51 & 26.57 & 26.60\\
$M_{UV}<-10$ & 26.48 & 26.51 & 26.59 & 26.68\\
\\
& \multicolumn{4}{c}{$z\sim3$}\\
$M_{UV}<-17$ & 26.45 & 26.51 & 26.64 & 26.70\\
$M_{UV}<-15$ & 26.52 & 26.58 & 26.72 & 26.77\\
$M_{UV}<-13$ & 26.56 & 26.62 & 26.76 & 26.81\\
$M_{UV}<-10$ & 26.58 & 26.66 & 26.88 & 26.94\\
\\
& \multicolumn{4}{c}{$z\sim4$}\\
$M_{UV}<-17$ & 26.42 & 26.47 & 26.56 & 26.60\\
$M_{UV}<-15$ & 26.52 & 26.56 & 26.66 & 26.70\\
$M_{UV}<-13$ & 26.56 & 26.62 & 26.82 & 27.01\\
$M_{UV}<-10$ & 26.58 & 26.71 & 29.18 & 31.17\\
\\
& \multicolumn{4}{c}{$z\sim5$}\\
$M_{UV}<-17$ & 26.24 & 26.27 & 26.35 & 26.38\\
$M_{UV}<-15$ & 26.35 & 26.38 & 26.46 & 26.50\\
$M_{UV}<-13$ & 26.39 & 26.44 & 26.62 & 26.87\\
$M_{UV}<-10$ & 26.39 & 26.45 & 28.52 & 30.84\\
\\
& \multicolumn{4}{c}{$z\sim6$}\\
$M_{UV}<-17$ & 26.08 & 26.12 & 26.20 & 26.24\\
$M_{UV}<-15$ & 26.23 & 26.26 & 26.34 & 26.38\\
$M_{UV}<-13$ & 26.28 & 26.32 & 26.43 & 26.50\\
$M_{UV}<-10$ & 26.29 & 26.34 & 26.58 & 27.04\\
\\
& \multicolumn{4}{c}{$z\sim7$}\\
$M_{UV}<-17$ & 25.92 & 25.95 & 26.01 & 26.05\\
$M_{UV}<-15$ & 26.11 & 26.14 & 26.23 & 26.27\\
$M_{UV}<-13$ & 26.15 & 26.20 & 26.33 & 26.41\\
$M_{UV}<-10$ & 26.15 & 26.20 & 26.41 & 26.68\\
\\
& \multicolumn{4}{c}{$z\sim8$}\\
$M_{UV}<-17$ & 25.55 & 25.61 & 25.72 & 25.78\\
$M_{UV}<-15$ & 25.85 & 25.89 & 26.01 & 26.07\\
$M_{UV}<-13$ & 25.90 & 25.96 & 26.20 & 26.41\\
$M_{UV}<-10$ & 25.90 & 25.96 & 26.43 & 28.01\\
\\
& \multicolumn{4}{c}{$z\sim9$}\\
$M_{UV}<-17$ & 24.99 & 25.07 & 25.26 & 25.34\\
$M_{UV}<-15$ & 25.33 & 25.39 & 25.54 & 25.60\\
$M_{UV}<-13$ & 25.36 & 25.43 & 25.65 & 25.75\\
$M_{UV}<-10$ & 25.36 & 25.43 & 25.67 & 25.82
\enddata
\tablenotetext{a}{The derived luminosity densities represent a geometric mean of the LF results derived treating the parametric models as the truth and recovering the LF results using a median of the other parametric models.}
\tablenotetext{b}{See also Figure~\ref{fig:sfrdens} to see these luminosity densities (integrated to $-13$ mag) presented as the equivalent SFR densities a function of redshift.}
\end{deluxetable}

We will structure the comparisons we make to previous work as an
increasing function of redshift:\\

\noindent \textbf{$\mathbf{z\sim2}$-5:} To the present, there have
been only two studies which have reported LF results on the extreme
faint end of the $UV$ LF at $z\sim2$-5, i.e., at $>-16$ mag, for
Lyman-break galaxies, one by Alavi et al.\ (2016) based on lensed
samples of $z\sim1$-3 galaxies identified behind two HFF clusters
Abell 2744 and MACS0717 and Abell 1689 and a second by Parsa et
al.\ (2016) using a deep ($\sim$30.5 mag) photometric redshift
selection over the HUDF.  The faint constraints reported by Alavi et
al.\ (2016) and Parsa et al.\ (2016) lie above and below our own
constraints and reach very different conclusions.  Alavi et
al.\ (2016) report a faint-end slope from their LF results of
$-1.72\pm0.04$ and $-1.94\pm0.06$ at $z\sim2$ and $z\sim3$,
respectively, while Parsa find $-1.31\pm0.04$ and $-1.31\pm0.04$ at
approximately same redshifts.

As illustrated in Figure~\ref{fig:leverage}, the leverage available
from our lensed HFF samples should allow us to quantify the faint-end
slope for the $UV$ LF $z\sim2$ and $z\sim 3$ with great precision, if
an accurate account can be made for various sources of systematic
errors.  Given the consistency of our own blank-field and lensed
measurements of the faint end of the $z\sim2$ and $z\sim3$ LF, what
might drive the lower and higher volume density results obtained by
Parsa et al.\ (2016) and Alavi et al.\ (2016)?  For the faint end of
the Parsa et al.\ (2016) probe, the answer is not entirely clear, as
their LF determinations are in good agreement with both the Bouwens et
al.\ (2021a) blank-field LF results and our lensed results brightward
of $-16.5$ mag and appear to drive the differences in our faint-end
slope inferences.\footnote{In particular, if we fix $M^*$ to $-$19.78
mag, the value found by Parsa et al.\ (2016) and derive $\alpha$ using
the blank-field constraints from Bouwens et al.\ (2021a), we derive
$\alpha=-1.32$, almost identical to what Parsa et al.\ (2016) find.}
Faintward of $-16.5$ mag, differences between the Parsa et al.\ (2016)
LF measurements and our own are more significant.  One potential
concern for the LF determinations of Parsa et al.\ (2016) is the
inclusion of sources to $\sim$30.5-31.0 mag, i.e., at essentially the
detection limit of the HUDF.  At such faint magnitudes, segregating
sources into different redshift bins is more challenging, especially
given that the $UV$ observations Parsa et al.\ (2016) report utilizing
(essential for probing the position of the Lyman break) have a
$5\sigma$ depth of 28 mag, i.e., $\sim$10$\times$ brighter than the
$z\sim1$-4 sources being included in their LFs.

\begin{figure*}
\epsscale{1.17}
\plotone{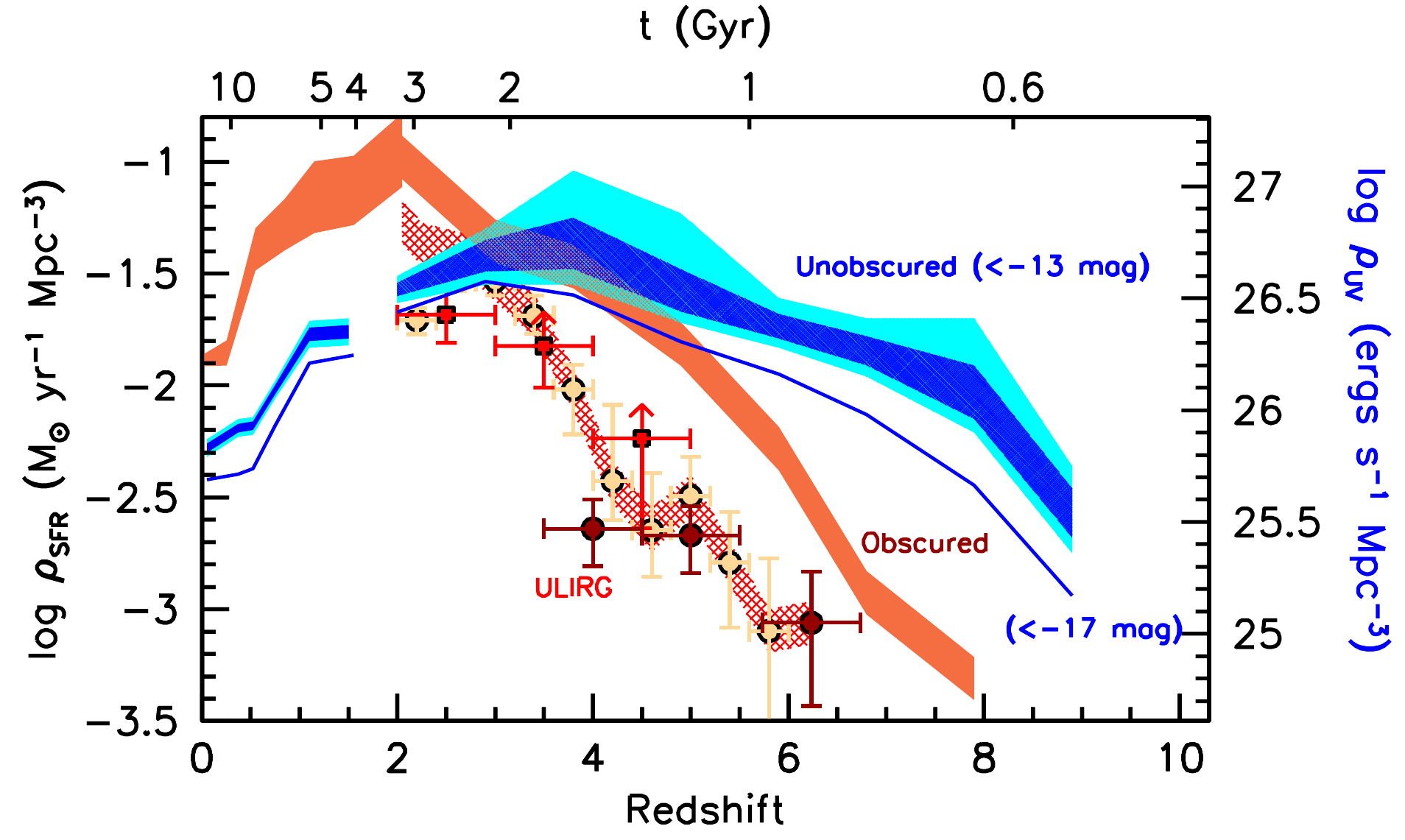}
\caption{68\% and 95\% confidence intervals on the inferred $UV$
  luminosity density and unobscured star formation rate density
  (\textit{blue and cyan shaded regions}) at $z\sim2$-9 derived from
  our new $UV$ LF results integrated down to $-$13 mag
  (\S\ref{sec:lumdens}).  Also shown are the obscured SFR density
  results (\textit{orange-shaded region}) derived by Bouwens et
  al.\ (2020) from $z\sim 8$ to $z\sim2$ by applying the ASPECS IRX
  results to $z\sim4$-8 samples, making use of published ULIRG results
  (\textit{shown with the hatched red region}) by Wang et al.\ (2019:
  \textit{dark red points}), Franco et al.\ (2020: \textit{red
    points}), and Dudzevi{\v{c}}i{\={u}}t{\.{e}} et al.\ (2020:
  \textit{solid brownish-yellow circles}), and adopting the $z=2$-3
  dust extinction estimates from Reddy \& Steidel (2009).  The shaded
  orange region from $z=0$ to $z=2$ shows the constraints on the
  obscured SFR density from Magnelli et al.\ (2013), while the blue
  and cyan shaded regions at $z<2$ show the constraints from Wyder et
  al.\ (2005) and Moutard et al.\ (2020).  The blue lines show the
  $UV$ luminosity densities and SFR densities we derive from our LF
  results brightward of $-$17 mag at $z>2$ and Wyder et al.\ (2005)
  and Moutard et al.\ (2020) find at $z<2$.  As Bouwens et al.\ (2009a,
  2016b), Dunlop et al.\ (2017), and Zavala et al.\ (2021) have
  previously concluded, the bulk of the SFR density is unobscured at
  $z>4$ and obscured at $z<4$.  There is a transition between the two
  regimes at $z=4$.
\label{fig:sfrdens}}
\end{figure*}

\begin{figure*}
  \epsscale{1.17}
  \plotone{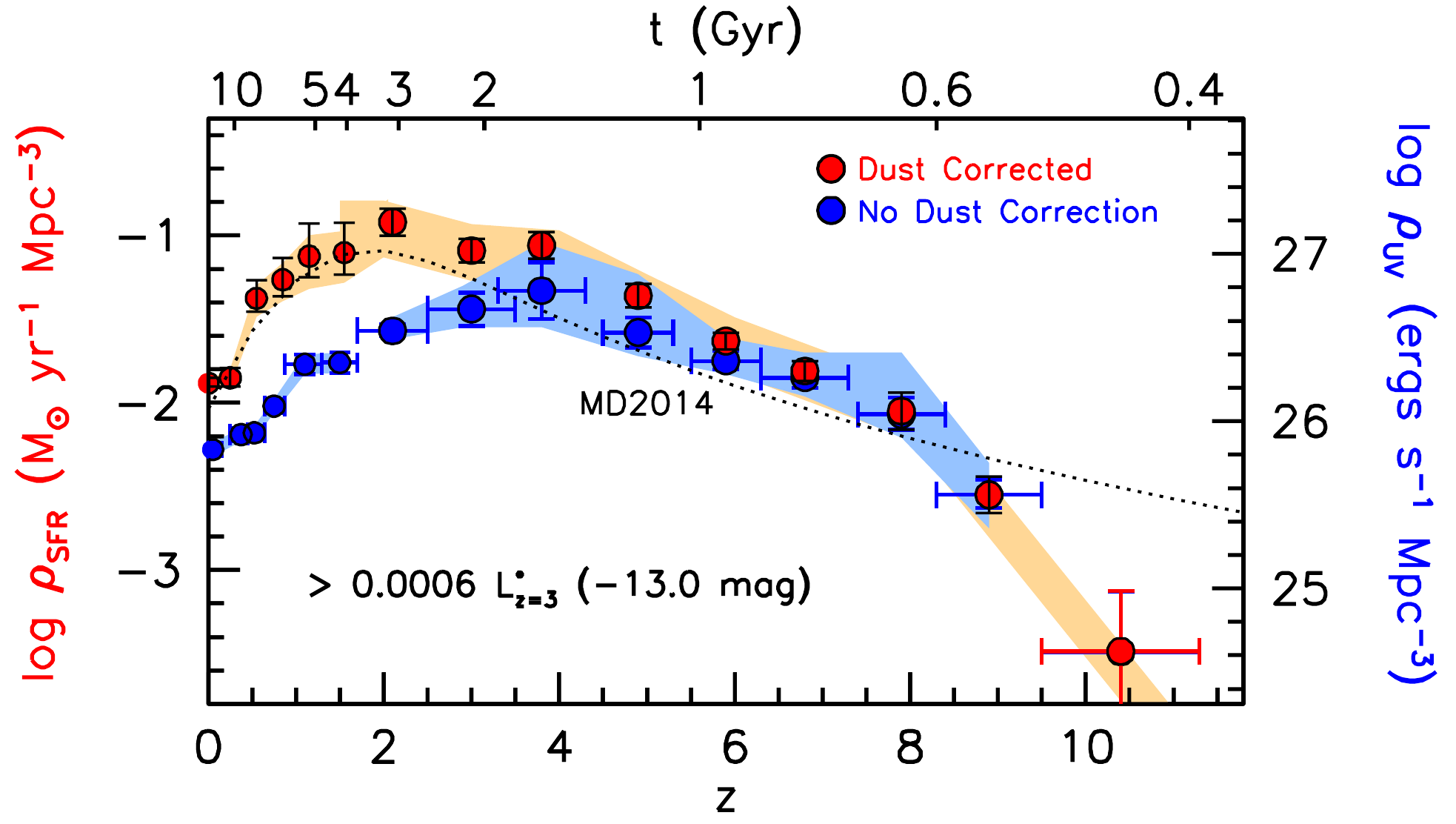}
  \caption{Unobscured and dust-corrected SFR density of the universe
    (\textit{blue} and \textit{red} solid circles, respectively, with
    $1\sigma$ error bars) derived from our new $UV$ LF results at
    $z=2$-9 and integrating down to $-13$ mag (as in
    Figure~\ref{fig:sfrdens}).  The right axis gives the equivalent
    $UV$ luminosity density vs. redshift.  The light blue shaded
    contours indicate approximate 95\% confidence intervals on the
    unobscured SFR and $UV$ luminosity densities, while the light red
    shaded contours illustrate the overall trends in the evolution of
    the SFR density.  The present determinations of the SFR density is
    higher than similar determinations Madau \& Dickinson (2014:
    \textit{dotted line}) due to our integrating $\sim$4 mag further
    down the $UV$ LF (to a faint-end limit of $-13$ mag vs. $-17$ mag
    used by Madau \& Dickinson 2014), use of new constraints on the
    obscured SFR density at $z\geq4$ from ALMA and Herschel (shown in
    Figure~\ref{fig:sfrdens}), and use of the Magnelli et al.\ (2013)
      constraints at $z\leq 2$. \label{fig:sfrtotal}}
\end{figure*}

Alavi et al.\ (2016)'s $z\sim2$-3 LF results are higher than our own
for the faintest sources ($>$$-$15 mag).  This almost certainly
results from the completeness corrections Alavi et al.\ (2016)
calculate extrapolating the Shibuya et al.\ (2015) size-luminosity
relation to lower luminosities, and in fact making use of a similar
size luminosity relation we find a $3\times$ higher volume density for
the faintest sources, largely reproducing Alavi et al.\ (2016)'s
results.  It is worth emphasizing that use of the size-luminosity
relation from Shibuya et al.\ (2015) for completeness calculations was
very reasonable, given the lack of information that existed five years
ago regarding the sizes of the faintest sources at $z\sim2$-3.

Finally, we include some constraints on the $z\sim2$-5 LFs of galaxies
by Weisz et al.\ (2014) that rely on resolved stellar population
analyses of nearby dwarf galaxies, abundance matching, and evolving
the luminosities of these dwarf galaxies backwards in time (see also
Boylan-Kolchin et al.\ 2015).  While such analyses are obviously very
interesting to pursue and should be useful in providing indicative
constraints on the volume density of faint sources during the first
few billion years of the universe, they also necessarily involve a
number of significant assumptions regarding the precise history of the
stars that make up these analyses (e.g., mergers, galaxy disruption,
etc.).  Given the uncertainties, it is encouraging how consistent the
Weisz et al.\ (2014) LF inferences and our own results are.\\

\noindent \textbf{$\mathbf{z\sim6}$-9:} At lower luminosities, our new
$z\sim6$-7 LF results from the HFFs are in excellent agreement with
the $z\sim6$-7 results from Atek et al.\ (2015b) using the first three
HFF clusters, the $z\sim5$-10 results from Castellano et al.\ (2016)
based on the first two HFF clusters, our previous results obtained
from the first four HFF clusters (Bouwens et al.\ 2017b), the
$z\sim6$-7 results from Atek et al.\ (2018) based on the full HFF
program, the Yue et al.\ (2018) results based on the first four HFF
clusters, and the LF results obtained by both McLure et al.\ (2013)
and Oesch et al.\ (2013) at $z\sim9$.  The $z=6$-9 LF results of
Bhatawdekar et al.\ (2019) also appear to be in reasonable agreement
with our own LF results, particularly if the comparison is made
against their results modeling the sizes of faint galaxies as ``disk
galaxies'' (with a mean size of 0.15$''$).  Given the very small sizes
adopted in our analysis, we would have expected the ``point-source''
results from Bhatawdekar et al.\ (2019) to be most consistent with our
own, but the point-source results from Bhatawdekar et al.\ (2019) are
$\sim$1.5$\times$ lower; it is not clear why this would be the case.

\begin{deluxetable*}{cccccc}
\tablewidth{14.cm}
\tablecolumns{6}
\tabletypesize{\footnotesize}
\tablecaption{$UV$ Luminosity Densities and Star Formation Rate Densities to $-13.0$ AB mag (0.0006 $L_{z=3} ^{*}$: \S\ref{sec:lumdens})\label{tab:sfrdens}}
\tablehead{
\colhead{Lyman} & \colhead{} & \colhead{$\textrm{log}_{10} \mathcal{L}$} & \multicolumn{3}{c}{$\textrm{log}_{10}$ SFR density} \\
\colhead{Break} & \colhead{} & \colhead{(ergs s$^{-1}$} & \multicolumn{3}{c}{($M_{\odot}$ Mpc$^{-3}$ yr$^{-1}$)} \\
\colhead{Sample} & \colhead{$<z>$} & \colhead{Hz$^{-1}$ Mpc$^{-3}$)\tablenotemark{a}} & \colhead{Unobscured} & \colhead{Obscured\tablenotemark{b,c}} & \colhead{Total}}
\startdata
U$_{275}$ & 2.1 & 26.54$\pm$0.03 & $-$1.61$\pm$0.03 & $-$1.07$\pm$0.10 & $-$0.96$\pm$0.08\\
U$_{336}$ & 3.0 & 26.69$\pm$0.07 & $-$1.46$\pm$0.07 & $-$1.36$\pm$0.10 & $-$1.11$\pm$0.07\\
B & 3.8 & 26.74$\pm$0.12 & $-$1.41$\pm$0.12 & $-$1.47$\pm$0.10 & $-$1.14$\pm$0.08\\
V & 4.9 & 26.54$\pm$0.09 & $-$1.62$\pm$0.09 & $-$1.81$\pm$0.10 & $-$1.40$\pm$0.07\\
i & 5.9 & 26.38$\pm$0.05 & $-$1.78$\pm$0.05 & $-$2.28$\pm$0.10 & $-$1.66$\pm$0.05\\
z & 6.8 & 26.27$\pm$0.06 & $-$1.89$\pm$0.06 & $-$2.93$\pm$0.10 & $-$1.85$\pm$0.06\\
Y & 7.9 & 26.08$\pm$0.12 & $-$2.07$\pm$0.12 & $-$3.31$\pm$0.10 & $-$2.05$\pm$0.11\\
J & 8.9 & 25.54$\pm$0.11 & $-$2.61$\pm$0.11 & --- & $-$2.61$\pm$0.11
\enddata
\tablenotetext{a}{From Table~\ref{tab:lumd}.}
\tablenotetext{b}{From Table 8 from ASPECS HUDF analysis of the
  infrared excess (Bouwens et al.\ 2020).  The obscured SFR density
  from Bouwens et al.\ (2020) explicitly includes the ULIRG results
  from Wang et al.\ (2019), Franco et al.\ (2020), and
  Dudzevi{\v{c}}i{\={u}}t{\.{e}} et al.\ (2020).}
\tablenotetext{c}{In light of the considerable challenges in deriving
  star formation rates from far-IR SEDs resulting from the uncertain
  SED shapes, uncertain contribution from AGN, and uncertainties in
  the selection volume, we assume a fiducial 0.1 dex uncertainty in
  the obscured SFR densities at $z\geq2$.}
\end{deluxetable*}

Our LF results at $z=6$-8 also show good overall agreement with the
results from Livermore et al.\ (2017) brightward of $-17$ mag.  At
lower luminosities, the differences are larger.  We refer interested
readers to \S6.2 of Bouwens et al.\ (2017a) and \S6.2-6.3 of Bouwens
et al.\ (2017b) for a discussion of the differences.  Our results also
show a broad similarity to the $z\sim6$-9 LF results of Ishigaki et
al.\ (2018), but we note a slight excess in the volume density of
sources they find in their $z\sim6$-7 LF results and $z\sim9$ results
(\textit{upper and lower right panels} of Figure~\ref{fig:comp69}) at
$\sim-19$ mag and a slight deficit in their $z\sim8$ LF results at
$\sim-18$ to $-17$ mag.  The slightly higher volume densities that
Ishigaki et al.\ (2018) find in their LF results at $z\sim9$ may
derive from their probing a slightly lower redshift range with their
$z\sim9$ selection than we do in our own determinations.  Use of the
photometric redshift estimates from Ishigaki et al.\ (2018: their
Table 8) substantiates this assertion, as Ishigaki et al.\ (2018)
probe a mean redshift of $z\sim8.5$ with their $z\sim9$ selection and
we probe a mean redshift of $z\sim8.9$.  The slight deficit Ishigaki
et al.\ (2018) report at $-18$ and $-17$ mag in their $z\sim8$ LF
results appears to derive from the limited number of sources Ishigaki
et al.\ (2018) have in the faintest bins, i.e., 1, 1, and 3,
respectively.

\subsection{UV Luminosity and SFR Densities\label{sec:lumdens}}

Given the impressively deep $UV$ LF results we have over the redshift
range $z=2$-9, it is clearly interesting to use these measurements to
map out the evolution of the $UV$ luminosity density of galaxies to
very faint luminosities from $z\sim9$ to $z\sim2$.

To maximize the utility of this exercise, we have elected to compute
luminosity density to the faint-end limits $-17$ mag, $-15$ mag, $-13$
mag, and $-10$ mag.  We adopted those faint-end limits due to their
frequent use in both blank-field LF studies and reionization
calculations (e.g., Robertson et al.\ 2013, 2015; Bouwens et
al.\ 2015b; Ishigaki et al.\ 2018).

To compute the luminosity densities implied by our new LF results, we
simply compute the luminosity density implied by various
parameterizations and marginalize across the likelihood distribution.
We compute both 68\% and 95\% confidence intervals on the luminosity
densities and have tabulated our results in Table~\ref{tab:lumd}.  The
results are also presented in Figure~\ref{fig:sfrdens}.  From the results presented in both the table and
figures, it is clear that probes to $-13$ mag contribute meaningfully
to the $UV$ luminosity density vis-a-vis probes to $-$17 mag,
increasing the total $UV$ luminosity by 0.1 dex (st $z\sim2$) and by
0.4 dex (at $z\sim8$).

Remarkably, the present observational results allow us to constrain
the luminosity densities to an uncertainty of $<$0.05 dex at
$z\sim2$-3 and $z\sim6$-7 and $<$0.12 dex at $z\sim4$-5 and
$z\sim8$-9, equivalent to a $<$13\% and $<$30\% relative uncertainty,
respectively.  The $UV$ luminosity densities we derive brightward of
$-10$ mag are much less well constrained.  While the 95\% confidence
intervals we derive from our $z\sim2$-3, $z\sim6$-7, and $z\sim9$
results span a $\lesssim$0.5 dex range, these same intervals span
$\gtrsim2$ dex range at $z\sim4$-5 and $z\sim8$.

It is interesting to convert our new determinations of the $UV$
luminosity density into equivalent star formation rate densities using
the conversion factors in Madau \& Dickinson (2014).  Assuming a
Chabrier (2003) IMF, a constant star formation rate, and metallicity
$Z=0.002$ $Z_{\odot}$, the conversion factor $\cal{K}$$_{FUV}$ is $0.7
\times 10^{-29} M_{\odot}\,\textrm{year}^{-1} \,\textrm{erg}^{-1}
\,\textrm{s}\, \textrm{Hz}$.  The specific value adopted for the
metallicity does not have a huge impact on this conversion factor
($<$1.5\%: Madau \& Dickinson 2014).  The equivalent SFR densities to
the integrated $UV$ luminosity densities to $-$13 mag are also shown
on the left vertical axis of Figure~\ref{fig:sfrdens}.

We have compared the implied unobscured SFR density from our $UV$ LF
results in Figure~\ref{fig:sfrdens} and have compared it to obscured
SFR density inferred from the Bouwens et al.\ (2020) ASPECS HUDF
study, the obscured SFR density results at $z=0$-2 from Magnelli et
al.\ (2013) and the unobsucred SFR density results from $z=0$ to
$z=1.5$ from Wyder et al.\ (2005) and Moutard et al.\ (2020).  We also
present the unobscured SFR density inferred at $z\geq2$ integrated
down to a brighter limit, $-17$ mag, similar to what Madau \&
Dickinson (2014) utilize in their SFR density figures.

For convenience, Table~\ref{tab:sfrdens} presents our new SFR density
estimates along with the estimates of the obscured SFR densities
Bouwens et al.\ (2020) derived combining their own inferences of the
obscured star formation from their ASPECS HUDF results with the ULIRG
results of Wang et al.\ (2019), Franco et al.\ (2020a), and
Dudzevi{\v{c}}i{\={u}}t{\.{e}} et al.\ (2020).  We adopt a fiducial
0.1 dex uncertainty in the obscured SFR densities at $z\geq2$ given
the considerable challenges in deriving star formation rates from
far-IR SEDs resulting from the uncertain SED shapes, uncertain
contribution from AGN, and uncertainties in the selection volume.
Figure~\ref{fig:sfrtotal} compares the unobscured SFR densities to the
total SFR densities.

It is clear from these results that unobscured star formation
dominates the SFR density in the high-redshift universe and obscured
star formation dominates the SFR density at intermediate and low
redshift.  As has been found before (e.g., Bouwens et al.\ 2009a,
2016b; Dunlop et al.\ 2017; Zavala et al.\ 2021), we find that the
cross-over point between these two regimes is at $z\sim4$.
Additionally, it is interesting to note the impact that the faint-end
limit can have on the transition redshift between the two regimes.  In
the Bouwens et al.\ (2020) ASPECS study, the transition redshift was
$z\sim5$ using a brighter faint-end limit of $-$17 mag, but using a
faint-end limit of $-$13 mag, enabled by our new HFF results, we find
a transition redshift $z\sim4$.

\subsection{Faint End Form of the LF and Existence of A Possible Turn-over\label{sec:mt}}

Thanks to the faintness of lensed HFF samples, our new constraints put
us in position to set constraints where the $UV$ LF might turn over at
the faint end.  This question is relevant both because it provides
insight into the efficiency of star formation in lower-mass galaxies
(Behroozi et al.\ 2013) and because it allows for a more accurate
accounting for the contribution of ionizing photons from especially
faint star-forming galaxies.

As in Bouwens et al.\ (2017b), we can constrain the brightest
magnitude where a turn-over in the LF can possibly occur using
Eq.~\ref{eq:mt} and the overall likelihood distribution we have
derived on the three parameters $M^*$, $\alpha$, and $\delta$.  By
marginalizing over the results we have obtained on the $UV$ LF at
$z=2$-9 based on the four different families of parameterized lensing
models, we can derive 68\% and 95\% confidence intervals on the $UV$
luminosity of the turn over.

\begin{figure}
\epsscale{1.15}
\plotone{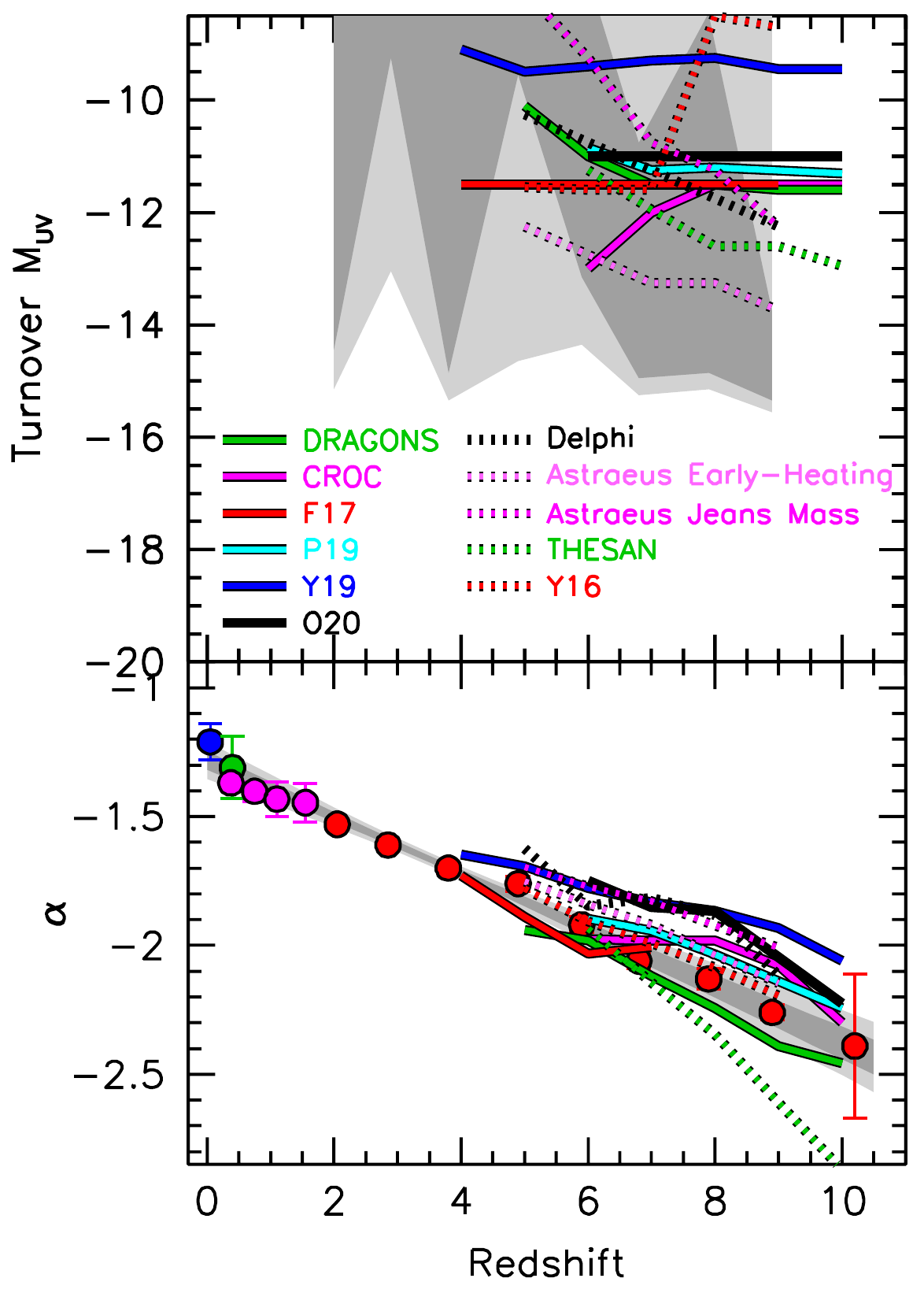}
\caption{(\textit{upper panel}) 68\% and 95\% confidence intervals
  (\textit{dark and light grey shaded regions, respectively}) on the
  $UV$ luminosity of the turn-over in the $UV$ LF obtained from our
  analysis (\S\ref{sec:lfresults}) of the lensed $z=2$-9 HFF samples
  (Bouwens et al.\ 2022b).  The turn-over luminosities in the various
  theoretical LFs (\S\ref{sec:theory}) are also shown as a function of
  redshift.  Our new LF results rule out the presence of a turn-over
  in the $UV$ LF brightward of $\approx -15.5$ mag (95\% confidence)
  over the entire redshift range $z=2$-9.  At $z\sim3$, our LF results
  rule out the existence of such a turn-over brightward of $-13.1$ mag
  (95\% confidence).  (\textit{lower panel}) Comparison of our derived
  redshift trend for the faint-end slope $\alpha$
  (Figure~\ref{fig:schevol}: \textit{dark and light grey shaded
    regions indicate the 68\% and 95\% confidence regions,
    respectively}) with that seen in the theory LFs
  (\S\ref{sec:theory}) as a function of redshift.  Both the luminosity
  of the turn-overs $M_T$ and the faint-end slopes of the theory LFs
  appear to be in excellent overall agreement with our observational
  constraints.\label{fig:mt}}
\end{figure}

We present in the upper panel of Figure~\ref{fig:mt} the constraints
we are able to obtain on the position of the turn-over in the $UV$ LF
for star-forming galaxies from $z\sim9$ to $z\sim2$.  We find that our
results rule out the presence of a turn-over brightward of $-15.5$ mag
(95\% confidence) for all $z=2$-9 samples we consider.  We are able to
obtain our tightest constraints on the luminosity of a possible
turn-over at $z\sim3$, where our results rule out the presence of a
turn-over brightward of $-13.1$ mag (95\% confidence).  We note that
Alavi et al.\ 2014 had previously presented evidence for the $UV$ LF
at $z\sim2$ extending so faint.  At $z\sim6$, our results rule out the
presence of a turn-over brightward of $-14.3$ mag.

Interestingly enough, our $z\sim9$ LF results seem consistent with a
turn-over at $-$15 mag.  While indeed this would be interesting if
this were the case, it is challenging to establish the robust presence
of a turn-over in the LF at $z\sim9$ due to the small number of
sources expected faintward of $-16$ mag, and therefore our results
cannot even establish the presence of a turn-over at $2\sigma$
significance.  To make matters even more challenging, another
complicated factor is the impact incompleteness in faint samples could
have on the results.  If faint sources have larger sizes than adopted
here based on several recent observational probes (Bouwens et
al.\ 2017a, 2022a; Kawamata et al.\ 2018; Yang et al.\ 2022), this
would result in faint sources being much less complete in our
selections than in our simulations and cause us to systematically
underestimate the volume density of faint galaxies.  This would cause
the actual luminosity of a turn-over in the $UV$ LF to be
substantially fainter than what we infer.

In general, the present constraints on the luminosity of a turn-over
in the LF are consistent with most previous results in the literature.
Atek et al.\ (2015b, 2018), Castellano et al.\ (2016), Livermore et
al.\ (2017), Bouwens et al.\ (2017b), and Yue et al.\ (2018) all agree
that the HFF results provide strong evidence against the $UV$ LF
showing a turn-over brightward of $-$15 mag.  At $z\geq 6$ and
faintward of $-$15 mag, there has been a wide variety of differing
conclusions drawn about whether firm constraints can be set regarding
the existence of a turn-over and how faint those constraints extend
(e.g., Castellano et al.\ 2016; Livermore et al.\ 2017; Bouwens et
al.\ 2017b; Atek et al.\ 2018; Yue et al.\ 2018).

\begin{figure*}
\epsscale{1.16}
\plotone{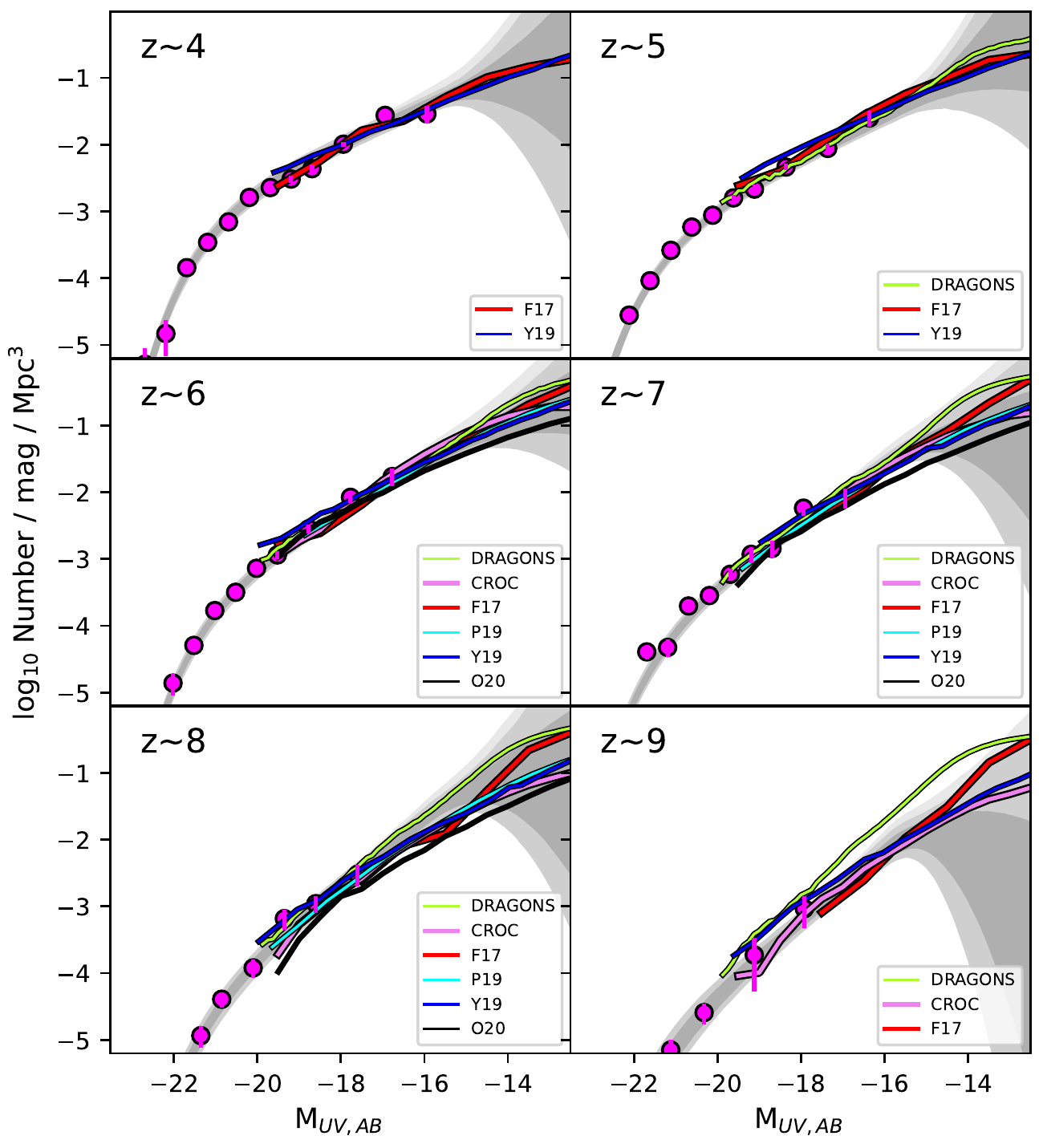}
\caption{Comparison of the 68\% and 95\% likelihood contours we derive
  for the $z=4$-9 LFs (\textit{dark and light grey shaded regions,
    respectively}) with the predictions of the DRAGONS (Liu et
  al.\ 2016), CROC (Gnedin 2016), Finlator et al.\ (2017), Park et
  al.\ (2019), Yung et al.\ (2019), and Ocvirk et al.\ (2020).  The
  solid magenta circles show the blank-field results obtained by
  Bouwens et al.\ (2021a).  The LF results are only shown faintward of
  $-20$ mag to focus on the faint-end form of the models and not the
  behavior of the models at the bright end where treatment of dust
  extinction can play a dominant role.  No comparison is made to
  $z\sim2$-3 due to the general lack of model LF predictions for these
  redshift intervals.  In general, we find excellent consistency
  between our new observational results and the different expectations
  from the theoretical models. \label{fig:lf29theory}}
\end{figure*}

\begin{figure*}
\epsscale{1.15}
\plotone{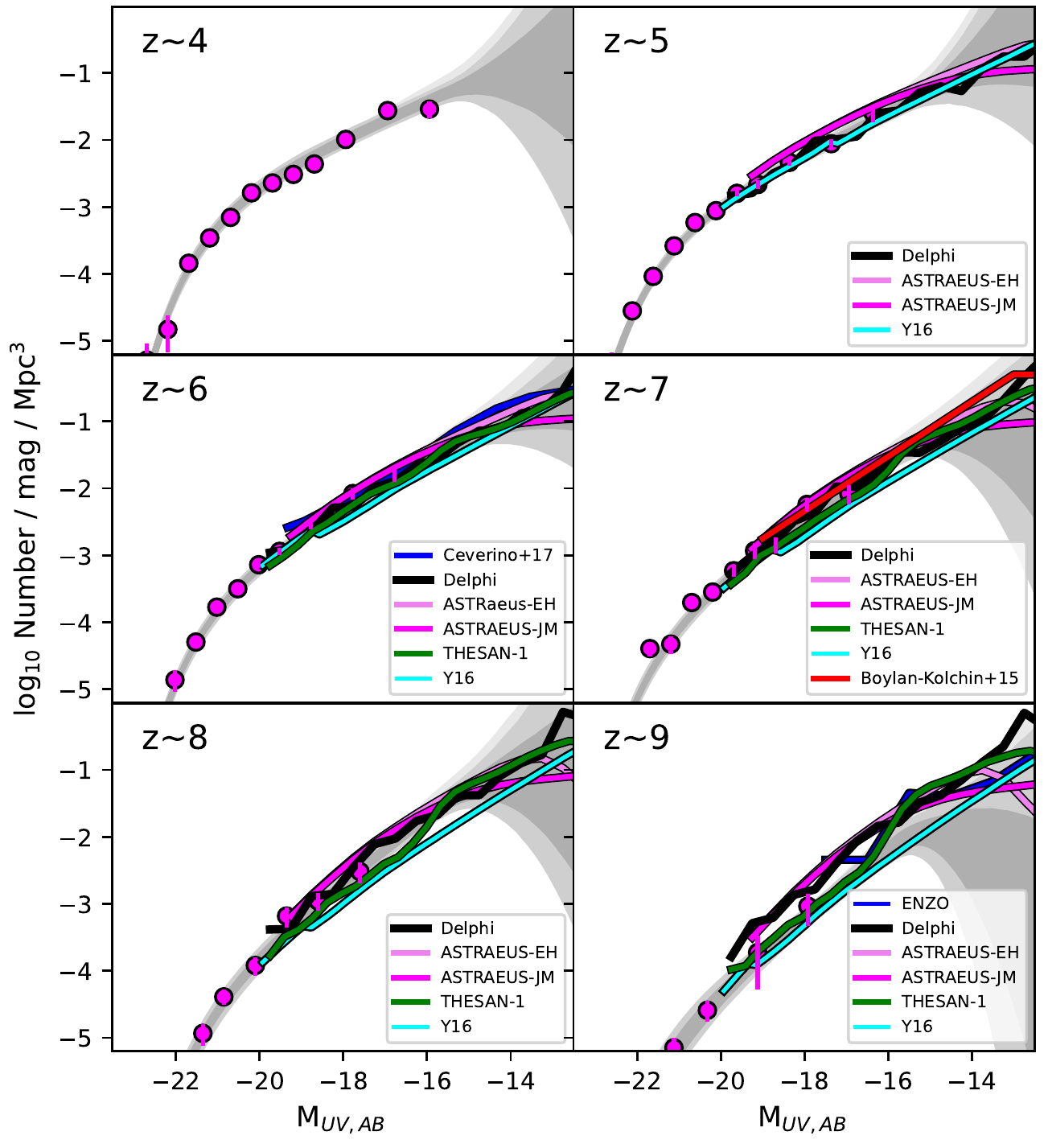}
\caption{Similar to Figure~\ref{fig:lf29theory}, but for the Delphi
  model (Dayal et al.\ 2014, 2022), FirstLight (Ceverino et
  al.\ 2017), ASTRAEUS-EarlyHeating (Hutter et al.\ 2021),
  ASTRAEUS-JeanMass (Hutter et al.\ 2021), ENZO (O'Shea et al.\ 2015),
  THESAN project (Kannan et al.\ 2022), and Yue et al.\ (2016: Y16).
  Also shown with the red line is the abundance matching constraints
  on the LF at $z\sim7$ from the Boylan-Kolchin et al.\ (2015)
  analysis.\label{fig:lf29theory2}}
\end{figure*}

\subsection{Comparison with Theoretical Models for the $UV$ LF\label{sec:theory}}

Finally, it is useful for us to compare the current constraints on the
evolution of the $UV$ LF with that available from a number of recent
theoretical models and cosmological hydrodynamical simulations.  While
we had previously looked at this in \S6.3 of Bouwens et al.\ (2017b),
here we have the advantage that we can compare the simulation/theory
results with our new $UV$ LF results over a much more extended
baseline in both redshift and cosmic time, reaching from $z\sim9$ to
$z\sim4$.  While no comparisons are made at $z\sim2$ and $z\sim3$ due
to the lack of published model LF results in these two redshift
intervals, it would be presumably possible to derive such LF results
in the near future based on a number of on-going simulation efforts,
e.g., the IllustrisTNG (Pillepich et al.\ 2018; Springel et al.\ 2018;
Naiman et al.\ 2018; Nelson et al.\ 2018) and NewHorizons (Dubois et
al.\ 2021) simulations.

We consider the following theoretical models:\\

\noindent \textit{DRAGONS [Liu et al.\ 2016]:} The LF results from Liu
et al.\ (2016) rely on the Dark-ages Reionization And Galaxy-formation
Observables from Numerical Simulations
(DRAGONS)\footnote{http://dragons.ph.unimelb.edu.au} project which
build semi-numerical models of galaxy formation on top of halo trees
derived from N-body simulations done over different box sizes to probe
a large dynamical range.  The semi-numerical models include gas
cooling physics, star formation prescriptions, feedback and merging
prescriptions, among other components of the model.  The turn-over in
the LF results of Liu et al.\ (2016) at $\sim -11.5$ mag correspond to
the approximate halo masses $\sim10^8$ $M_{\odot}$ where the gas
temperature is $10^4$ K.  Above this temperature, atomic cooling
processes become efficient.\\\vspace{-0.2cm}

\noindent \textit{CROC [Gnedin 2014, 2016]:} The model LF results for
the Cosmic Reionization On Computers (CROC) were computed using
gravity + hydrodynamical simulations executed with the Adaptive
Refinement Treement (ART) code (Kravtsov 1999; Kravtsov 2002; Rudd et
al.\ 2008).  A wide variety of physical processes, including gas
cooling and heating, molecular hydrogen chemistry, star formation,
stellar feedback, radiative transfer of ionizing and UV light from
stars is included in these simulations and done 20h$^{-1}$ Mpc boxes
at a variety of resolutions.  There is a flattening in the effective
slope of CROC LFs to fainter magnitudes, with a peak at $\sim-12$ mag.
However, the peak at $\sim-11.5$ mag is reported to depend on the
minimum particle size in the simulations and thus not to be a robust
result of the simulation.\\\vspace{-0.2cm}

\noindent \textit{Finlator et al.\ 2015, 2016, 2017 [F17]:} The
Finlator et al.\ (2015, 2016, 2017) LF results are derived from a
cosmological simulation of galaxy formation in a $(7.5h^{-1})^3$
Mpc$^3$ volume of the universe including both gravity and
hydrodynamics.  It is implemented in the GADGET-3 code (Springel
2005).  Gas cooling has been added to this code through collisional
excitation of hydrogen and helium (Katz et al.\ 1996).  Metal line
cooling is implemented using the collisional ionization equilibrium
tables from Sutherland \& Dopita (1993).  Star formation is included
using the Kennicutt-Schmidt law, with supernovae feedback implemented
following the ''ezw'' prescription from Dav{\'e} et al.\ (2013) and
metal enrichment from supernovae as in Oppenheimer \& Dav{\'e} (2008).
Less efficient gas cooling at lower halo masses results in a
flattening of the Finlator et al.\ (2015, 2016, 2017) LF results at
the faint end, with the turn-over in the LF occuring at $\sim$$-$11.5
mag.\\\vspace{-0.2cm}

\noindent \textit{Park et al.\ 2019 [P19]:} The Park et al.\ (2019) LF
results are based on a flexible, physically motivated modeling of star
formation in galaxy halos.  In their model, Park et al.\ (2019) take
the star formation efficiency (SFE), the SFE scaling with halo mass,
and a turn-over mass to be free parameters which they then fit to the
LF constraints from Bouwens et al.\ (2015a), Bouwens et al.\ (2017b),
and Oesch et al.\ (2018a).  In their fits, Park et al.\ (2019) allow
for the turn-over mass to be between $10^8$ $M_{\odot}$ and $10^{10}$
$M_{\odot}$.  Given that the tuning of the Park et al.\ (2019) LF
model to match the observations of Bouwens et al.\ (2015a) and Bouwens
et al.\ (2017b), it is not especially surprising that their model fits
our new observational constraints quite well.  The approximate
turn-over luminosity in the Park et al.\ (2019) results occurs at
$\sim$$-$11.3 mag.\\\vspace{-0.2cm}

\noindent \textit{Yung et al.\ 2019 [Y19]:} The Yung et al.\ (2019) LF
results are based on a recent version of the Santa Cruz semi-analytic
model (Somerville et al.\ 2015), which includes not only merger trees
constructed by a standard Press-Schechter formalism (Lacey \& Cole
1993), but also gas cooling, star formation, chemical evolution, and
SNe-driven winds, photoionization feedback, and a critical molecular
hydrogen surface density necessary for star formation.  Yung et
al.\ (2019) produced their results to provide semi-analytical model
forecasts for JWST and rely on halos with circular velocities
$V_{\text{vir}} \approx 20 - 500$ km/s.  Yung et al.\ (2019) report
that SNe feedback play the dominant role in flattening the LF at the
faint end.  The turn-over at $M_{UV,AB}\sim-9$ mag is imposed as a
result of the atomic cooling limit in halos with $V_{\text{vir}}
\approx 20$ km/s and is thus not a resolution effect.\\\vspace{-0.2cm}

\noindent \textit{CoDa2 [Ocvirk et al.\ 2016, 2020 -- O20]:} The
Cosmic Dawn (CoDa) simulations use the RAMSES-CUDATON code (Ocvirk et
al.\ 2016) to execute a full modeling of both gravity + hydrodynamics
+ radiative transfer for a large $\sim$(100 Mpc)$^3$ volume of the
universe. The simulations include standard prescriptions for star
formation and supernovae explosions following standard recipes (Ocvirk
et al.\ 2008; Governato et al.\ 2009, 2010). One key feature of the
Cosmic Dawn simulations is the inclusion of radiative transfer into
the simulations through the ATUN code (Aubert \& Teyssier 2008), in
the sense that hydrodynamics and radiative transfer are now fully
coupled. As a result, the effects of photoionization heating on
low-mass galaxies are fully included in the CoDa simulations. Ocvirk
et al.\ (2016, 2020) report that radiative feedback plays a big role
in suppressing star formation in low mass galaxies and modulating the
very faint-end ($M_{AB} > -11$) of the LF, resulting in a faint-end
turn-over to the $UV$ LF at $\approx-11$ mag.\\\vspace{-0.2cm}

\noindent \textit{FirstLight [Ceverino et al.\ 2017]:} The model LF
results from the FirstLight project are based on zoom-in simulations
of galaxies with circular velocity between 50 km$\,$s$^{-1}$ and 250
km$\,$s$^{-1}$.  The galaxy simulation results are executed using the
ART gravity+hydrodynamics code (Kravtsov et al.\ 1997; Kravtsov 2003).
This code also include gas cooling (atomic hydrogen, helium, metal,
and molecular hydrogen), photoionization heating, star formation,
radiative feedback, and SNe feedback.  Ceverino et al.\ (2017) report
that stellar feedback drives a flattening of their LF results at the
faint end, i.e., $M_{UV,AB}>-14$ mag, with an approximate turn-over
luminosity $\approx-11.5$ mag.\\\vspace{-0.2cm}

\noindent \textit{Renaissance [O'Shea et al.\ 2015]:} The
``Renaissance'' simulations (O'Shea et al.\ 2015) are zoom-in
simulations of a $(28.4 \textrm{Mpc}/h)^3$ volume of the universe,
powered by the Enzo code (Bryan et al.\ 2014).  This code
self-consistently follows the evolution of gas and dark matter,
includes $H_2$ formation and destruction from photodissociation, and
includes star formation and supernovae physics.  Ionizing and UV
radiation are produced as given by Starburst99 (Leitherer et
al.\ 1999).  Individual dark-matter particles in the simulations have
masses of $2.9\times 10^4$ $M_{\odot}$ and thus the smallest resolved
halos in the simulation have masses of $2\times10^6$ $M_{odot}$
($\sim$70 particles/halo).  A detailed description of the
implementation of the physics and sub-grid recipes is provided in Chen
et al.\ (2014) and Xu et al. (2013, 2014).  In the ``Renaissance''
simulations, flattening in the UV LF directly results from the
decreasing fraction of baryons converted to stars in the lowest mass
halos, as a result of radiative feedback and less efficient gas
cooling, with an approximate turn-over luminosity at $\approx -8.5$
mag.  The presented results from O'Shea et al.\ (2015) are at
$z\sim12$ where results are available and compared with our $z\sim9$
LF results.\\\vspace{-0.2cm}

\noindent \textit{Delphi [Dayal et al.\ (2014, 2022)]:} \code{Delphi}
uses a binary merger tree approach (Parkinson et al.\ 2008) to jointly
track the build-up of dark matter halos and their baryonic components
(gas, stellar, metal and dust mass). This model follows the assembly
histories of $z \sim 4.5$ galaxies with halo masses ${\rm log}(M_h/
\Msun)=8-14$ up to $z \sim 40$. The Star formation efficiency in any
halo is the minimum between that required to eject the rest of the gas
and an upper maximum threshold value. The flattening of the UV LF with
decreasing redshift is driven by a flattening of the halo mass
function coupled with a decrease in the gas mass as a result of the
Supernova feedback experienced by a galaxy over its entire assembly
history.  The approximate turn-over luminosity ranges from $-$12 mag
to $-$10 mag.\\\vspace{-0.2cm}

\noindent \textit{ASTRAEUS [Hutter et al.\ 2021] Jeans Mass + Early
  Heating:} The Astraeus framework couples an N-body simulation ($160$
comoving Mpc; dark matter mass resolution of $6.2 \times 10^6 h^{-1}\,
\msun$) with a modified version of the {\it Delphi} model for galaxy
formation and a semi-numerical scheme for reionization. The authors
introduce a filtering mass below which baryonic fluctuations can be
suppressed due to reionization heating and explore six models for such
reionization feedback. Here, we explore two models: (i) the
Early-heating model where reionization feedback is time-delayed and
has a weak to intermediate impact; and (ii) the Jeans mass model which
results in an instantaneous and maximum radiative feedback. The
flattening at the faint end of the UV LF is a result of the impact of
feedback (both Supernova and radiative) and the simulation resolution,
with a turn-over luminosity occurring between $\sim$$-$12 and $-$14
mag for the early heating simulation and between $-$8 and $-12$ mag
for the Jeans mass simulation.\\\vspace{-0.2cm}

\noindent \textit{THESAN PROJECT [Kannan et al.\ 2022; Garaldi et
    al.\ 2022; Smith et al.\ 2022]:} The galaxy $UV$ LF results from
the THESAN project are based on a radiation-magneto-hydrodynamics
simulation of a large volume of the universe (95.5 cMpc on a side)
that models both the large scale statistical properties of the
intergalactic medium and the galaxies that lie within the volume.  The
flagship simulation resolves baryonic and dark matter masses down to
5.8$\times$10$^5$ $M_{\odot}$ and 3.1$\times$10$^6$ $M_{\odot}$,
respectively.  The simulations are executed using efficient code
AREPO-RT (Kannan et al.\ 2019), a radiation hydrodynamics extension to
the hydrodynamics code AREPO (Springel 2010).  Star formation, black
accretion, SNe winds and other subgrid physics are implemented using
the recipes developed and tested as part of the IllustrisTNG
simulations (Vogelsberger et al.\ 2014).  The $UV$ LF predicted by the
highest-mass component of the THESAN project show a faint-end
turn-over at $\sim$$-$12 mag, which is largely set by the resolution
limit of the lowest-mass galaxies in the simulation.\\\vspace{-0.2cm}

\noindent \textit{Yue et al.\ 2016 [Y16]:} Yue et al.\ (2016) make
use of a semi-analytic formalism to predict the evolution of the $UV$
LF.  Yue et al.\ (2016) start with the halo mass function, break up
the star formation history of each halo into segments according to
which the halo grows in mass by a factor of two, and then assume that
the SFR must be such to maintain a constant stellar mass-halo mass
relation which they calibrate to the $z\sim5$ LF of Bouwens et
al.\ (2016).  This approach is very similar to what Mason et
al.\ (2015) employ in predicting the evolution of the $UV$ LF (see
also Trenti et al.\ 2010 and Tacchella et al.\ 2013).  Yue et
al.\ (2016) then look into the impact that radiative feedback would
have during the reionization era.  Yue et al.\ (2016) then consider
the star formation to be quenched in galaxies below some fixed
circular velocity.  Here we show their results where the quenching
occurs below a circular velocity of 30 km/s, where the turn-over in
the $UV$ LF occurs at $\sim$$-$11.5 mag.\\\vspace{-0.2cm}

\noindent
Of course, the aforementioned simulation efforts of the galaxy $UV$ LF
at $z\geq 4$ are not exhaustive and are restricted to those studies
which probe the faint-end form of the $UV$ LF, i.e., $\ll$$-$16 mag,
and thus do not include other very sophisticated modeling efforts such
as conducted by ASTRID (Bird et al.\ 2022), Bluetides (Wilkins et
al.\ 2017), and FLARES (Vijayan et al.\ 2021).

Finally, and as in Bouwens et al.\ (2017b), we again compare with the
empirical constraints on the form of the $UV$ LFs at $z\sim7$ from
Boylan-Kolchin et al.\ (2015):\\\vspace{-0.1cm}

\noindent \textit{Boylan-Kolchin et al.\ 2015:} Boylan-Kolchin et
al.\ (2015) constrain the faint-end of the $z\sim7$ LF by leveraging
sensitive probes of the color-magnitude relationship of nearby dwarf
galaxies to estimate their luminosity at $z\sim7$.  By comparing the
inferred luminosity distribution for these dwarfs with their implied
numbers extrapolating $z\sim7$ LFs to $-10$ mag, Boylan-Kolchin et
al.\ (2015) infer a break in the LF at $z\sim-13$ mag and transition
from a faint-end slope of $\sim-2$ to $\sim$$-1.2$.\\\vspace{-0.2cm}

We compare the predicted $UV$ LFs from these theoretical and empirical
models with our new observational constraints on the faint-end form of
the $UV$ LFs in Figures~\ref{fig:lf29theory} and
\ref{fig:lf29theory2}.  We also present the approximate turn-over
luminosities and faint-end slopes derived from the theory LFs in
Figure~\ref{fig:mt} as a function of redshift.  The dark and light
grey shaded regions show the 68\% and 95\% confidence constraints by
marginalizing over our LF fit results.

In general, we find good overall agreement between our observational
findings and the general form of the model and empirical $UV$ LFs.
The faint-end slopes for all models show roughly the same trend with
redshift as we see in the observations, although several models lie
above and below the trends we derive here.

Additionally, the approximate $UV$ luminosity of the turn-over in the
theory LFs, which lie in the range $\sim-13.5$ to $\sim-9$ mag, is
consistent with our observational constraints (Figure~\ref{fig:mt}),
which generally constrain the turn-over to be fainter than $-15.5$ or
$-14.3$ mag depending on the redshift.  At $z\sim3$, we obtain the
tightest constraints on the $UV$ luminosity of the turn-over.  We find
the turn-over to be fainter than $-13.1$ mag (95\% confidence).
Unfortunately, none of the theoretical models provide predicted LFs at
$z\sim3$, but those which do so at $z\sim4$, i.e., Finlator et
al.\ (2017) and Yung et al.\ (2019), are consistent with showing a
turn-over at $\sim$$-$13.1 or fainter.

\section{Summary}

Here we explore the use of a substantial sample of $>$2500 lensed
$z=2$-9 galaxies behind the six clusters in the HFF program to
characterize the faint-end form of the $UV$ LF at $z\geq 2$,
quantifying the faint-end slope of these LFs, establishing the
prevalence of extremely faint galaxies, and setting constraints on a
possible turn-over at the faint end of these LFs.

The construction of the 2534 source sample of lensed $z\sim2$-9
galaxies is described in detail in a companion paper (Bouwens et
al.\ 2022b) and leverages deep HST observations from 0.25$\mu$m to
1.6$\mu$m and a composite Lyman-break + photometric redshift
selection.  This sample includes 765, 1176, 68, 59, 274, 125, 51, and
16 sources at $z\sim2$, 3, 4, 5, 6, 7, 8, and 9, respectively.  Fewer
galaxies could be reliably identified at $z\sim4$ and $z\sim5$ due to
confusion with breaks in the SEDs of foreground cluster galaxies.

To maximize the robustness of the source magnification factors
utilized in our analysis, we used the median magnification factors
derived from the latest parametric lensing models made available for
each of the HFF clusters, i.e., version 3 or 4.  As demonstrated in
Figures 5-6 of Bouwens et al.\ (2022b), use of the median provides us
with a much more reliable way of estimating the magnification factors,
allowing us to make use of sources with magnifications $>$40 (and in
some cases to $\sim$100).  A description of our calculation of these
magnification factors and presentation of our lensed $z=2$-9 sample is
also provided in the companion paper (Bouwens et al.\ 2022b).

Even with the use of the median magnification factors, the true
magnification of individual sources is uncertain, particularly in high
magnification regions, and must be carefully accounted for in
determinations of the $UV$ LF.  To overcome the challenges posed by
uncertainties in the magnification maps -- we made use of a
forward-modeling methodology developed in Bouwens et al.\ (2017b) to
constrain the faint-end shape of the $UV$ LFs at $z=2$-9 in the
presence of these uncertainties.

We applied this methodology to the lensed $z=2$-9 samples from Bouwens
et al.\ (2022b) and derived constraints on the faint-end slope
$\alpha$ and normalization $\phi^*$ of the $UV$ LF as well as a
curvature parameter $\delta$ to capture the potential flattening of
the $UV$ LF faintward of $-16$ mag (Bouwens et al.\ 2017b).  To
maximize the robustness, individual parameters in the LF were
estimated using a Markov Chain Monte Carlo process.  The selection
volumes used in our LF determinations were estimated assuming
point-source sizes for the lensed population, consistent with the
observational findings from Bouwens et al.\ (2017a, 2017c, 2022a) and
Kawamata et al.\ (2018).

We first considered the use of the HFF lensed samples to derive
constraints on the faint-end slope of the $UV$ LF.  Remarkably, the
faint-end slope $\alpha$ results we recover from the lensed samples
are completely consistent with the slopes found from blank-field
studies (e.g., Bouwens et al.\ 2021a) over the entire redshift range
$z\sim9$ to $z\sim2$.  This is the first time such consistent
faint-end slope results have been found over such an extended range in
redshift and strongly suggests that systematic uncertainties are now
finally understood.

Next, we made full use of both blank-field LF constraints and faint
lensed samples from the HFFs to obtain the most accurate constraints
available to date on the overall shape of the $UV$ LF from $z=9$ to
$z=2$.  We find a flattening in the faint-end slope $\alpha$ from
$z\sim9$ ($-$2.28$\pm$0.10) to $z\sim2$ ($-$1.53$\pm$0.03), i.e.,
$d\alpha/dz=-0.11\pm0.01$, limited evolution in the characteristic
luminosiy from $z\sim9$ to $z\sim3$, and a monotonic increase in the
normalization $\phi^*$ of the $UV$ LF with cosmic time.  These newly
derived parameters and evolution are consistent with what has been
found in other recent blank-field studies (e.g., Bouwens et
al.\ 2015a, 2021a; Parsa et al.\ 2016), strengthening earlier
conclusion supporting a link between the build-up of galaxies and
their dark matter halos.  \S\ref{sec:schparm} updates previous fitting
formula results for the redshift evolution of the Schechter parameters
leveraging our new LF determinations.

Additionally, our new results allow us to constrain the evolution of
the $UV$ luminosity density (integrated to $-$13 mag) from $z\sim9$ to
$z\sim2$, with $<$0.05 dex ($<$13\%) uncertainties in the luminosity
density at $z\sim2, 3, 6, 7$ and $<$0.12 dex ($<$30\%) uncertainties
at $z\sim4$, 5, 8, and 9 (Table~\ref{tab:lumd}).  Our computed
luminosity densities to $-13$ mag are 0.1 dex ($z\sim2$) to 0.4 dex
($z\sim8$) higher than to $-$17 mag, showing how significantly faint
galaxies contribute to the total SFR density.  If the escape fraction
and Lyman-continuum photon production efficiency of faint galaxies is
similar to that bright galaxies, we might expect their contribution to
the total reservoir of ionizing photons needed to derive cosmic
reionization is similarly large.

We have similarly computed the unobscured SFR density brightward of
$-$13 mag using our LF results and then make a comparison to the
evolution of the obscured SFR density (Figure~\ref{fig:sfrdens}:
\S\ref{sec:lumdens}).  We find that the bulk of the star formation at
$z>4$ is unobscured and at $z<4$, it is mostly obscured.  Redshift
$z=4$ marks the transition between these two regimes.  While there
were previous reports of this by Bouwens et al.\ (2009a, 2016b), Dunlop
et al.\ (2017), and Zavala et al.\ (2021), the present deep probe
provides the deepest account to date of both the UV luminosity density
and unobscured star formation (Figure~\ref{fig:sfrtotal}) in comparing to
the obscured SFR density (here taken from Bouwens et al.\ 2020).
Accounting for both is clearly essential for an accurate
characterization of the full extent of the star formation history of
the universe.

Our new LF determinations also allow us to set firm constraints on the
possible luminosity of a turn-over at the faint end of the $z=2$-9
LFs.  Our results (\S\ref{sec:mt}) rule out the existence of a
turn-over in the $UV$ LF brightward of $-15.5$ mag (95\% confidence)
over the redshift range $z=2$-9, consistent with previous results from
Atek et al.\ (2015b, 2018), Castellano et al.\ (2016), Livermore et
al.\ (2017), Bouwens et al.\ (2017b), Yue et al.\ (2018), Ishigaki et
al.\ (2018), and Bhatawdekar et al.\ (2019).

At $z\sim3$ and $z\sim6$, our results allow us to set even tighter
constraints on the presence of a turn-over in the $UV$ LF, ruling out
such a turn-over brightward of $-13.1$ mag and $-14.3$ mag,
respectively.  Figure~\ref{fig:mt} compares the constraints we can set
on a possible turn-over in the $z=2$-9 LFs with model LFs from a
variety of theoretical models, and excellent overall agreement is
found.

In the future, it should be possible to significantly extend these
results to even low luminosities and to higher redshifts taking
advantage of increased sensitivity and wavelength range of JWST.
Given how interesting current constraints are already in constraining
the faint end of the $UV$ LF at $z>2$ and constraining the physical
processes that govern star formation in lower-mass galaxies, future
results with JWST seem likely to be extremely exciting.

\acknowledgements

We are greatly appreciative to Pratika Dayal, Kristian Finlator,
Nicholas Gnedin, Anne Hutter, Chuanwu Liu, Brian O'Shea, Rahul Kannan,
and Bin Yue for sending us the predictions they derive for the LF
results at high redshift.  We especially appreciate discussions with
Pratika Dayal, Kristian Finlator, Anne Hutter, and Pierre Ocvirk
regarding different turn-over mechanisms that would impact the faint
end of the $UV$ LFs.  We also appreciate feedback on the galaxy
formation model descriptions from Pratika Dayal, Andrea Ferrara,
Andrei Mesinger, Pierre Ocvirk, and Aaron Yung.  We are thankful to
the feedback from our anonymous referee and other scientists in the
community such as Eros Vanzella, which greatly improved this
manuscript.  This work utilizes gravitational lensing models produced
by PIs Bradac, Natarajan \& Kneib (CATS), Merten \& Zitrin, Sharon,
Williams, Keeton, Bernstein and Diego, and the GLAFIC group. This lens
modeling was partially funded by the HST Frontier Fields program
conducted by STScI. STScI is operated by the Association of
Universities for Research in Astronomy, Inc. under NASA contract NAS
5-26555. The lens models were obtained from the Mikulski Archive for
Space Telescopes (MAST).  We acknowledge the support of NASA grants
HST-AR-13252, HST-GO-13872, HST-GO-13792, HST-AR-15027, HST-GO-16037,
and NWO grants 600.065.140.11N211 (vrij competitie) and TOP grant
TOP1.16.057.  RSE acknowledges financial support from European
Research Council Advanced Grant FP7/669253.

\appendix

\section{A.  Lensing-Model Specific LF Constraints}

In deriving constraints on the overall shape of the $UV$ LF, it is
necessary to cope with uncertainties in the magnification of
individual sources.  To cope with these uncertainties, one family of
magnification model here is treated as representing the ``truth'' and
then the median magnification of the parametric magnification models
is used to derive the distribution of sources in $UV$ luminosity.
That distribution can then be compared to that recovered from the
observations.

To gain insight into the overall uncertainties in the faint end form
of the $UV$ LF, it is interesting to compare the results one derives
treating various lensing models as the truth.  It is the purpose of
this appendix to illustrate the approximate scope of these
uncertainties.

In Table~\ref{tab:lfparm}, we present our LF parameter determinations
alternatively treating each family of magnification models as the
truth and then running a bunch of MCMC simulations to converge on a
best LF.  The uncertainties we quote on each LF parameter include not
only the formal uncertainties from the MCMC simulations, but also the
computed scatter in the parameters allowing for a systematically low
or high selection volume.

\begin{figure*}
\epsscale{1.15}
\plotone{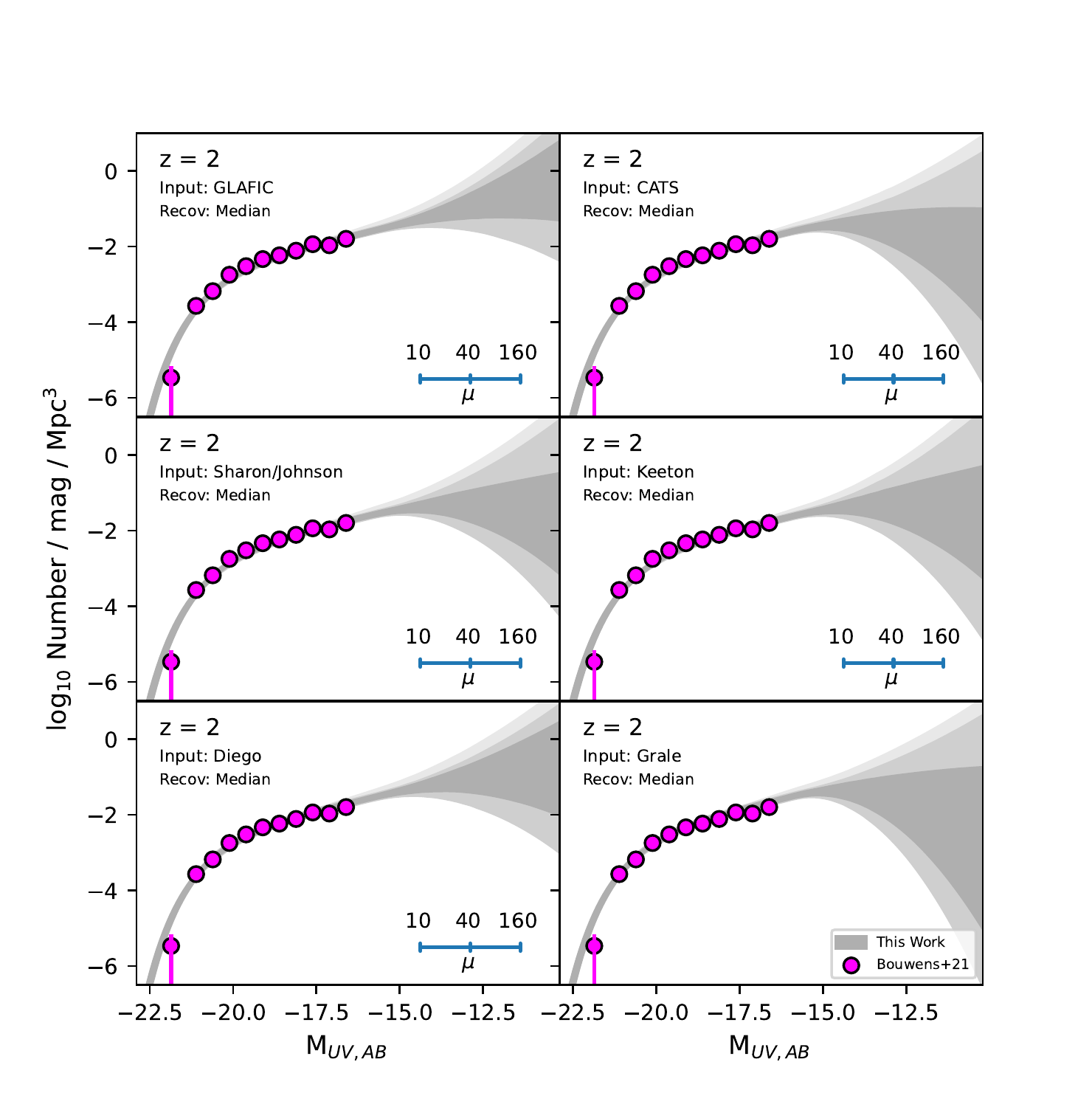}
\caption{The 68\% and 95\% likelihood contours (\textit{dark and light
    grey shaded regions, respectively}) we derive on the shape of the
  $UV$ LFs at $z\sim2$ based on our lensed HFF samples and the
  presented constraints on the LF from blank-field studies.  Each
  panel shows the likelihood contours derived using one of six lensing
  v4 models as the truth in our forward-modeling procedure (Bouwens et
  al.\ 2017b) and then recover LF results using the median
  magnification from the other parametric lensing models.  The light
  shaded regions shown here are similar to Figure~\ref{fig:lf29p}.
  The blank-field constraints are from Bouwens et al.\ (2021a) and
  include essentially HST observations acquired to date over legacy
  fields.\label{fig:lf2}}
\end{figure*}

\begin{figure*}
\epsscale{1.15}
\plotone{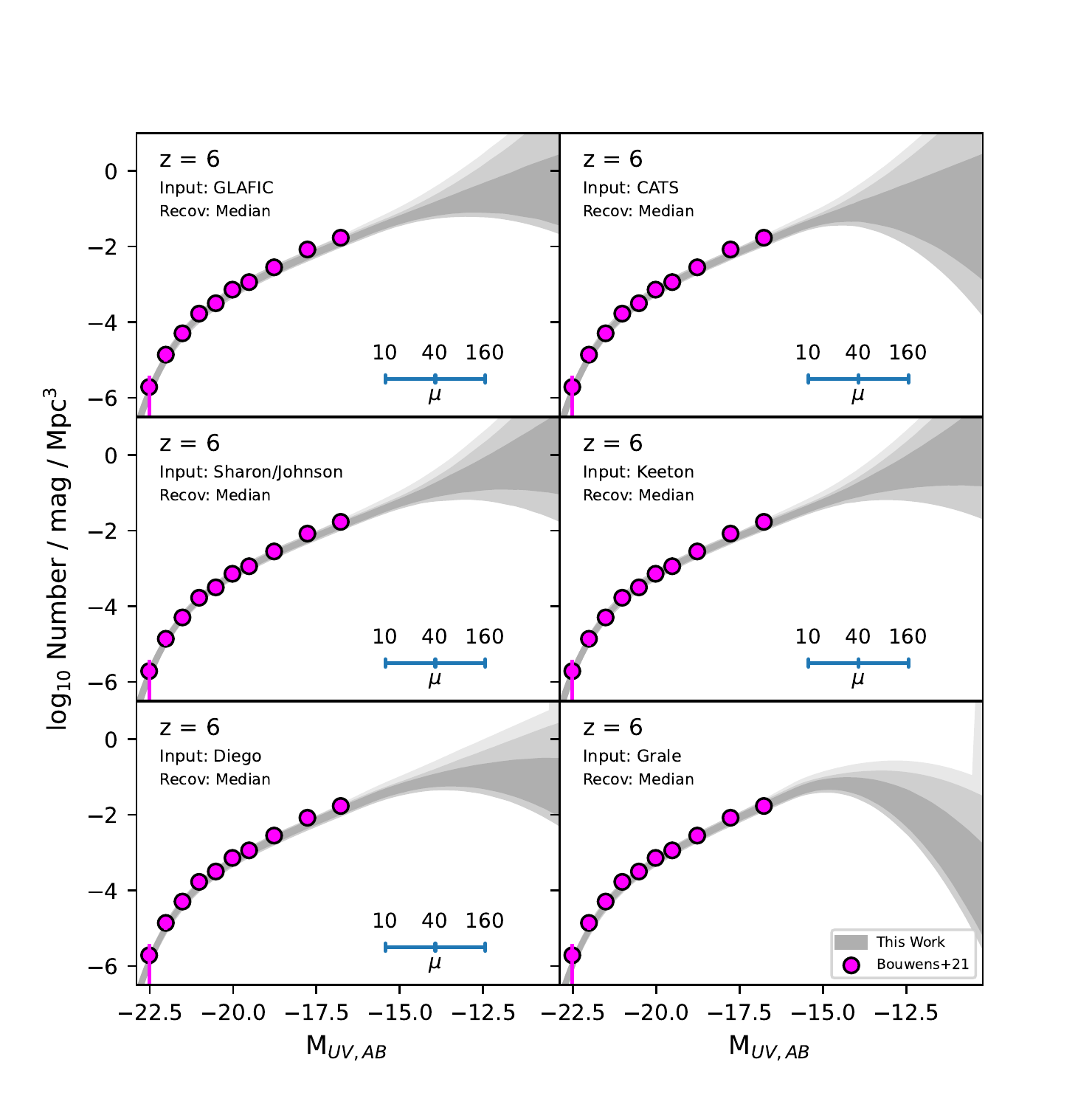}
\caption{Identical to Figure~\ref{fig:lf2}, but for $z\sim6$ LF
  results.\label{fig:lf6}}
\end{figure*}

Secondly, we include Figures~\ref{fig:lf2} and \ref{fig:lf6} showing
the range of $z\sim2$ and $z\sim6$ LF constraints obtained treating
different lensing models as the truth.  These figures show the 68\%
and 95\% confidence regions we derive for the $z\sim2$ and $z\sim6$ LF
results, alternatively treating one of the v4 magnification models as
representing the truth.  These figures are similar to Figure 8 from
Bouwens et al.\ (2017b).

\begin{deluxetable*}{ccccccccc}
  \tablecolumns{9}
  \tablewidth{17cm}
\tabletypesize{\footnotesize}
\tablecaption{Parameterized $z\sim2$-9 LF Results Using HFF Samples + Blank-Field Constraints vs. Adopted Lensing Model\label{tab:lfparm}}
\tablehead{
\colhead{Adopted ``True''} & \colhead{} & \colhead{$\phi^*$ $(10^{-3}$} & \colhead{} & \colhead{} & \colhead{} & \colhead{$\phi^*$ $(10^{-3}$} & \colhead{} & \colhead{}\\
\colhead{Magnification Model} & \colhead{$M_{UV} ^{*}$} & \colhead{Mpc$^{-3}$)} & \colhead{$\alpha$} & \colhead{$\delta$\tablenotemark{a}} & \colhead{$M_{UV} ^{*}$} & \colhead{Mpc$^{-3}$)} & \colhead{$\alpha$} & \colhead{$\delta$\tablenotemark{a}}}
\startdata
& \multicolumn{4}{c}{$z\sim2$} & \multicolumn{4}{c}{$z\sim6$}\\
\textsc{CATS} & $-$20.29$\pm$0.08 & 3.5$\pm$0.5 & $-$1.52$\pm$0.03 & 0.15$\pm$0.11  &  $-$20.88$\pm$0.07 & 0.56$\pm$0.11 & $-$1.88$\pm$0.03 & 0.12$\pm$0.12 \\
Sharon/Johnson & $-$20.30$\pm$0.08 & 3.4$\pm$0.4 & $-$1.53$\pm$0.02 & 0.10$\pm$0.10  &  $-$20.87$\pm$0.06 & 0.59$\pm$0.10 & $-$1.86$\pm$0.03 & 0.02$\pm$0.08 \\
\textsc{GLAFIC} & $-$20.29$\pm$0.08 & 3.4$\pm$0.4 & $-$1.52$\pm$0.03 & $-$0.01$\pm$0.08  &  $-$20.89$\pm$0.06 & 0.54$\pm$0.11 & $-$1.88$\pm$0.02 & 0.06$\pm$0.07 \\
Keeton & $-$20.30$\pm$0.08 & 3.2$\pm$0.4 & $-$1.53$\pm$0.02 & 0.10$\pm$0.11  &  $-$20.86$\pm$0.06 & 0.60$\pm$0.10 & $-$1.86$\pm$0.03 & 0.01$\pm$0.08 \\
\textsc{grale} & $-$20.31$\pm$0.08 & 4.0$\pm$0.5 & $-$1.53$\pm$0.03 & 0.20$\pm$0.14  &  $-$20.98$\pm$0.08 & 0.45$\pm$0.09 & $-$1.98$\pm$0.04 & 0.36$\pm$0.08 \\
Diego & $-$20.29$\pm$0.08 & 3.7$\pm$0.5 & $-$1.52$\pm$0.03 & 0.04$\pm$0.09  &  $-$20.91$\pm$0.08 & 0.49$\pm$0.10 & $-$1.89$\pm$0.03 & 0.13$\pm$0.06 \\
 & \\
& \multicolumn{4}{c}{$z\sim3$} & \multicolumn{4}{c}{$z\sim7$}\\
\textsc{CATS} & $-$20.85$\pm$0.13 & 2.2$\pm$1.0 & $-$1.60$\pm$0.04 & $-$0.04$\pm$0.07  &  $-$21.14$\pm$0.10 & 0.20$\pm$0.07 & $-$2.06$\pm$0.07 & 0.33$\pm$0.22 \\
Sharon/Johnson & $-$20.85$\pm$0.12 & 2.2$\pm$0.8 & $-$1.60$\pm$0.04 & $-$0.06$\pm$0.05  &  $-$21.15$\pm$0.09 & 0.19$\pm$0.05 & $-$2.06$\pm$0.06 & 0.26$\pm$0.17 \\
\textsc{GLAFIC} & $-$20.82$\pm$0.12 & 2.5$\pm$0.9 & $-$1.59$\pm$0.04 & $-$0.08$\pm$0.04  &  $-$21.11$\pm$0.11 & 0.22$\pm$0.09 & $-$2.03$\pm$0.08 & 0.16$\pm$0.16 \\
Keeton & $-$20.85$\pm$0.09 & 2.2$\pm$0.5 & $-$1.60$\pm$0.03 & $-$0.07$\pm$0.04  &  $-$21.14$\pm$0.09 & 0.19$\pm$0.05 & $-$2.05$\pm$0.06 & 0.21$\pm$0.15 \\
\textsc{grale} & $-$20.96$\pm$0.16 & 1.9$\pm$1.0 & $-$1.65$\pm$0.06 & 0.05$\pm$0.09  &  $-$21.21$\pm$0.08 & 0.17$\pm$0.04 & $-$2.13$\pm$0.05 & 0.56$\pm$0.19 \\
Diego & $-$20.88$\pm$0.16 & 2.2$\pm$1.3 & $-$1.62$\pm$0.05 & $-$0.04$\pm$0.06  &  $-$21.18$\pm$0.09 & 0.18$\pm$0.05 & $-$2.09$\pm$0.06 & 0.31$\pm$0.17 \\
 & \\
& \multicolumn{4}{c}{$z\sim4$} & \multicolumn{4}{c}{$z\sim8$}\\
\textsc{CATS} & $-$20.94$\pm$0.07 & 1.7$\pm$0.3 & $-$1.70$\pm$0.03 & $-$0.07$\pm$0.22  &  $-$20.95$\pm$0.23 & 0.084$\pm$0.078 & $-$2.24$\pm$0.10 & 0.37$\pm$0.27 \\
Sharon/Johnson & $-$20.94$\pm$0.07 & 1.6$\pm$0.3 & $-$1.69$\pm$0.03 & $-$0.15$\pm$0.16  &  $-$20.89$\pm$0.23 & 0.097$\pm$0.089 & $-$2.20$\pm$0.10 & 0.22$\pm$0.27 \\
\textsc{GLAFIC} & $-$20.94$\pm$0.07 & 1.6$\pm$0.3 & $-$1.70$\pm$0.03 & $-$0.22$\pm$0.11  &  $-$20.88$\pm$0.21 & 0.10$\pm$0.07 & $-$2.19$\pm$0.10 & 0.16$\pm$0.25 \\
Keeton & $-$20.94$\pm$0.07 & 1.6$\pm$0.3 & $-$1.69$\pm$0.03 & $-$0.19$\pm$0.16  &  $-$20.87$\pm$0.23 & 0.10$\pm$0.09 & $-$2.18$\pm$0.11 & 0.16$\pm$0.27 \\
\textsc{grale} & $-$20.99$\pm$0.10 & 1.3$\pm$0.4 & $-$1.72$\pm$0.04 & 0.08$\pm$0.18  &  $-$20.96$\pm$0.16 & 0.094$\pm$0.041 & $-$2.27$\pm$0.06 & 0.50$\pm$0.22 \\
Diego & $-$20.95$\pm$0.08 & 1.7$\pm$0.4 & $-$1.70$\pm$0.03 & $-$0.11$\pm$0.12  &  $-$20.99$\pm$0.21 & 0.077$\pm$0.059 & $-$2.27$\pm$0.09 & 0.55$\pm$0.28 \\
 & \\
& \multicolumn{4}{c}{$z\sim5$} & \multicolumn{4}{c}{$z\sim9$}\\
\textsc{CATS} & $-$21.13$\pm$0.10 & 0.73$\pm$0.13 & $-$1.78$\pm$0.04 & $-$0.05$\pm$0.21  &  $-$21.15$\pm$0.04 & 0.018$\pm$0.007 & $-$2.29$\pm$0.10 & 0.53$\pm$0.28 \\
Sharon/Johnson & $-$21.14$\pm$0.10 & 0.72$\pm$0.13 & $-$1.78$\pm$0.04 & $-$0.08$\pm$0.19  &  $-$21.15$\pm$0.08 & 0.018$\pm$0.013 & $-$2.29$\pm$0.12 & 0.54$\pm$0.28 \\
\textsc{GLAFIC} & $-$21.14$\pm$0.09 & 0.72$\pm$0.13 & $-$1.79$\pm$0.04 & $-$0.04$\pm$0.23  &  $-$21.15$\pm$0.06 & 0.020$\pm$0.012 & $-$2.26$\pm$0.12 & 0.48$\pm$0.24 \\
Keeton & $-$21.13$\pm$0.10 & 0.72$\pm$0.13 & $-$1.78$\pm$0.05 & $-$0.05$\pm$0.18  &  $-$21.14$\pm$0.09 & 0.018$\pm$0.012 & $-$2.26$\pm$0.11 & 0.57$\pm$0.40 \\
\textsc{grale} & $-$21.15$\pm$0.09 & 0.74$\pm$0.14 & $-$1.80$\pm$0.04 & 0.03$\pm$0.20  &  $-$21.15$\pm$0.06 & 0.021$\pm$0.018 & $-$2.34$\pm$0.15 & 0.52$\pm$0.34 \\
Diego & $-$21.13$\pm$0.09 & 0.74$\pm$0.13 & $-$1.78$\pm$0.04 & $-$0.09$\pm$0.22  &  $-$21.15$\pm$0.07 & 0.019$\pm$0.012 & $-$2.31$\pm$0.12 & 0.57$\pm$0.26 \\
& 
\enddata
\tablenotetext{a}{Best-fit curvature in the shape of the $UV$ LF faintward of $-16$ mag (\S3.2).}
\end{deluxetable*}

\section{B.  Sensitivity of faint-end slope determinations to the Probed $UV$ Luminosity Range}

In deriving constraints on the faint-end slope $\alpha$ of the $UV$ LF
from lensed sources behind the HFF clusters and comparing this slope
with faint-end slope determinations derived from blank-field studies,
an important question is how comparable these two faint-end slopes
are.  If the faint end of the $UV$ LF showed a modest departure from a
power-law form, one would expect slight differences in the derived
slopes.

To evaluate how large such differences might be, we make use of some
of the model $UV$ LFs discussed in \S\ref{sec:theory} and then
characterize the differences in derived faint-end slope $\alpha$ based
on the magnitude range over which the $UV$ LF is derived.  For this
exercise, faint-end slope determinations from blank-field and lensing
cluster studies are assumed to occur from $-16.5$ mag to $-18.5$ mag
and from $-18.0$ mag to $-15.0$ mag, respectively.  In characterizing
the impact that the luminosity range can have in deriving the
faint-end slope $\alpha$, a characteristic luminosity $M_{UV} ^{*} =
-21$ is assumed and the impact that this would have on the power-law
slope over the luminosity ranges described is removed ($\Delta\alpha
\sim 0.02$).

\begin{figure}
  \epsscale{0.6}
  \plotone{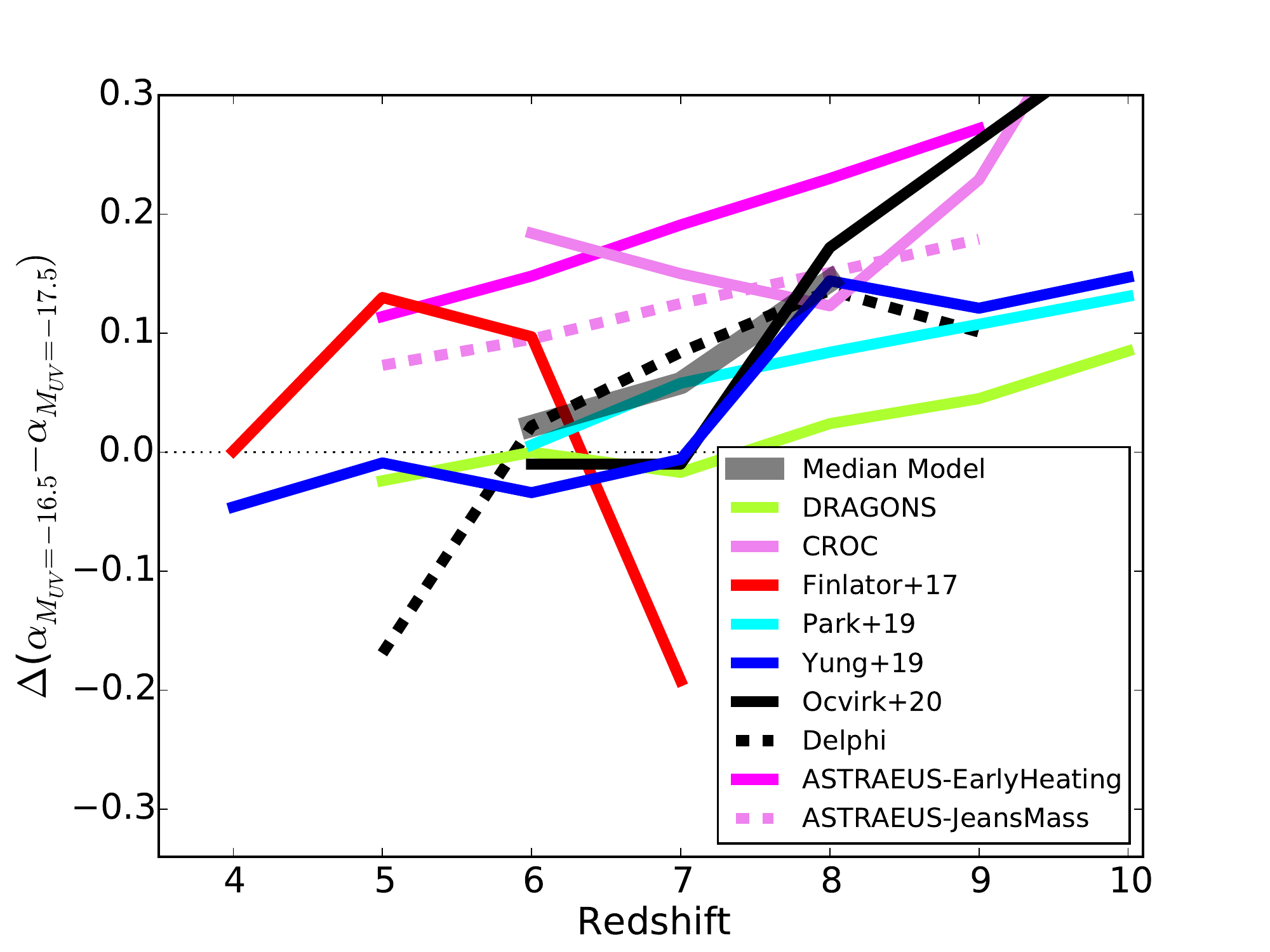}
\caption{Differences between the effective faint-end slope $\alpha$ of
  the $UV$ LF measured over the $UV$ luminosity range $M_{UV}$ $-18.0$
  mag to $-15.0$ mag and over the range $M_{UV}$ $-18.5$ to $-16.5$
  mag.  The former luminosity range is more indicative of faint-end
  slope determinations from lensing cluster studies like the present
  one and the latter range is more indicative of determinations from
  blank-field studies that utilize HUDF samples (e.g., B21a).  The
  presented results are based on the model LFs shown in
  Figures~\ref{fig:lf29theory}-\ref{fig:lf29theory2} and are shown as
  a function of redshift.  The median difference $\Delta \alpha$ seen
  across all of the model LFs is shown as the thick shaded black line
  and appears to lie in the range
  $\sim0.00$-0.15.\label{fig:alpha_dep}}
\end{figure}

Figure~\ref{fig:alpha_dep} shows the differences in the faint-end
slope determinations depending on which luminosity range is utilized.
The thick shaded line shows the median result.  While there is a
modest amount of scatter, model to model, the model results suggest
LFs might show slightly shallower slopes, i.e.,
$\Delta\alpha\sim0.0$-0.2 if probed from a lensing cluster study than
from blank fields.  Nevertheless, given how small the difference is
and that it is dependent on which LF model we consider, we will not
consider it further, but to note that a $\Delta \alpha \sim 0.1$
offset may be relevant for such a comparison.

\end{document}